\documentclass[12pt,notitlepage]{article}
\usepackage[margin=1in]{geometry}

\usepackage{amsmath}
\usepackage{amssymb}
\usepackage{enumerate}
\usepackage{float}
\usepackage[utf8]{inputenc}
\usepackage{comment}
\usepackage{tabularx}
\usepackage{xcolor}
\usepackage{bm}
\usepackage{setspace}
\usepackage{hyperref}
\usepackage{tocloft}
\usepackage{graphicx}
\usepackage{booktabs}
\usepackage{multirow}
\usepackage{array}
\usepackage{subcaption}
\usepackage{tikz}
\usetikzlibrary{angles,quotes,calc}
\usetikzlibrary{decorations.pathreplacing}
\usepackage{mathtools}
\usepackage{adjustbox}

\newlistof{appsec}{appsec}{Online Appendix Contents}

\newcommand{\listofappendixcontents}{%
	\listof{appsec}{Online Appendix}%
}

\newcommand{\appsection}[1]{%
	\section{#1}%
	\addcontentsline{appsec}{section}{\protect\numberline{\thesection}#1}%
}

\newcommand{\appsubsection}[1]{%
	\subsection{#1}%
	\addcontentsline{appsec}{subsection}{\protect\numberline{\thesubsection}#1}%
}

\newcommand{\appsubsubsection}[1]{%
	\subsubsection{#1}%
	\addcontentsline{appsec}{subsubsection}{\protect\numberline{\thesubsubsection}#1}%
}

\allowdisplaybreaks

\usepackage[bottom]{footmisc}
\newtheorem{ass}{Assumption}

\renewcommand{\thesection}{\arabic{section}}
\renewcommand{\thesubsection}{\arabic{section}.\arabic{subsection}}
\renewcommand{\thesubsubsection}{\arabic{section}.\arabic{subsection}.\arabic{subsubsection}}

\newcolumntype{L}[1]{>{\raggedright\let\newline\\\arraybackslash\hspace{0pt}}m{#1}}
\newcolumntype{C}[1]{>{\centering\let\newline\\\arraybackslash\hspace{0pt}}m{#1}}

\usepackage{natbib}
\newcommand{\E}{\mathbb{E}}

\newcommand{\Var}{\mathbb{V}}
\newcommand{\Cov}{\mathbb{C}}

\newcommand{\Prob}{\mathbb{P}}

\makeatletter
\DeclareFontFamily{U}{mathx}{\hyphenchar\font45}
\DeclareFontShape{U}{mathx}{m}{n}{
	<-> mathx10
}{}
\DeclareSymbolFont{mathx}{U}{mathx}{m}{n}

\DeclareMathAccent{\widecheck}{0}{mathx}{"71}
\makeatother

\hypersetup{
	pdftitle={RDDExtrapolation},    
	pdfauthor={Austin Feng and Francesco Ruggieri},     
	pdfnewwindow=true,      
	colorlinks=true,       
	linkcolor=black,          
	citecolor=blue,        
	filecolor=black,      
	urlcolor=blue           
}

\begin{document}
	
	\title{
		\makebox[\textwidth][c]{
			\begin{tabular}{c}
				Structural Extrapolation in Regression Discontinuity Designs\\
				with an Application to School Expenditure Referenda\thanks{We thank St\'ephane Bonhomme, Morten Grindaker, Lancelot Henry de Frahan, Hugo Lopez, Marco Loseto, Federico Mainardi, Magne Mogstad, Kirill Ponomarev, and Alex Torgovitsky for helpful comments. We are grateful to the Becker Friedman Institute for Economics at the University of Chicago for granting us access to the L2 Voter Registration data.}
			\end{tabular}
		}
	}
	
	\author{Austin Feng\thanks{Carnegie Mellon University, Department of Statistics \& Data Science. Email: \href{mailto:austinfe@andrew.cmu.edu}{austinfe@andrew.cmu.edu}} \qquad \qquad Francesco Ruggieri\thanks{University of Chicago, Kenneth C. Griffin Department of Economics. Email: \href{mailto:ruggieri@uchicago.edu}{ruggieri@uchicago.edu}}}
	
	\date{\today}
	\maketitle
	
	\medskip
	
	\begin{abstract}
		We propose a structural approach to extrapolate average partial effects away from the cutoff in regression discontinuity designs (RDDs). Our focus is on applications that exploit closely contested school district referenda to estimate the effects of changes in education spending on local economic outcomes. We embed these outcomes in a spatial equilibrium model of local jurisdictions in which fiscal policy is determined by majority rule voting. This integration provides a microfoundation for the running variable---the share of voters who approve a ballot initiative---and enables identification of structural parameters using RDD coefficients. We then leverage the model to simulate the effects of counterfactual referenda over a broad range of proposed spending changes. These scenarios imply realizations of the running variable away from the threshold, allowing extrapolation of RDD estimates to nonmarginal referenda. Applying the method to school expenditure ballot measures in Wisconsin, we document substantial heterogeneity in housing price capitalization across the approval margin.
	\end{abstract}
	
	\setcounter{page}{0}\thispagestyle{empty}
	
	\newpage
	
	\section{Introduction}
	
	Regression discontinuity designs (RDDs) are widely used to estimate causal parameters in settings where the probability of exposure to a treatment changes discontinuously at a known deterministic threshold. Under a mild continuity assumption, RDDs nonparametrically point identify a cutoff-specific average treatment effect (ATE) or local average treatment effect (LATE), depending on whether this probability shifts from zero to one or by a smaller amount within the unit interval, respectively (\citealt{hahnetal2001}). A well-known limitation of RDDs concerns their external validity: the average effect identified for agents at the margin between different levels of treatment exposure may not generalize to nonmarginal units, potentially narrowing the policy relevance of the empirical analysis.
	
	To address this limitation, the existing literature has proposed a variety of approaches that can be grouped into two main strands (\citealt{cattaneoetal2021}). The first strand has developed methods for extrapolating effects away from the threshold imposing restrictions on the joint distribution of the outcome, the running variable, and one or more cutoffs (\citealt{donglewbel2015}, \citealt{bertanhaimbens2020}, \citealt{cattaneoetal2021}). The second strand has incorporated information from auxiliary variables observed alongside the RD design but not intrinsic to it (\citealt{meallirampichini2012}, \citealt{wingcook2013}, \citealt{angristrokkanen2015}, \citealt{Rokkanen2015}). While theoretically valid for extrapolating treatment effects, these statistical approaches offer limited traction for policy-relevant counterfactuals that require manipulating the underlying determinants of the running variable and the outcome.
	
	In this paper, we propose a structural approach that hinges on developing an economic model of the running variable. We focus on the empirically relevant setting of school expenditure referenda, which local public finance and education economists frequently exploit to estimate the effects of school capital investments on student test scores and housing prices (\citealt{cfr2010}, \citealt{darolia2013}, \citealt{hongzimmer2016}, \citealt{martorelletal2016}, \citealt{abottetal2020}, \citealt{baron2022}, \citealt{rohlinetal2022}, \citealt{baronetal2024}, \citealt{bilaschon2024}). School districts in which a ballot measure is narrowly approved or rejected may systematically differ from those in which a similar initiative passes or fails by a large margin, due to different political preferences, housing market characteristics, or the prior burden of property taxes. Consequently, the average treatment effect at the cutoff may not coincide with the same parameter away from the threshold.
	
	Our extrapolation strategy proceeds in two steps. First, we interpret the average partial effects identified by the RDD within a spatial equilibrium model that jointly determines population, housing prices, tax rates, and school spending across local jurisdictions. In this model, households vote on their preferred combination of school spending and property taxes, providing an economic foundation for the running variable: the share of voters who approve a ballot initiative to increase education expenditures. By mapping the RDD coefficients to the model, we recover structural parameters consistent with the reduced-form treatment effects at the cutoff. In the second step, we leverage the model to simulate counterfactual referenda spanning a broad range of proposed spending changes. These policy alternatives generate realizations of the running variable away from the threshold, enabling extrapolation of average partial effects across the approval margin.
	
	We apply our method to revisit the capitalization of school expenditure authorizations into housing prices in the years following referendum approval, a margin long studied in public finance as a measure of how prospective homebuyers value the tradeoff between improved public services and higher property taxes\footnote{See \cite{rossyinger1999} for a review of the earlier literature.}. Our empirical analysis focuses on Wisconsin, where school district referenda are regularly used to authorize both operational and capital spending (\citealt{baron2022}). In the first part of the paper, we estimate that the average arc elasticity of housing prices with respect to education expenditures is approximately equal to one at the referendum approval threshold. In the second part, we apply our extrapolation strategy, grounded in a spatial equilibrium framework, to estimate capitalization effects for referenda that were either decisively approved or rejected. We find substantial heterogeneity away from the cutoff. To the right of the threshold, the average elasticity rises steadily, implying that the positive capitalization effect estimated at the margin extends to referenda backed by a larger share of voters. To the left, the elasticity declines and becomes negative for ballot measures that garnered limited support. This pattern indicates that some rejected proposals would have weakened local housing demand through negative net sorting, driven primarily by outmigration of households with a relatively low willingness to pay for enhanced education services. Taken together, these results suggest that housing market responses to locally determined changes in government spending may vary systematically with the degree of voter agreement over the proposed policies.
	
	We view the economic approach developed in this paper as particularly useful for empirical settings in which the running variable is not policy-relevant and lacks a clear interpretation in terms of economic primitives. In our application, the approval vote share reflects a complex aggregation of individual preferences over education spending, shaped by local sorting and housing market dynamics. Because manipulating the vote share does not correspond to a well-defined policy intervention, we instead extrapolate average effects away from the cutoff by shifting a policy lever: the proposed change in government spending. This variable affects both the referendum outcome and the endogenous variables in the model, making it a valid basis for extrapolation. A salient feature of our approach is that it allows for a clear separation between the variation used to identify average effects at the cutoff and the variation exploited for extrapolation. While the running variable may be policy-relevant in some contexts, this distinction becomes essential when it is not---as in our case.
	
	Our paper contributes to three distinct but related literatures. First, we propose a method for extrapolating average partial effects away from the cutoff in regression discontinuity designs. Departing from existing statistical approaches, we microfound the running variable by embedding it in a model of economic behavior, thereby establishing a formal link between RDD coefficients and structural parameters. In this respect, our paper is related to \cite{mehta2019}, which develops a framework for extrapolation in settings where the cutoff is chosen by a planner seeking to maximize the net benefits of a policy intervention. Our approach differs in that we specify a structural model in which a policy variable can be explicitly manipulated, rather than assuming that a feature of the RDD reflects an optimal design choice by the policymaker. Our framework enables researchers to estimate causal parameters associated with a specified range of counterfactual changes in education spending.
	
	Second, we contribute to the literature that links program evaluation methods with models of economic behavior to recover policy-relevant parameters (\citealt{heckmanvytlacil2001}, \citealt{heckmanvytlacil2005}, \citealt{heckmanvytlacil2007part2}, \citealt{mst2018}, \citealt{walters2018}, \citealt{roseshemtov2021}, \citealt{mogstadtorgowalters2024}) and welfare measures (\citealt{chetty2009}, \citealt{suarezzidar2016}, \citealt{kleven2021}, \citealt{tebalditorgovitskyyang2023}, \citealt{lobel2025}). Third, we extend prior work in public finance that estimates the effects of local changes in government spending on housing prices (\citealt{oates1969}, \citealt{gyourkotracy1989}, \citealt{reback2005}, \citealt{hilbermayer2009}, \citealt{cfr2010}, \citealt{hilber2017}) and population composition (\citealt{ferreyra2007}, \citealt{banzhafwalsh2008}, \citealt{abramitzky2009}, \citealt{bilaschon2024}). We show that the capitalization of education spending into housing prices varies across the approval margin, implying that conclusions drawn from marginally approved referenda may not generalize to nonmarginal settings. This underscores the value of structural extrapolation for evaluating counterfactuals of practical policy interest.
	
	The remainder of the paper is structured as follows. Section~\ref{sec_background} provides institutional background on the use of local referenda in U.S. school districts and reviews how prior research in local public finance has leveraged these settings to estimate causal parameters. In Section~\ref{sec_effects_cutoff}, we implement a regression discontinuity design to estimate the average effect of marginally approved school expenditure authorizations on housing prices in Wisconsin. Section~\ref{sec_model} introduces a spatial equilibrium model in which residents select into participation in school funding referenda. In Section~\ref{sec_identification}, we establish a formal link between the reduced-form RDD estimates and the structural parameters of the model, providing conditions under which these parameters are identified. Section~\ref{sec_extrapolation} describes how the model structure can be used to extrapolate average effects away from the threshold. In Section~\ref{sec_simulation}, we present a Monte Carlo simulation that assesses the finite-sample performance of our approach for recovering the model's structural parameters from RDD coefficients. Section~\ref{sec_empappl} applies the proposed extrapolation method to estimate the capitalization effects of expenditure changes for nonmarginal referenda in Wisconsin. Section~\ref{sec_concl} concludes.

	\section{Background}\label{sec_background}
	
	School districts fund a significant portion of their operations with revenue from property taxes, accounting for more than 80 percent of their receipts from local sources (\citealt{nces_table23510_2023}). However, state constitutions often impose caps on tax rates, annual growth in tax revenue, or annual growth in assessed property values, thereby constraining the extent to which school districts and other local governments can tax their base (\citealt{lincolntaxlimits}). In several states, school districts can bypass these constraints if a majority of voters approves a spending initiative in a local referendum. These ballot initiatives are often intended to fund large capital expenditures, such as school construction or renovation projects (\citealt{fischer_duncombe_syverson_2023}). If a referendum is approved, a school district will typically issue general obligation bonds and repay the principal and interest over a predetermined number of years using extra property tax revenue.
	
	Beginning with the seminal contribution of \cite{cfr2010}, researchers in empirical public finance and education economics have leveraged school bond referenda to estimate the effects of increased school expenditures on housing prices, student achievement, and other educational outcomes. Identifying causal parameters in this context is inherently challenging: school district property tax rates are likely to be systematically related to unobserved determinants of both educational and housing market outcomes. For example, households that place a high value on public education may be more likely to sort into well-funded districts (\citealt{poterba1997}) and to invest more heavily in their children's academic success outside of school (\citealt{guryanhurstkearney2008}). Similarly, cross-district heterogeneity in property tax rates may reflect unobserved differences in housing market fundamentals: areas with better natural amenities may both command higher housing prices and enable greater fiscal extraction by local governments (\citealt{bruecknerneumark2014}, \citealt{diamond2017}). Because property tax rates are equilibrium outcomes of collective choice, simple comparisons of conditional means generally lack a causal interpretation. Regression discontinuity designs that exploit variation in the outcome of school spending referenda address these concerns by comparing jurisdictions that narrowly approved or narrowly rejected their respective ballot measures.
	
	In this paper, we focus on the effect of school district referendum approval on housing prices. Research in local public finance has long recognized that the extent to which property tax changes are capitalized into housing prices is an informative measure of the efficiency of local public goods provision (\citealt{bickerdike1902}, \citealt{marshall1948}, \citealt{oates1969}, \citealt{brueckner1982}, \citealt{cushing1984}, \citealt{barrowrouse2004}, \citealt{figliolucas2004}, \citealt{cfr2010}, \citealt{bilaschon2024}). Evidence of positive capitalization following an expenditure increase is typically interpreted as an indication that prospective homebuyers value the associated improvement in public services more than the change in the tax burden required to fund it. Our setting is the state of Wisconsin, where school districts routinely hold referenda to authorize both operational and capital expenditures (\citealt{baron2022}).
	
	\section{The Effect of School Expenditure Authorizations on Housing Prices for Marginally Approved Referenda}\label{sec_effects_cutoff}
	
	In this section, we leverage a regression discontinuity design to estimate the effect of school expenditure authorizations on housing prices for marginally approved referenda in Wisconsin.
	
	\subsection{Data}\label{sec_data_avgval}
	
	The Wisconsin Department of Public Instruction collects and publishes comprehensive data on all school district referenda held in the state since 1990 (\citealt{wiscreferenda}). This dataset includes, among other variables, information on the approval vote share, which we use as the running variable in the RDD.
	
	To construct the outcome of interest, we follow an approach similar to that of \cite{bilaschon2024} and \cite{dyndisc2025}. Specifically, we rely on a repeat-sales house price index developed by \cite{contatlarson2024}, which covers all Census tracts located within Core-Based Statistical Areas\footnote{The term ``Core-Based Statistical Area'' refers collectively to both Metropolitan Statistical Areas and Micropolitan Statistical Areas (\citealt{censusglossary}).} in the United States from 1989 to 2021. The index is normalized to 100 in 1989 for all tracts, allowing for within-tract temporal comparisons but not cross-sectional ones. To allow for level comparisons across school districts, we incorporate data on the average value of owner-occupied single-family homes at the Census tract level, as reported in the U.S. Census Bureau's 2000 Decennial Census\footnote{The collection of this variable was discontinued beginning with the 2010 Decennial Census.}. For each tract, we compute a calibration factor as the ratio of the 2000 Census home value to the 2000 value of the house price index from \cite{contatlarson2024}, and apply this factor to the full time series of the index. The resulting measure of housing prices allows for both cross-sectional and intertemporal comparisons. Next, we compute the centroid of each Census tract and assign it to the corresponding elementary, secondary, or unified school district based on the 2010 TIGER/Line shapefiles provided by the U.S. Census Bureau\footnote{We use 2010 tract and school district boundaries because the house price index constructed by \cite{contatlarson2024} is based on 2010 Census tracts.} (\citealt{censustiger}). Finally, for each district, we calculate a population-weighted average of housing prices across its constituent Census tracts\footnote{Although we compute population-weighted averages, this choice is not consequential, as Census tracts are designed to contain approximately 4,000 inhabitants (\citealt{censusglossary}).}. This yields the outcome variable used in the RDD.
	
	\begin{figure}[H]
		\begin{center}
			\caption{Density of the Approval Vote Share Margin}\label{fig_runningvar}
			\vspace{0mm}
			\includegraphics[width=0.8\textwidth,height=\textheight,keepaspectratio]{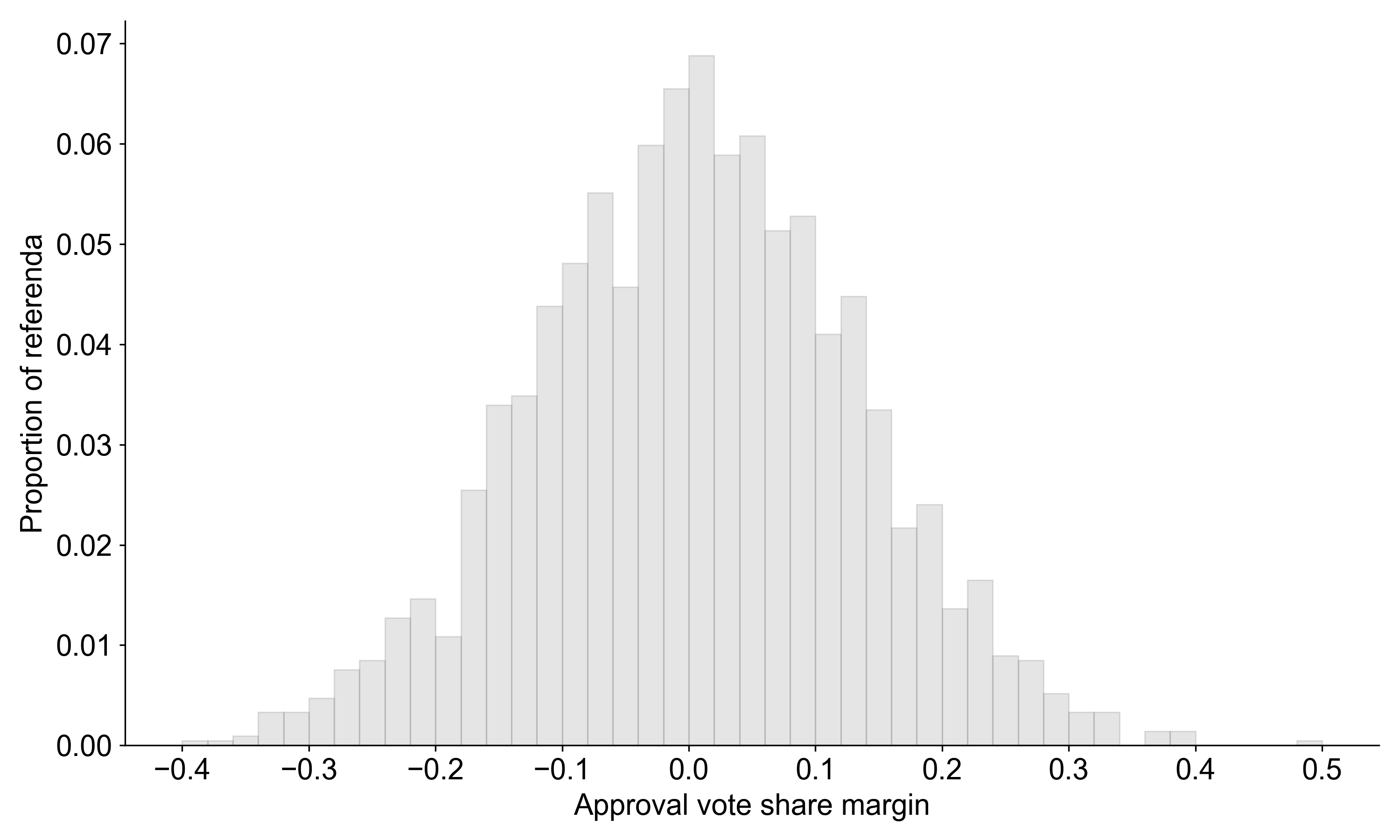}
		\end{center}
		\vspace{-3mm}
		\begin{footnotesize}
			\begin{spacing}{1}
				\noindent
				\textsc{Notes}: This figure displays a histogram of the approval vote share margin, defined as the difference between the share of votes in favor of the proposed expenditure measure and the 50 percent approval threshold, for 3,528 referenda held by Wisconsin school districts between 1990 and 2022.
			\end{spacing}
		\end{footnotesize}
	\end{figure}

	\vspace{-3mm}

	The matched sample includes 3,528 referenda, of which 58.8 percent were approved. The average approval vote share margin is 2.22 percentage points, with an average of 10.5 percentage points among approved referenda and –9.59 percentage points among those that were rejected. Figure \ref{fig_runningvar} displays a histogram of the approval vote share margin. To assess the validity of the design, we test for discontinuities in the density of the running variable at the cutoff using the local polynomial density estimators developed by \cite{cattaneojanssonma2020}. The null hypothesis of equal densities on either side of the cutoff is not rejected ($p$-value = 0.82), suggesting that manipulation of the running variable around the threshold is unlikely to be a concern in this setting.
	
	\subsection{Identification and Estimation of Average Effects at the Cutoff}\label{sec_identification_cutoff}
	
	Let $P_j$ and $S_j$ denote, respectively, the housing price and the approval vote share margin in school district $j$. Each referendum proposes a deterministic and binding expenditure increase $\Delta G_j > 0$, which is known to residents prior to voting. Define the approval indicator as $D_j \equiv \mathbb{I} \left [ S_j > 0 \right ]$, where $D_j = 1$ if the referendum is approved. Initially, we adopt a potential outcomes model in which $P_j \left ( d \right )$ denotes the potential housing price in district $j$ under treatment status $d \in \left \{ 0,1 \right \}$.
	
	Our primary target parameter is the average treatment effect of referendum approval on log housing prices at the threshold:
	{\setlength{\abovedisplayskip}{7pt plus 2pt minus 2pt}
	\setlength{\belowdisplayskip}{7pt plus 2pt minus 2pt}
	\begin{equation}
		\mathrm{ATE} \left ( 0 \right ) \equiv \E \left [ \log P_j \left ( 1 \right ) - \log P_j \left ( 0 \right ) | S_j = 0 \right ]
	\end{equation}}
	This parameter is nonparametrically point identified under a standard continuity assumption (\citealt{hahnetal2001}). 
	\begin{ass}[Continuity at the Cutoff]\label{ass_cont}
		\sloppy For each $d \in \left \{ 0,1 \right \}$, the function $s \mapsto \E \left [ \log P_j \left ( d \right ) | S_j = s \right ]$ is continuous at $s=0$.
	\end{ass}
	Under Assumption \ref{ass_cont}, $\mathrm{ATE} \left ( 0 \right )$ is identified via the sharp regression discontinuity estimand
	{\setlength{\abovedisplayskip}{7pt plus 2pt minus 2pt}
		\setlength{\belowdisplayskip}{7pt plus 2pt minus 2pt}
	\begin{equation}\label{eq_sharp_rd_estimand}
		\theta \left ( 0 \right ) \equiv \lim_{s \downarrow 0} \E \left [ \log P_j | S_j = s \right ] - \lim_{s \uparrow 0} \E \left [ \log P_j | S_j = s \right ]
	\end{equation}}

	However, $\theta \left ( 0 \right )$ is difficult to interpret when proposed expenditure changes $\Delta G_j$ vary across referenda. A binary treatment may obscure meaningful variation in the intensity of the underlying policy intervention. To recover a more interpretable, elasticity-like parameter, we normalize the sharp RD estimand by the average realized change in expenditures at the cutoff. Specifically, we consider the variable $D_j \times \Delta \log G_j$, which equals the proposed change in log spending for approved referenda and is zero otherwise. Since $\Delta G_j$ is known to voters prior to the election and is binding upon approval, the resulting first stage is deterministic. The corresponding estimand takes the form of a fuzzy regression discontinuity estimand with a known first-stage shift:
	\begin{equation}\label{eq_fuzzy_rd_estimand}
		 \theta^\mathrm{G} \left ( 0 \right ) \equiv \frac{\lim_{s \downarrow 0} \E \left [ \log P_j | S_j = s \right ] - \lim_{s \uparrow 0} \E \left [ \log P_j | S_j = s \right ]}{\lim_{s \downarrow 0} \E \left [ D_j \times \Delta \log G_j | S_j = s \right ]}
	\end{equation}
	where the lower-limit expectation in the denominator is omitted, as it equals zero by construction. To interpret this parameter, we redefine potential outcomes as functions of the realized change in spending, writing $P_j \left ( d \times \Delta G_j \right )$ for $d \in \left \{ 0,1 \right \}$. We now adapt the continuity assumption to this setting.
	\begin{ass}[Continuity at the Cutoff]\label{ass_cont2}
		\sloppy For each $d \in \left \{ 0,1 \right \}$, the functions $s \mapsto \E \left [ \log P_j \left ( d \times \Delta G_j \right ) | S_j = s \right ]$ and $s \mapsto \E \left [ \Delta \log G_j | S_j = s \right ]$ are continuous at $s=0$.
	\end{ass}
	Under Assumption \ref{ass_cont2}, the estimand $\theta^\mathrm{G} \left ( 0 \right )$ identifies a weighted average of housing price arc elasticities with respect to proposed changes in school expenditures among jurisdictions at the approval threshold:
	\begin{equation}\label{eq_avg_ape}
		\mathrm{WAVE} \left ( 0 \right ) \equiv \E \left [ \omega_j \times \frac{\log P_j \left ( \Delta G_j \right ) - \log P_j \left ( 0 \right )}{\Delta \log G_j} \bigg | S_j = 0 \right ]
	\end{equation}
	where weights are defined as $\omega_{j} \equiv \Delta \log G_j \slash \E \left [ \Delta \log G_j | S_j = 0 \right ]$, ensuring they integrate to one at the cutoff. This result is proved in Appendix \ref{appx_proof_estimand}.

	For estimation, we implement local polynomial regression. Given a random sample $\left \{ \left [ S_{j}, P_{j}, \Delta G_{j} \right ]' \right \}_{j=1}^{n}$ and a bandwidth $h_n > 0$, let $\mathcal{S} \left ( h_n \right ) = \left [ -h_n, h_n \right ]$ be a discontinuity window implied by realizations of the running variable around the zero cutoff. Let $\mathcal{S}^{-} \left ( h_n \right ) = \left [ -h_n, 0 \right )$ and $\mathcal{S}^{+} \left ( h_n \right ) = \left [ 0, h_n \right ]$ indicate, respectively, the left and right discontinuity half-windows. For any outcome $A$, we estimate intercepts via local linear regression:
	\begin{align*}
		\left [ \widehat{\mu}^{(0)}_{A+,1} \left ( h_n \right ), \widehat{\mu}^{(1)}_{A+,1} \left ( h_n \right ) \right ]' & \equiv \arg \min_{b_0,b_1 \in \mathbb{R}} \sum_{j=1}^{n} \mathbb{I} \left [ S_{j} \in \mathcal{S}^{+} \left ( h_n \right ) \right ] \left ( \log A_{j} - b_0 - b_1 S_{j} \right )^2 k_{h_n} \left ( S_{j} \right ) \\
		\left [ \widehat{\mu}^{(0)}_{A-,1} \left ( h_n \right ), \widehat{\mu}^{(1)}_{A-,1} \left ( h_n \right ) \right ]' & \equiv \arg \min_{b_0,b_1 \in \mathbb{R}} \sum_{j=1}^{n} \mathbb{I} \left [ S_{j} \in \mathcal{S}^{-} \left ( h_n \right ) \right ] \left ( \log A_{j} - b_0 - b_1 S_{j} \right )^2 k_{h_n} \left ( S_{j} \right )
	\end{align*}
	where $k_{h_n} \left ( S_j \right ) = \left ( 1 - | S_j | \slash h_n \right ) \slash h_n$ is the triangular kernel. Assuming standard regularity conditions hold (see Assumptions 1 and 2 in \citealt{cct2014}), we estimate $\mathrm{WAVE} \left ( 0 \right )$ with
	\begin{equation}
		\widehat{\theta}^{\mathrm{G}} \left ( 0, h_n \right ) \equiv \frac{\widehat{\mu}^{(0)}_{P+,1} \left ( h_n \right ) - \widehat{\mu}^{(0)}_{P-,1} \left ( h_n \right )}{\widehat{\mu}^{(0)}_{\Delta G+,1} \left ( h_n \right )}
	\end{equation}
	We compute $\widehat{\theta}^{\mathrm{G}} \left ( 0, h_n \right )$ using a bandwidth selected to minimize the estimator’s mean squared error, following \citet{ik2012}. To account for estimation with observations away from the cutoff, we apply standard bias correction and construct nonparametric confidence intervals with the method developed by \citet{cct2014}. We compute standard errors using the nearest-neighbor variance estimator from the same paper, adopting the default tuning parameter $j^* = 3$.
	
	\subsection{Results}
	
	We estimate the effect of school district expenditure authorization on housing prices measured five years after each referendum. Because of the temporal lag between treatment assignment and outcome measurement, additional referenda may occur during the intervening period. The difficulty of identifying interpretable causal parameters in settings where jurisdictions are subject to repeated treatment assignments over time was first highlighted by \citet{cfr2010} and has since become an important concern in empirical local public finance. A growing body of research in applied econometrics has developed identification strategies tailored to such environments, commonly referred to as dynamic regression discontinuity designs (\citealt{cfr2010}, \citealt{hsushen2024}, \citealt{dyndisc2025}). In this paper, we do not adopt these dynamic RD approaches. Our focus lies instead on developing a framework to extrapolate average partial effects away from the approval cutoff. Consequently, we interpret our estimand as an intent-to-treat effect generated by the discontinuity in the referendum approval margin.
	
\begin{figure}[H]
		\begin{center}
			\caption{Binned Averages of Log Housing Prices by Approval Vote Share Margin}\label{fig_rdplot_price}
			\vspace{0mm}
			\includegraphics[width=0.8\textwidth,height=\textheight,keepaspectratio]{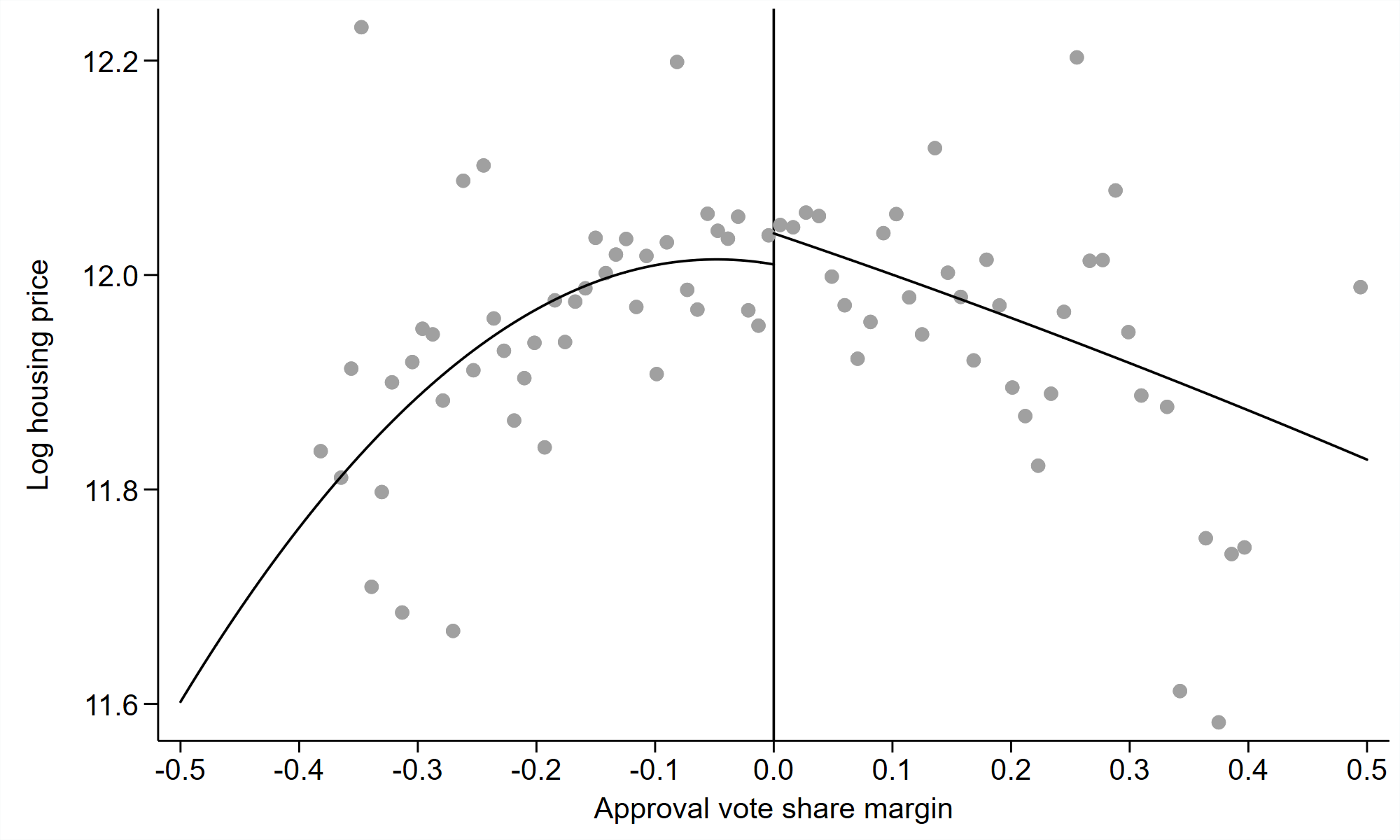}
		\end{center}
		\vspace{-3mm}
		\begin{footnotesize}
			\begin{spacing}{1}
				\noindent
				\textsc{Notes}: This figure displays nonparametric estimates of average log housing prices at the school district level, binned by the approval vote share margin. Fitted values are obtained from global quadratic regressions estimated separately on each side of the cutoff. The number and spacing of bins are selected using spacing estimators, following the data-driven procedure proposed by \cite{cct2015_rdplot}. The approval vote share margin is defined as the difference between the share of votes in favor of the proposed expenditure measure and the 50 percent approval threshold. Housing prices are measured five years after each referendum. The sample comprises 2,122 referenda held by Wisconsin school districts between 1990 and 2015.
			\end{spacing}
		\end{footnotesize}
	\end{figure}

	We begin by describing the empirical distribution of log housing prices at the school district level, conditional on the approval vote margin. Figure~\ref{fig_rdplot_price} displays nonparametric estimates of average housing prices within bins implied by the running variable. The figure reveals a concave relationship: districts with referenda that are either overwhelmingly approved or rejected tend to exhibit lower average housing prices relative to those near the cutoff. To the right of the threshold, average housing prices are modestly higher, consistent with the positive capitalization of marginally approved expenditure authorizations into property values.
	
	\vspace{-3mm}
	
	We then turn to formally estimating the effect of referendum approval on housing prices. Panel A of Table~\ref{table_rd_avgval_5} reports local linear estimates of $\mathrm{ATE}(0)$, the average treatment effect at the approval threshold. In the baseline specification, marginal approval increases housing prices by an estimated 10.1 percent over a five-year horizon. Controlling for referendum year indicators slightly reduces the estimate to 9.7 percent. The inclusion of interaction indicators between school district type (elementary, secondary, or unified) and referendum content (operational or capital expenditures) does not meaningfully affect the estimate, which rises to 10.5 percent.
	
	Panel B rescales the estimates from Panel A by the average realized change in school district expenditures at the cutoff, yielding an estimate of the weighted average of arc elasticities defined in equation~\eqref{eq_avg_ape}. Across all specifications, a one-percent increase in authorized school spending---induced by referendum approval---raises housing prices by approximately one percent over five years.
	
	\begin{table}[H]
		\begin{center}
			\caption{Estimated Effects of Referendum Approval on Housing Prices}\label{table_rd_avgval_5}
			\vspace{0mm}
			\begin{tabular}{>{\raggedright\arraybackslash}p{5cm}*{3}{>{\centering\arraybackslash}p{1.8cm}}}
	\toprule\toprule
	& (1) & (2) & (3) \\
	\midrule
	\addlinespace
	\multicolumn{4}{l}{\textit{Panel A: Estimates of }$\text{ATE} \left ( 0 \right )$} \\
	\addlinespace
	$D_j$ 	& 0.101 & 0.097 & 0.105 \\
			& (0.045) & (0.039) & (0.035) \\
	\addlinespace
	\midrule
	\addlinespace
	\multicolumn{4}{l}{\textit{Panel B: Estimates of }$\text{WAVE} \left ( 0 \right )$} \\
	\addlinespace
	$D_j \times \Delta \log G_j$ 	& 1.032 & 0.979 & 1.064 \\
								& (0.485) & (0.426) & (0.386) \\
	\addlinespace
	\midrule
	\addlinespace
	Year & No & Yes & Yes \\
	District-Referendum Type & No & No & Yes \\
	Bandwidth & 0.097 & 0.084 & 0.080 \\
	\bottomrule\bottomrule
\end{tabular}
		\end{center}
		\vspace{-1mm}
		\begin{footnotesize}
			\begin{spacing}{1}
				\noindent
				\textsc{Notes:} This table reports local linear estimates of the average effect of approving school district expenditure referenda on housing prices in Wisconsin from 1990 to 2015. Panel A presents bias-corrected estimates of the average treatment effect at the cutoff; the corresponding estimand is defined in equation~\eqref{eq_sharp_rd_estimand}. Panel B presents bias-corrected estimates of a cutoff-specific, weighted average of arc elasticities with respect to expenditure changes, with the estimand given in equation~\eqref{eq_fuzzy_rd_estimand}. In both panels, estimation relies on a triangular kernel, with bandwidths selected to minimize the mean squared error of the estimator (\citealt{ik2012}), following the procedure developed by \cite{cct2014}. Standard errors are computed using the nearest-neighbor variance estimator proposed by \cite{cct2014}, with the default tuning parameter $j^* = 3$. The ``Year'' and ``District-Referendum Type'' rows indicate whether the specification includes referendum year indicators and interaction indicators between school district type (elementary, secondary, or unified) and referendum content (operational or capital expenditures), respectively. The sample comprises 2,122 referenda.
			\end{spacing}
		\end{footnotesize}
	\end{table}
	
	\vspace{-3mm}
	
	Overall, the results indicate that marginally approved referenda lead to a positive capitalization of school expenditure authorizations into local housing prices, consistent with prior evidence from \citet{cfr2010} and \citet{bilaschon2024}. However, the sign and magnitude of this effect need not generalize to nonmarginal referenda, potentially limiting the relevance of these estimates for evaluating the welfare implications of changes in school spending.  The concave pattern in Figure~\ref{fig_rdplot_price} suggests that jurisdictions with decisive referendum outcomes---whether approvals or rejections---may differ systematically, in both observable characteristics and unobserved determinants of housing demand, from those near the cutoff. Because the approval vote share reflects a complex aggregation of individual preferences over education spending relative to the property tax burden, we next develop a model of its determinants to ground the extrapolation of average partial effects in economic fundamentals.
	
	\section{An Economic Model of the Running Variable}\label{sec_model}
	
	In this section, we present an economic model of the running variable, i.e., the share of voters who support a ballot initiative to increase education spending in their school district.
	
	Whether a resident prefers school spending to increase or remain unchanged is likely determined by several factors. First, the strength of their preferences over public services relative to housing consumption. Ceteris paribus, a higher valuation of publicly provided education increases the likelihood that a resident supports the ballot initiative. Second, the extent to which changes in fiscal policy are capitalized into housing prices, thereby affecting the attractiveness of the district to potential movers. All else equal, a less elastic housing supply amplifies the local price response to a change in government spending. Thus, both household preferences and local housing market conditions shape an individual's propensity to support or oppose a referendum. Importantly, this analysis must account for the openness of school districts: residents from other areas may move in if they value the expenditure authorization, just as local residents may exit if they do not. This consideration naturally motivates a spatial equilibrium model with multiple horizontally differentiated locations.
	
	In line with a long tradition of modeling equilibrium across local jurisdictions\footnote{See \citealt{ellickson1971}, \citealt{hamilton1975}, \citealt{stiglitz1977}, \citealt{westhoff1977}, \citealt{brueckner1979}, \citealt{brueckner1979jpub}, \citealt{brueckner1979jue}, \citealt{roseackerman1979}, \citealt{brueckner1983}, \citealt{epplefilimonromer1984}, \citealt{eppleromer1991}, \citealt{eppleplatt1998}, \citealt{epplesieg1999}, \citealt{brueckner2000}, \citealt{eppleromersieg2001}, \citealt{calabreseetal2006}, \citealt{epplegordonsieg2010}, \citealt{calabreseeppleromano2012}, \citealt{brueckner2023}.}, we consider a metropolitan area in which households choose where to live, housing prices adjust locally, and public goods are provided through majority voting. The model is repeated static: agents optimize myopically, with no forward- or backward-looking behavior. Each period yields an allocation of households, government spending, tax rates, and housing prices across jurisdictions. For notational simplicity, we omit time subscripts throughout.
	
	We consider a unit mass of households indexed by $i$, each of whom chooses to reside in one of a discrete set of school districts indexed by $j \in \mathcal{J}$, or outside the metropolitan area altogether. Jurisdiction boundaries are fixed, and the model abstracts from commuting and the labor market. Income is treated as an endowment, reflecting the assumption that school fiscal policy does not influence firm location decisions. Consequently, the value of geographic proximity between residential and workplace locations is subsumed within the district-specific amenity component of household utility.
	
	\subsection{Households}
	
	The household residential choice problem builds on the framework of \cite{eppleplatt1998}, with one important modification. Specifically, we augment households' utility function with an additive idiosyncratic preference shock for locations. This assumption aligns with standard approaches in urban economics that incorporate random utility components in neighborhood choice models (\citealt{bayerferreiramcmillan2007}, \citealt{berlinwall2015}, \citealt{almagrodominguez2025}), as well as in models of worker and firm location in public finance (\citealt{bussogregorykline2013}, \citealt{klinemoretti2014}, \citealt{suarezzidar2016}, \citealt{stsmus2019}) and labor economics (\citealt{moretti2011}, \citealt{moretti2013}, \citealt{diamond2016}, \citealt{diamondgaubert2022}).
	
	In addition, we abstract from the vertically differentiated structure of local governments that characterizes most U.S. metropolitan areas (\citealt{berry2008}, \citealt{berry2009}, \citealt{ruggieri2024}). This simplification is justified by our focus on school districts and the impact of their fiscal policy changes. Accordingly, we assume that other jurisdictions providing non-school public services do not adjust their policies in response to changes in school district expenditures.
	
	In district $j$, households' utility is log-additive in exogenous location amenities $A_j$, housing floor space $H$, a composite numeraire consumption good $X$, and public K-12 education expenditures $G_j$. To capture congestion in the consumption of public education services, we follow \citet{stsmus2019} and scale $G_j$ by $N_j^\chi$, where $N_j$ denotes the mass of residents in district $j$ and $\chi \in \left [ 0,1 \right ]$ governs the degree of rivalry in utility from public education. When $\chi = 0$, households perceive education as a purely nonrival good and derive utility from aggregate expenditures. When $\chi = 1$, utility depends solely on per-capita expenditures, reflecting fully rival consumption of education services despite their public provision.
	
	The price of the numeraire good is normalized to one and households are endowed with income $Y_i$. As in \cite{bussogregorykline2013}, they demand one unit of housing inelastically and rent housing space at rate $P_j$. They also pay property taxes to finance the provision of education services, with the property tax rate in school district $j$ denoted by $\tau_j$. Formally, in any location $j$, household $i$ demands housing space and the numeraire to maximize their utility subject to a budget constraint:
	\begin{align}
		& \max_{H, X} \left \{ A_{ij} + \alpha_{i} \log \frac{G_j}{N_j^\chi} + \beta_i \log H + \gamma_i \log X \right \} \notag \\
		& \ \ \text{s.t.} \quad X + P_j H \left ( 1+\tau_j \right ) \leq Y_i \quad \text{and} \quad H = 1 \label{eq_utility_max}
	\end{align}
	Household $i$'s indirect utility stemming from this utility maximization problem is
	\begin{equation}
		V_{ij} = \alpha_i \log G_j - \alpha_i \chi \log N_j + \gamma_i \log \left [ Y_i - P_j \left ( 1 + \tau_j \right ) \right ] + A_{ij} \label{eq_indirect_utility}
	\end{equation}
	We model the amenity component of utility as the sum of a location-specific mean and a random variable that follows a Type-I Extreme Value distribution with scale parameter $\theta$,
	\begin{equation}
		A_{ij} = \overline{A}_{j} + U_{ij} \quad \text{ with } \quad U_{ij} \sim \text{Gumbel} \left ( 0,\theta \right ) \label{eq_amenity_shock}
	\end{equation}
	Households sort into the school district that yields the highest indirect utility or opt to reside outside the metropolitan area, in which case their utility is normalized to zero. As in \cite{mcfadden1974}, the parametric assumption on the idiosyncratic component of utility implies a closed-form expression for the probability that household $i$ chooses location $j$:
	\begin{equation}
		N_{ij} = \frac{\exp \left ( v_{ij} \slash \theta \right )}{1 + \sum_{\ell \in \mathcal{J}} \exp \left ( v_{i\ell} \slash \theta \right )}
	\end{equation}
	where the nonstochastic component of utility is \sloppy $v_{ij} \equiv \overline{A}_{j} + \alpha_i \log G_j - \alpha_i \chi \log N_j + \gamma_i \log \left [ Y_i - P_j \left ( 1 + \tau_j \right ) \right ]$. Letting $\delta_i \equiv \left [ \alpha_i, \gamma_i, Y_i \right ]'$ be a random vector whose joint probability distribution and support are denoted with $F$ and $\mathcal{D}$, respectively, the expected mass of households who sort into location $j$ is $N_{j} = \int_{\mathcal{D}} N_{ij} \left ( \delta_i \right ) d F \left ( \delta_i \right )$.

	\subsection{Housing Market}
	
	In each district, housing space is supplied competitively. Firms in the construction sector produce with homogeneous technology that exhibits decreasing returns to scale (\citealt{klinemoretti2014}, \citealt{suarezzidar2016}). Thus, the marginal cost of housing space is strictly increasing in the output. For rental rates of housing above the average cost, the housing supply function is
	\begin{equation}
		\log H^{\text{S}}_j = \lambda + \eta \log P_j + B_j\label{eq_housing_supply}
	\end{equation}
	where $\lambda$ is a deterministic constant, $\eta > 0$ denotes the elasticity of housing supply, and $B_j$ is a random variable that captures idiosyncratic productivity shocks in the construction sector. Moreover, the utility maximization and location choice problems jointly yield the aggregate demand for housing in location $j$,
	\begin{equation}
		\log H^{\text{D}}_{j} = \log \int_{\mathcal{D}} N_{ij} \left ( \delta_i \right ) d F \left ( \delta_i \right ) = \log N_j \label{eq_housing_mkt_clearing}
	\end{equation}
	The market-clearing rental rate of housing is such that aggregate housing expenditures in equilibrium are $\log P_j + \log H_j = \frac{1+\eta}{\eta} \log N_j - \frac{\lambda}{\eta} - \frac{B_j}{\eta}$.

	\subsection{Provision of Local Public Education Services}
	
	In this section, we embed a school expenditure authorization regression discontinuity design (\citealt{cfr2010}) into our spatial equilibrium model. We do so by modeling school districts as holding referenda on whether to change their education expenditures by $\Delta G_j$. The size of the proposed hike is determined outside the model, and so is the timing of these local referenda. When a school district holds a referendum, its residents approve or reject a new expenditure level $G$ and set a property tax rate $\tau$ to fund it. Each jurisdiction runs a balanced budget, 
	\begin{equation}
		G_j = \tau_j P_j H_j \label{budget}
	\end{equation}
	Clearly, for any level of $G_j$, $\tau_j$ is pinned down by population and total rental payments.
	
	The remainder of this section delves into the collective action process that aggregates preferences to determine a district's expenditure-tax mix. First, we demonstrate how households trade off higher levels of public spending on education with higher gross-of-tax housing prices. Second, we model participation in local referenda and explain the relevance of selective turnout for the determination of public expenditures. This set of arguments allows us to construct a microfoundation for the running variable in the regression discontinuity design.
	
	Households have heterogeneous preferences for public expenditures and private consumption of nonhousing goods and services. Formally, the level of education spending preferred by household $i$ who lives in school district $j$ is the one that maximizes their indirect utility,
	\begin{equation}
		G_{ij} \equiv \arg \max_{G_j} v_{ij} = \arg \max_{G_j} \Big \{ \alpha_{i} \log G_j - \alpha_i \chi \log N_j + \gamma_i \log \left [ Y_i - P_j \left ( 1 + \tau_j \right ) \right ] \Big \}
	\end{equation}
	The first-order condition associated with this maximization problem\footnote{In Appendix \ref{appx_soc}, we prove that the objective function is strictly concave in $\log G_j$ and thus $G_{ij}$ is a global maximizer.} is
	\begin{equation}
		\underbrace{\alpha_{i}}_{\text{marginal benefit}} = \underbrace{\alpha_i \chi \frac{d \log N_j}{d \log G_j} \Bigg |_{G_j = G_{ij}} + \gamma_i \rho_{ij} \frac{d \log P_j}{d \log G_j} \Bigg |_{G_j = G_{ij}} + \gamma_i \rho_{ij} \frac{d \log \left ( 1 + \tau_{j} \right )}{d \log G_j} \Bigg |_{G_j = G_{ij}}}_{\text{marginal cost}}\label{foc_spend}
	\end{equation}
	where $\rho_{ij} \equiv \frac{P_j \left ( 1 + \tau_j \right )}{Y_i - P_j \left ( 1 + \tau_j \right )}$ denotes the share of gross-of-tax housing expenditures relative to disposable income available for nonhousing consumption. Intuitively, the marginal benefit of an increase in education spending is its marginal utility. On the other hand, the marginal cost of an increase in education spending is the marginal disutility that stems from an increase in the local gross-of-tax rental rate of housing required to finance it. If education expenditures exhibit some degree of rivalry and thus $\chi \neq 0$, this cost also includes the marginal disutility of congestion.
	
	Clearly, $P_j$ and $\tau_j$ are endogenous variables and their values are constrained by two restrictions, namely housing market clearing and balanced budget in equations \eqref{eq_housing_mkt_clearing} and \eqref{budget}, respectively. Following \cite{eppleromer1991}, these equations define a Government Possibility Frontier (GPF), a relationship between government spending and the gross-of-tax rental rate of housing along which any spending change is such that the two constraints hold. In the remainder of this section, the maintained assumption is that voters internalize the effect of a change in a jurisdiction's expenditure on that location's housing market and government budget. However, they take as given the housing market in other communities and the fiscal policy chosen by other local governments. As a consequence, the relevant variables for a resident-voter in school district $j$ are $\left \{ G_j, P_j, \tau_j \right \}$. By assumption, $\left \{ G_{\ell}, P_{\ell}, \tau_{\ell} \right \}_{\ell \neq j}$ are held constant in the derivations that follow.
	
	Suppose that district $j$ is holding a referendum on whether to increase education expenditures by $\Delta G_j > 0$. We assume that residents compare the status quo with the alternative in a utilitarian framework. Specifically, any household $i$ compares $V_{ij} \left ( 0 \right )$, their current indirect utility, with $V_{ij} \left ( \Delta G_j \right )$, the potential indirect utility should the referendum be approved. The counterfactual utility incorporates the anticipated equilibrium responses of housing prices and property tax rates. Formally, the Government Possibility Frontier defined by equations \eqref{eq_housing_mkt_clearing} and \eqref{budget} traces out $P_j$ and $\tau_j$ as functions of $G_j$.
	
	The comparison of individual utilities by residents is the first step in constructing a microfoundation for the running variable. A natural addition involves the consideration of participation in local referenda. As thoroughly described by \cite{berry2009}, turnout in local elections in the United States is typically low\footnote{Drawing on a complete census of school district tax and bond referenda held in California, Ohio, Texas, and Wisconsin from 2000 to 2015, \cite{koganlavertupeskowitz2018} finds that average turnout does not exceed 30 percent in any of the four states and falls below 20 percent of the voting-age population in California and Texas.}, especially when referenda are scheduled not to coincide with general elections in November (\citealt{koganlavertupeskowitz2018}). Perhaps unsurprisingly, the few participants are extremely selected, with turnout disproportionately driven by white, affluent, and elderly voters (\citealt{berry2024report}). In addition, special interest groups play a sizable role in driving the outcome of local consultations (\citealt{anzia2014}).
	
	Motivated by this evidence, we propose an economic model of the individual decision to participate in the referendum. Specifically, we posit that individual $i$ chooses to vote in the referendum held by jurisdiction $j$ if the perceived benefit from participating exceeds the associated cost. The benefit is modeled as a household- and location-specific function of the proposed change in government expenditure, denoted $R_{ij} \left ( \Delta G_j \right )$. The cost of participation is an unobserved random variable $C_{ij}$ with support on the positive real line. It captures both monetary and non-monetary costs of voting, including the time and effort required to acquire information about the referendum and the opportunity cost of casting a ballot. The participation decision is then formally expressed as
	\begin{equation}
		T_{ij} \left ( \Delta G_j \right ) = \mathbb{I} \left [ C_{ij} \leq R_{ij} \left ( \Delta G_j \right ) \right ] 
	\end{equation}
	As a result, the individual probability of turnout is $\Prob \left ( T_{ij} = 1 \right ) = F_C \left ( R_{ij} \right )$, where $F_{C}$ denotes the cumulative distribution function of $C_{ij}$. A jurisdiction's turnout is defined as the ratio of the expected mass of voters to the expected mass of residents:
	\begin{equation}
		T_j \left ( \Delta G_j \right ) \equiv \frac{\overbrace{\int_{\mathcal{D}} N_{ij} \left ( \delta_i \right ) \Prob \left ( T_{ij} \left ( \Delta G_j \right ) = 1 | \delta_i \right ) d F \left ( \delta_i \right )}^{\text{expected mass of resident voters in $j$}}}{\underbrace{\int_{\mathcal{D}} N_{ij} \left ( \delta_i \right ) d F \left ( \delta_i \right )}_{\text{expected mass of residents in $j$}}}
	\end{equation}
	Next, we define a Bernoulli random variable $W_{ij}$ that equals one if household $i$ approves the proposed change in government spending:
	\begin{equation}
		W_{ij} \left ( \Delta G_j \right ) \equiv \mathbb{I} \left [ v_{ij} \left ( \Delta G_j \right ) \geq v_{ij} \left ( 0 \right ) \right ]
	\end{equation}
	The expected approval vote share in jurisdiction $j$ is given by the ratio of the expected mass of approving voters to the expected mass of voters. The proposed expenditure change is authorized if the approval vote share exceeds a predetermined threshold, which we set to the true institutional value of 50 percent\footnotemark. Thus, we define the approval vote share margin as
	\begin{equation}
		S_j \left ( \Delta G_j \right ) \equiv \frac{\overbrace{\int_{\mathcal{D}} N_{ij} \left ( \delta_i \right ) \Prob \left ( T_{ij} \left ( \Delta G_j \right ) = 1 | \delta_i \right ) W_{ij} \left ( \left ( \Delta G_j \right ); \delta_i \right )  d F \left ( \delta_i \right )}^{\text{expected mass of resident voters approving in $j$}}}{\underbrace{\int_{\mathcal{D}} N_{ij} \left ( \delta_i \right )\Prob \left ( T_{ij} \left ( \Delta G_j \right ) = 1 | \delta_i \right ) d F \left ( \delta_i \right )}_{\text{expected mass of resident voters in $j$}}} - 0.5 \label{eq:voteshare} \footnotetext{In Wisconsin, as in most states, the approval threshold is 50 percent. However, ten states impose supermajority requirements for referenda authorizing school facility investments (\citealt{bilaschon2024}). For example, California requires 55 percent, Washington 60 percent, and Idaho 67 percent.}
	\end{equation}
	
	\subsection{Definition of Equilibrium}\label{sec_def_equilibrium}
	
	An equilibrium consists of a finite set of school districts indexed by $j \in \mathcal{J}$; a unit mass of households indexed by $i$, each endowed with strictly positive income $Y_i$; a partition of households across jurisdictions such that each location has strictly positive population $N_j$; a set of stochastic location amenities $\left \{ \overline{A}_j \right \}_{j}$; a set of stochastic productivity shocks in the residential construction sector $\left \{ B_j \right \}_{j}$; a vector of rental rates $\left \{ P_j \right \}_{j}$ and property tax rates $\left \{ \tau_j \right \}_{j}$; an allocation of public education spending $\left \{ G_j \right \}_{j}$; an allocation of housing space $\left \{ H_{i} \right \}_{i}$ and numeraire consumption good $\left \{ X_{i} \right \}_{i}$ such that the following conditions are satisfied:
	\begin{enumerate}[(1)]
		\item In every school district, households choose housing space and the numeraire consumption good to maximize utility subject to a budget constraint, as given in equation~\eqref{eq_utility_max}.

		\item Each household resides in the jurisdiction that yields the highest indirect utility, as defined in equation~\eqref{eq_indirect_utility}, with idiosyncratic location preference shocks parameterized according to equation~\eqref{eq_amenity_shock}.

		\item The supply of housing units in each location follows the specification in equation~\eqref{eq_housing_supply}.
		
		\item The housing market clears in every jurisdiction, as described in equation~\eqref{eq_housing_mkt_clearing}.
		
		\item Each jurisdiction satisfies a balanced budget constraint, as given by equation~\eqref{budget}.
		
		\item Each jurisdiction's level of government spending is determined according to majority-rule voting among residents. If a referendum is held, a proposed spending increase $\Delta G$ is authorized if the approval vote share margin, defined in equation~\eqref{eq:voteshare}, is positive.
	\end{enumerate}

	\section{Identification of Structural Parameters}\label{sec_identification}
	
	In the previous section, we nested the running variable of a school expenditure authorization RD design into a spatial equilibrium model of local jurisdictions. In this section, we link the reduced-form average partial effects at the cutoff from Section \ref{sec_effects_cutoff} using the model structure. By doing so, we derive a system of equations that allows us to infer the structural parameters of the model. 
	
	We restrict our focus to a version of the model in which household preferences and income are finitely heterogeneous. Specifically, we assume that the unit mass of households can be partitioned into a finite set of observable types indexed by $k \in \mathcal{K} = \left \{ 1, \dots, \overline{k} \right \}$, each with positive mass $\sigma^k$. The random vector $\delta_i \equiv \left [ \alpha_i, \gamma_i, Y_i \right ]'$ is then discrete and has support $\left \{ \left [ \alpha^k, \gamma^k , y^k \right ]' \right \}_{k}$. As a result, the spatial equilibrium features $| \mathcal{J} | \times | \mathcal{K} |$ expected population masses $\left \{ N^k_j \right \}_{j,k}$.
	
	\subsection{Household Preferences and Elasticity of Housing Supply}\label{sec_ident_params_rdd}
	
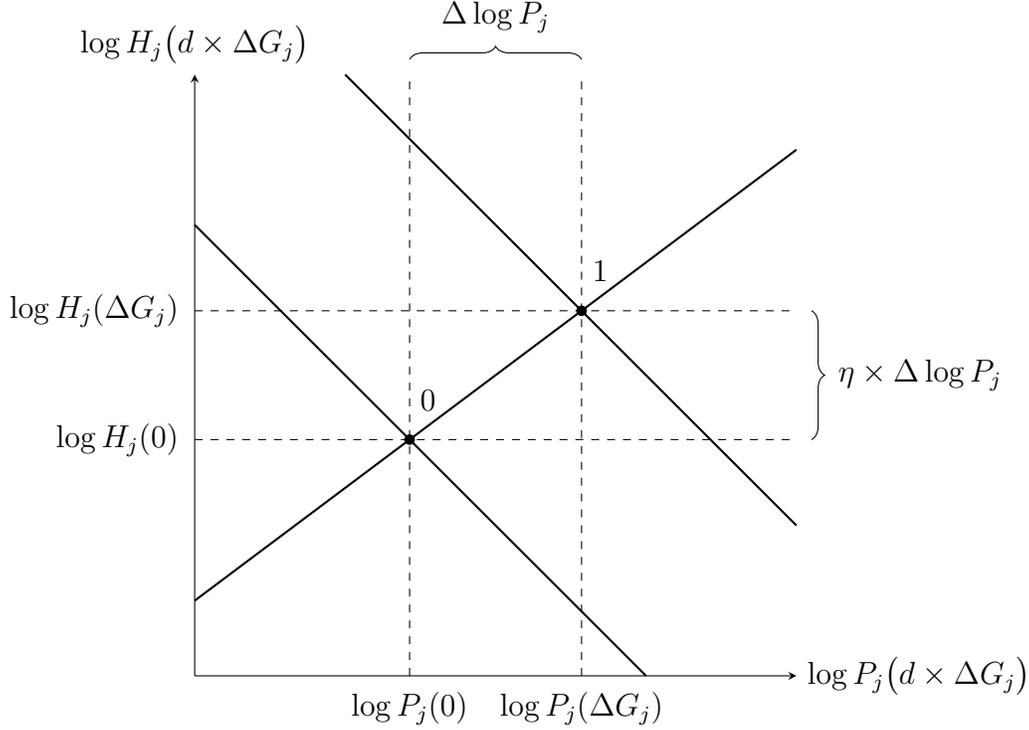
\begin{figure}[H]
		\begin{center}
			\caption{Equilibria in the Local Housing Market}\label{fig_ident_eta}
			\vspace{0mm}
			\begin{tikzpicture}[scale=1,>=stealth,decoration={brace,amplitude=5pt}]
				\draw[->] (0,0) -- (8,0) node[right] {$\log P_j\bigl(d\times\Delta G_j\bigr)$};
				\draw[->] (0,0) -- (0,8) node[above] {$\log H_j\bigl(d\times\Delta G_j\bigr)$};
				
				\draw[thick,domain=0:8] plot (\x,{0.75*\x+1});
				
				\draw[thick,domain=0:6] plot (\x,{-\x+6});
				\draw[thick,domain=2:8] plot (\x,{-\x+10});
				
				\coordinate (P0) at (2.857,3.143);
				\coordinate (P1) at (5.143,4.857);
				
				\fill (P0) circle (2pt) node[above=15pt,right=0pt] {0};
				\fill (P1) circle (2pt) node[above=15pt,right=0pt] {1};
				
				\draw[dashed,thin] (0,3.143) -- (8,3.143);
				\draw[dashed,thin] (0,4.857) -- (8,4.857);
				
				\draw[dashed,thin] (2.857,0) -- (2.857,8);
				\draw[dashed,thin] (5.143,0) -- (5.143,8);
				
				\node[below=2pt] at (2.857,0) {$\log P_j(0)$};
				\node[below=2pt] at (5.143,0) {$\log P_j(\Delta G_j)$};
				\node[left=2pt]  at (0,3.143) {$\log H_j(0)$};
				\node[left=2pt]  at (0,4.857) {$\log H_j(\Delta G_j)$};
				
				\draw[decorate] 
				(2.857,8.2) -- (5.143,8.2)
				node[midway, above=6pt] {$\Delta \log P_j$};
				
				\draw[decorate] 
				(8.2,4.857) -- (8.2,3.143)
				node[midway, right=6pt] {$\eta \times \Delta \log P_j$};
			\end{tikzpicture}
		\end{center}
	\vspace{-2mm}
	\begin{footnotesize}
	\begin{spacing}{1}
		\noindent
		\textsc{Notes}: This figure illustrates two equilibria in location $j$'s housing market. The horizontal axis measures the logarithm of potential rental rates and the vertical axis measures the logarithm of potential housing space. Point $0$ corresponds to the equilibrium under referendum rejection, with untreated potential outcomes $\log P_j(0)$ and $\log H_j(0)$ observed. Point $1$ corresponds to the equilibrium under referendum approval, which increases housing demand and leads to the treated potential outcomes $\log P_j(\Delta G_j)$ and $\log H_j(\Delta G_j)$ being observed. The slope of the chord connecting points 0 and 1, i.e., the ratio $\Delta \log H_j \slash \Delta \log P_j$, equals the elasticity of housing supply $\eta$.
	\end{spacing}
	\end{footnotesize}
	\end{figure}

	The approval of a referendum induces a change in local public spending by a known amount $\Delta G_j$. We seek to characterize the equilibrium response of each endogenous variable in the model to this policy shock. To this end, we compute arc elasticities that summarize the proportional response of outcomes to proportional changes in expenditures. For any endogenous variable $Z_\ell$ in location $\ell \in \mathcal{J}$, let $Z_\ell \left ( 0 \right )$ denote the potential outcome under the status quo (i.e., absent referendum approval), and let $Z_\ell \left ( \Delta G_j \right )$ denote the potential outcome under the approved expenditure change. The arc elasticity of $Z_\ell$ with respect to education spending is defined as
	\begin{equation}
		E_{Z_\ell} \left ( \Delta G_j \right ) \equiv \frac{\log Z_\ell \left ( \Delta G_j \right ) - \log Z_\ell \left ( 0 \right )}{\Delta \log G_j}
	\end{equation}
	While this elasticity captures the causal response of an individual outcome to the spending shock, the structure of the model allows us to go further. Rather than analyzing each outcome in isolation, the spatial equilibrium imposes a system of interdependent equations that jointly determine how all endogenous variables adjust to the shock. This structure provides a formal basis for linking elasticities across outcomes. Specifically, consider the following nonredundant equations that govern the behavior of the endogenous variables in equilibrium.
	\begin{enumerate}[(a)]
		\item The mass of type-$k$ households sorting into school district $\ell \in \mathcal{J}$:
		\begin{equation}
			N^k_\ell = \sigma^k \frac{\exp \left ( v^k_{\ell} \slash \theta^k \right )}{1 + \sum_{j' \in \mathcal{J}} \exp \left ( v^k_{j'} \slash \theta^k \right )} \label{eq_discrete_households}
		\end{equation}
		with $v^k_\ell \equiv \overline{A}_\ell + \alpha^k \log G_{\ell} - \alpha^k \chi \log N_{\ell} + \gamma^k \log \left [ y^k - P_\ell \left ( 1 + \tau_{\ell} \right ) \right ]$.
		\item The equilibrium rental rate of housing in district $\ell \in \mathcal{J}$: $\log P_\ell = \frac{1}{\eta} \log \sum_{k \in \mathcal{K}} N^{k}_\ell - \frac{\lambda}{\eta} - \frac{B_\ell}{\eta}$. Equivalently, the equilibrium quantity of housing space in jurisdiction $\ell \in \mathcal{J}$:
		\begin{equation}
			\log H_\ell = \lambda + \eta \log P_\ell + B_\ell \label{eq_discrete_housingsupply}
		\end{equation}
		\item The balanced budget run by school district $\ell \in \mathcal{J}$: $G_\ell = \tau_\ell P_\ell H_\ell$.
	\end{enumerate}
	For each of these equilibrium conditions, we compute arc elasticities and use them to derive a system of equations characterizing the response of the spatial equilibrium to the expenditure change $\Delta \log G_j$. For example, the housing supply equation \eqref{eq_discrete_housingsupply} implies
	\begin{equation}
		\frac{\Delta \log H_\ell}{\Delta \log G_j} = \eta \frac{\Delta \log P_\ell}{\Delta \log G_j}\label{eq_arcel_housingsupply}
	\end{equation}
	This relationship reflects the fact that a change in education spending affects housing demand through household mobility, while the supply of housing remains directly unaffected. The resulting shift in demand leads to price adjustments that can be used to infer the supply elasticity $\eta$. Figure~\ref{fig_ident_eta} illustrates the equilibrium in location~$j$'s housing market under both referendum rejection and approval, showing how differences in potential outcomes map into the structural parameter of interest.
	
	Although the arc elasticities in equation~\eqref{eq_arcel_housingsupply} are not observable, Section~\ref{sec_identification_cutoff} establishes that they are point identified in expectation using a regression discontinuity design centered at the approval threshold\footnote{The fuzzy RD estimand in equation~\eqref{eq_fuzzy_rd_estimand} identifies the target parameter in equation~\eqref{eq_avg_ape}: a weighted average of arc elasticities with respect to the proposed spending change. The weights are proportional to $\Delta \log G_j$ and integrate to one at the cutoff. We assume that marginally approved referenda feature similar proposed changes in education spending, so that $\Delta \log G_j$ is approximately constant near the threshold. As a result, the weights are approximately equal to one, and we can omit them from the subsequent analysis.}. Specifically, taking expectations of both sides of equation \eqref{eq_arcel_housingsupply} conditional on $S_j = 0$ and integrating over the joint probability distribution of unobservables (i.e., $\left \{ \overline{A}_j \right \}_{j \in \mathcal{J}}$ and $\left \{ B_j \right \}_{j \in \mathcal{J}}$), we obtain
	\begin{equation}
		\E \left [ \frac{\Delta \log H_\ell}{\Delta \log G_j} \bigg | S_j = 0 \right ] = \eta \times \E \left [ \frac{\Delta \log P_\ell}{\Delta \log G_j} \bigg | S_j = 0 \right ] \label{eq:totdiff_housing}
	\end{equation}
	Since both conditional expectations are identified, equation~\eqref{eq:totdiff_housing} can be used to recover the structural parameter $\eta$.
	
	\vspace{-3mm}
	
	Proceeding analogously with the household choice probability equation~\eqref{eq_discrete_households}, we derive the following approximation for the arc elasticity of the mass of type-$k$ households in district $j$ with respect to the proposed expenditure change:
	\begin{align}
		& \frac{\Delta \log N^k_j}{\Delta \log G_j} \approx \left ( 1 - \frac{N^k_{j}}{\sigma^k} \right ) \left ( \frac{\alpha^k}{\theta^k} - \frac{\chi \alpha^k}{\theta^k} \frac{\Delta \log N_j}{\Delta \log G_j} - \frac{\gamma^k \rho^k_j}{\theta^k} \frac{\Delta \log P_j}{\Delta \log G_j} - \frac{\gamma^k \rho^k_j}{\theta^k} \frac{\Delta \log \left ( 1 + \tau_j \right )}{\Delta \log G_j} \right ) \notag \\
		& - \sum_{\ell \neq j} \frac{N^k_{\ell}}{\sigma^k}  \left ( \frac{\alpha^k}{\theta^k} \frac{\Delta \log G_\ell}{\Delta \log G_j} - \frac{\chi \alpha^k}{\theta^k} \frac{\Delta \log N_\ell}{\Delta \log G_j} - \frac{\gamma^k \rho^k_\ell}{\theta^k} \frac{\Delta \log P_\ell}{\Delta \log G_j} - \frac{\gamma^k \rho^k_\ell}{\theta^k} \frac{\Delta \log \left ( 1 + \tau_\ell \right )}{\Delta \log G_j} \right )\label{eq_identification_pop}
	\end{align}
	This expression features two unknown structural parameters: $\alpha^k \slash \theta^k$ and $\gamma^k \slash \theta^k$. To identify them, we obtain a corresponding expression for a second district $j' \neq j$. Taking expectations of both equations conditional on $S_j = 0$ yields a system with two unknowns, from which we recover the parameters that enter type-$k$ households' indirect utilities.
	
	Analogous identification arguments based on the remaining equilibrium conditions are provided in Appendix~\ref{appx_identification}. These allow us to recover the full set of structural parameters: $| \mathcal{K} |$ preference parameters for education spending $\left \{ \alpha^k \slash \theta^k \right \}_{k}$; $| \mathcal{K} |$ preference parameters for nonhousing consumption $\left \{ \gamma^k \slash \theta^k \right \}_{k}$; the elasticity of housing supply $\eta$.
	
	Point identification requires that the number of equations be at least as large as the number of unknowns, which holds whenever $|\mathcal{J}| (|\mathcal{K}| + 2) \geq 2|\mathcal{K}| + 1$. In practice, the number of school districts exceeds the number of household types, resulting in an overidentified system.
	
	To conduct statistical inference on the structural parameters, we compute analytical standard errors using the delta method, which requires estimates of the pairwise covariances among the RDD coefficients (see Appendix \ref{appx_sec_inference}). To obtain these covariances, we extract each outcome's sample based on its own MSE-optimal bandwidth and stack them in pairs. For each pair, we estimate a model in which the local linear instrumental variables specifications are fully interacted with sample indicators, and we cluster heteroskedasticity-robust standard errors by referendum identifier. This procedure yields estimates of the covariance between the two coefficients associated with $D_j \times \Delta \log G_j$. Once the full variance-covariance matrix of RDD parameters is constructed, we apply Ledoit-Wolf shrinkage to its correlation matrix (\citealt{ledoitwolf2004}) in order to regularize the estimates and improve the stability of subsequent inference.
	
	\subsection{Selection into Voting}\label{sec_turnout_discrete}
	
	We maintain the assumption that the unobserved cost of participating in the referendum varies across both households and locations. Households of the same type are not assumed to face identical voting costs. We model the benefit of participation as the absolute value of the anticipated utility gain (or loss) from the proposed change in government spending. Intuitively, when $\Delta G_j$ represents a ``high-stakes'' proposal for a household, that household is more likely to turn out, all else equal. Formally, the benefit is parameterized as
	{\setlength{\abovedisplayskip}{6pt plus 2pt minus 2pt}
	\setlength{\belowdisplayskip}{6pt plus 2pt minus 2pt}
	\begin{equation}
		R_{ij} \left ( \Delta G_j \right ) \equiv \big | v^k_j \left ( \Delta G_j \right ) - v^k_j \left ( 0 \right ) \big |
	\end{equation}}
	where the superscript $k$ denotes the household's type, $k = k(i)$.
	
	The participation decision is then given by $T_{ij} \left ( \Delta G_j \right ) = \mathbb{I} \left [ C_{ij} \leq | v^k_j \left ( \Delta G_j \right ) - v^k_j \left ( 0 \right ) | \right ]$. Accordingly, the individual probability of turnout is $\Prob \left ( T_{ij} \left ( \Delta G_j \right ) = 1 \right ) = F_C \left ( | v^k_j \left ( \Delta G_j \right ) - v^k_j \left ( 0 \right ) | \right )$. Because the benefit does not vary across households of the same type, we compactly denote the probability of voting among type-$k$ households in jurisdiction $j$ as $T^k_j \left ( \Delta G_j \right ) \equiv F_C \left ( \big | v^k_j \left ( \Delta G_j \right ) - v^k_j \left ( 0 \right ) \big | \right )$. As in the general case, expected turnout in jurisdiction $j$ is defined as the ratio of the expected mass of voters to the expected mass of residents: $T_j \left ( \Delta G_j \right ) \equiv \sum_{k \in \mathcal{K}} N^k_j T^k_j \left ( \Delta G_j \right ) \slash \sum_{k \in \mathcal{K}} N^k_j$.
	
	Finally, we define the Bernoulli random variable $W^{k}_{j} \left ( \Delta G_j \right ) = \mathbb{I} \left [ v^{k}_{j} \left ( \Delta G_j \right ) \geq v^{k}_{j} \left ( 0 \right ) \right ]$ to indicate whether type-$k$ households prefer the proposed spending change in jurisdiction $j$. As in the general model, the expected approval vote share margin is defined as the difference between the expected mass of approving voters and the cutoff for passage:
	\begin{equation}
		S_j \left ( \Delta G_j \right ) \equiv \frac{\sum_{k \in \mathcal{K}} N^k_j T^k_j \left ( \Delta G_j \right ) W^k_j \left ( \Delta G_j \right )}{\sum_{k \in \mathcal{K}} N^k_j T^k_j \left ( \Delta G_j \right )} - 0.5 \label{eq_voteshare_finite}
	\end{equation}
	
	\subsection{Turnout Parameters}\label{sec_mle_turnout}
	
	Our strategy for extrapolating average effects away from the cutoff relies on the ability to recover the approval vote share associated with counterfactual values of the proposed change in education spending $\Delta G_j$. Because turnout influences both the size and composition of the electorate, it plays a central role in shaping the running variable. To capture systematic differences in participation across household types in U.S. local elections (\citealt{anzia2014}, \citealt{berry2024report}), we specify a structural model of turnout behavior and estimate its parameters to match these documented patterns.
	
	To this end, we parameterize the probability distribution of the unobserved cost of participation. Specifically, we assume that $C_{ij}$ is log-normally-distributed with a type-specific mean and variance:
	\begin{equation}
		\log C_{ij} \sim \mathcal{N} \Big ( \mu^k_0 + \mu^k_1 \Delta \log G_j , \left ( \sigma^k_0 \right )^2 \Big )
	\end{equation}
	The parameters $\left \{ \mu^k_0, \sigma^k_0 \right \}_{k \in \mathcal{K}}$ are constant across locations and capture intrinsic differences in participation costs across household types. The slope parameters $\left \{ \mu^k_1 \right \}_{k \in \mathcal{K}}$ are likewise location-invariant, but measure the sensitivity of each group's expected participation cost to the proposed change in spending. This structure induces correlation in $C_{ij}$ both across jurisdictions for households of a given type and across types within a given jurisdiction, although the latter arises solely through dependence on $\Delta \log G_j$.
	
	This specification incorporates two salient features of referendum participation. First, it allows turnout to respond to the size of the proposed spending change, reflecting the idea that larger school construction or renovation projects are more likely to mobilize nonmarginal voters. Second, it allows participation costs to vary across household types, accounting for the fact that some groups---such as elderly individuals without children---may be more inclined or able to participate in initiatives that influence the provision of local public services.
	
	Given this assumption, the probability of turnout among type-$k$ households is
	\begin{equation}
		T^k_j \left ( \Delta G_j \right ) = \Phi \left ( \frac{\log \big | v^k_j \left ( \Delta G_j \right ) - v^k_j \left ( 0 \right ) \big | - \left ( \mu^k_0 + \mu^k_1 \Delta \log G_j \right )}{\sigma^k_0} \right ) \label{eq_turnout_prob_normal}
	\end{equation}
	where $\Phi$ denotes the cumulative distribution function of a standard normal random variable.
	
	To identify the parameters governing the economic model of turnout, we proceed under the assumption that $T^k_j$ is observed. Although publicly available data report only aggregate turnout, we have access to the Labels \& Lists (L2) Voter Data, which compiles voter registration records from all fifty states and Washington, D.C., and supplements them with proprietary commercial data containing demographic and basic financial characteristics. We rely on this dataset to estimate participation rates for each household group in our sample of Wisconsin school district referenda.
	
	Since the structural parameters of the model are known, the benefit from participation, given by $| v^k_j \left ( \Delta G_j \right ) - v^k_j \left ( 0 \right )|$, is also known. In the remainder of this section, we exploit the parametric assumption on $C_{ij}$ to estimate the parameters that maximize the likelihood of observing the turnout rates implied by the model. Accordingly, we specify a measurement model for $T^k_j \left ( \Delta G_j \right )$. Letting $\widecheck{T}^k_j$ and $\widecheck{N}^k_j$ denote, respectively, the observed turnout and population count of type-$k$ households in jurisdiction $j$, we assume
	\begin{equation}
		\widecheck{T}^k_j \sim \text{Binomial} \left ( \widecheck{N}^k_j, T^k_j \right )
	\end{equation}
	That is, observed turnout is modeled as a binomial random variable, with the number of trials given by the observed population count and the probability of success equal to the model-implied turnout rate $T^k_j$.
	
	Given this specification, our parameter set of interest is $\mathcal{P} = \left \{ \left \{ \mu^k_0, \mu^k_1, \sigma^k_0 \right \}_{k \in \mathcal{K}} \right \}$ and we let $\vartheta$ denote the vector stacking all elements of $\mathcal{P}$. We index jurisdiction-referendum pairs by $j$\footnote{That is, $j$ indexes distinct referenda, allowing for multiple ballot measures within the same jurisdiction over time.}. The resulting likelihood function is then
	\begin{equation}\label{eq_likelihood}
		L \left ( \vartheta \right ) = \prod_{j=1}^{n} \prod_{k \in \mathcal{K}}
		\left(\!
		\begin{array}{c}
			\widecheck{N}^k_j \\
			\widecheck{T}^k_j
		\end{array}
		\!\right)
		T^k_j \left ( \vartheta \right )^{\widecheck{T}^k_j} \left ( 1 - T^k_j \left ( \vartheta \right ) \right )^{\widecheck{N}^k_j - \widecheck{T}^k_j}
	\end{equation}
	We estimate the parameters in $\mathcal{P}$ by maximizing the likelihood in \eqref{eq_likelihood} with respect to $\vartheta$.
	
	A distinctive feature of our setting is that the preference and housing supply parameters $\zeta \equiv \left [ \alpha^1, \alpha^2, \alpha^3, \alpha^4, \gamma^1, \gamma^2, \gamma^3, \gamma^4, \eta \right ]$ enter the likelihood function directly, as they influence both the decision to participate in the referendum and the choice to approve or reject the proposed expenditure change. Letting $\widehat{\zeta}$ denote the estimate of $\zeta$, the log-likelihood function conditional on $\widehat{\zeta}$ is given by the following expression, up to an additive constant:
	\begin{equation}
		\log L \left ( \vartheta; \widehat{\zeta} \right ) = \sum_{j=1}^{n} \sum_{k \in \mathcal{K}} \left [ \widecheck{T}^k_j \log T^k_j \left ( \vartheta; \widehat{\zeta} \right ) + \left ( \widecheck{N}^k_j - \widecheck{T}^k_j \right ) \log \left ( 1 - T^k_j \left ( \vartheta; \widehat{\zeta} \right ) \right ) \right ]
	\end{equation}
	
	Although $\widehat{\zeta}$ enters the likelihood function as a fixed constant, it is in fact the realized value of an estimator and thus subject to sampling variability. As a result, standard errors based solely on the curvature of the likelihood understate the true statistical uncertainty associated with $\widehat{\vartheta}$.
	
	To address this issue, we implement a parametric bootstrap procedure that incorporates the stochastic nature of $\widehat{\zeta}$ into inference for $\vartheta$. Specifically, we rely on the asymptotic normality of the estimator of $\zeta$, which is distributed approximately as $\mathcal{N} \left ( \widehat{\zeta}, \widehat{\Sigma}_{\zeta} \right )$ in finite samples. For each replication $m \in \left \{ 1, \dots, \overline{m} \right \}$, we draw $\widehat{\zeta}^{(m)}$ from this distribution, re-estimate $\vartheta$ by maximizing the likelihood conditional on $\widehat{\zeta}^{(m)}$, and obtain $\widehat{\vartheta}^{(m)}$ and its corresponding variance-covariance matrix $\widehat{\Sigma}^{(m)}_{\vartheta}$. We subsequently compute the within-replication variance-covariance matrix as the average of the estimated variance-covariance matrices:
	{\setlength{\abovedisplayskip}{6pt plus 2pt minus 2pt}
	\setlength{\belowdisplayskip}{6pt plus 2pt minus 2pt}
	\begin{equation}
		\overline{\widehat{\Sigma}}_{\vartheta} \equiv \frac{1}{\overline{m}} \sum_{m=1}^{\overline{m}} \widehat{\Sigma}^{(m)}_{\vartheta}
	\end{equation}}
	To capture simulation-induced dispersion in point estimates, we compute the between-replication variance-covariance matrix:
	\begin{equation}
		\widetilde{\widehat{\Sigma}}_{\vartheta} \equiv \frac{1}{\overline{m}-1} \sum_{m=1}^{\overline{m}} \left ( \widehat{\vartheta}^{(m)} - \overline{\widehat{\vartheta}} \right ) \left ( \widehat{\vartheta}^{(m)} - \overline{\widehat{\vartheta}} \right )' \quad \text{ with } \quad \overline{\widehat{\vartheta}} \equiv \frac{1}{\overline{m}} \sum_{m=1}^{\overline{m}} \widehat{\vartheta}^{(m)}
	\end{equation}
	To conclude, following \citet[pp.~76--77]{rubin1987}, we obtain the total variance-covariance matrix as
	\begin{equation}
		\widehat{\Sigma}_{\vartheta} = \overline{\widehat{\Sigma}}_{\vartheta} + \left ( 1 + \frac{1}{\overline{m}} \right ) \widetilde{\widehat{\Sigma}}_{\vartheta}
	\end{equation}
	This total variance accounts for both the uncertainty conditional on $\widehat{\zeta}$ and the additional variability introduced by treating $\widehat{\zeta}$ as an estimate rather than a known quantity.
	
	\section{Extrapolation of Average Partial Effects}\label{sec_extrapolation}
	
	In this section, we describe how the model structure and the microfoundation for the running variable allow us to extrapolate average effects away from the cutoff. The central feature of our approach is that we manipulate a policy variable---the proposed change in school spending, $\Delta G$---rather than the approval vote share, which reflects the interaction between $\Delta G$ and underlying economic fundamentals.
	
	\subsection{Structural Extrapolation Algorithm}\label{sec_extrap_algorithm}
	
	We maintain the assumption that all model parameters---including \sloppy $\left \{ \left \{ \mu^k_0, \mu^k_1, \sigma^k_0 \right \}_{k \in \mathcal{K}} \right \}$---are known. Given the model structure and these disciplining parameters, we can simulate counterfactual realizations of both the running variable and the outcomes of interest.
	
	As in any sharp regression discontinuity design, the conditional potential outcome mean $\E \left [ Z_\ell \left ( 0 \right ) | S_j = s \right ]$ is counterfactual for $s > 0$, while $\E \left [ Z_\ell \left ( \Delta G_j \right ) | S_j = s \right ]$ is counterfactual for $s \leq 0$, for any endogenous variable $Z_\ell$. To estimate these quantities, we simulate the data generating process for each school district–referendum pair across a range of hypothetical policy changes. This allows us to compute the counterfactual values of $\left\{ P_\ell, H_\ell, G_\ell, \tau_\ell, \left \{ N^k_\ell \right \}_{k} \right \}_{\ell}$ at realizations of the vote share away from the threshold.
	
	To illustrate our strategy, suppose that a proposed change $\Delta G_j$ yields a realization of the running variable $S_j \left ( \Delta G_j \right )$ greater than 0. We first compute the equilibrium that results from referendum approval and the associated increase in government spending. By construction, this equilibrium is extrapolated and observed in the simulation. We then compute the equilibrium that would arise under the same realization of $S_j$, but assuming the referendum did not result in any change in education spending. This pre-referendum equilibrium is, by definition, counterfactual. The same logic applies symmetrically when $S_j \left ( \Delta G_j \right ) \leq 0$, in which case the untreated equilibrium is observed and the treated equilibrium is counterfactual.
	
	Repeating this procedure over a grid of proposed expenditure changes allows us to recover extrapolated potential outcomes across the support of the running variable. We then average simulated outcomes within bins of $S_j \left ( \Delta G_j \right )$. The difference between treated and untreated equilibria defines the average treatment effect away from the cutoff.
	
	More specifically, the procedure proceeds as follows. For each jurisdiction–referendum pair $j$, consider a finely spaced grid $\mathcal{G}$ of positive values of $\Delta G_j$. For each $\Delta G_j \in \mathcal{G}$, we implement the following steps:
	\begin{enumerate}[(1)]
		\item For each household type $k$, compute the probability of participating in the referendum. Under the assumption that the unobserved cost of voting is log-normally distributed with type-specific parameters, this probability is given by equation \eqref{eq_turnout_prob_normal}. Because both the perceived benefit and the cost depend on $\Delta G_j$, turnout probabilities vary with the size of the proposed spending change.
		
		\item Compute the approval vote share margin $S_j \left ( \Delta G_j \right )$ using equation \eqref{eq_voteshare_finite}.
		
		\item Starting from the equilibrium observed in the data, update model outcomes $\left\{ P_\ell, H_\ell, \tau_\ell, \left \{ N^k_\ell \right \}_{k} \right\}_{\ell}$ as follows:
		\begin{enumerate}[(a)]
			\item Increase government spending in jurisdiction $j$ from $G_j$ to $G_j + \Delta G_j$, and solve for a new spatial equilibrium. All other jurisdictions are assumed not to hold referenda. This assumption reflects the SUTVA restriction used to identify structural parameters in Section \ref{sec_ident_params_rdd}. The resulting equilibrium captures the impulse response of the system to a spending shock in district $j$.
			
			\item Compute the equilibrium under referendum rejection (i.e., $\Delta G_j = 0$). Since government spending remains unchanged, this equilibrium does not vary across the grid $\mathcal{G}$ and serves as the common baseline for all counterfactual comparisons.
			
			\item Denote the endogenous variables from each equilibrium as $\left\{ P_\ell \left ( \Delta G_j \right ), H_\ell \left ( \Delta G_j \right ), G_\ell \left ( \Delta G_j \right ), \tau_\ell \left ( \Delta G_j \right ), \left \{ N^k_\ell \left ( \Delta G_j \right ) \right \}_{k} \right\}_{\ell}$ for both $\Delta G_j = 0$ and $\Delta G_j > 0$.
		\end{enumerate}
		
	\end{enumerate}
	This procedure is repeated for all jurisdiction–referendum pairs $j = 1,2, \dots, n$. For each endogenous variable, the output is a matrix of extrapolated values of dimension $n \times | \mathcal{G} | \times 2$, corresponding to the number of referenda, the grid of proposed spending changes, and the two policy scenarios (no change and proposed change), respectively. The algorithm also produces an $n \times | \mathcal{G} |$ matrix of simulated values of the running variable $S_j \left ( \Delta G_j \right )$.
	
	For any $s \in \left [ -0.5,0.5 \right ]$, the average arc elasticity of outcome $Z_\ell$ with respect to $G_j$, conditional on an approval vote share margin of $S_j = s$, is defined as
	\begin{equation}
		\textsc{AVE}_{Z_\ell} \left ( s \right ) \equiv \E \bigg [ \frac{\log Z_\ell \left ( \Delta G_j \right ) - \log Z_\ell \left ( 0 \right )}{\Delta \log G_j} \bigg  | S_j \left ( \Delta G_j \right ) = s \bigg ] 
	\end{equation}
	Since $S_j$ is a continuous random variable, a natural estimator for $\textsc{AVE} \left ( s \right )$ is
	\begin{equation}
		\widehat{\text{AVE}}_{Z_\ell} \left ( b \right ) \equiv \frac{\sum_{\Delta G_j \in \mathcal{G}} \frac{\log Z_\ell \left ( \Delta G_j \right ) - \log Z_\ell \left ( 0 \right )}{\Delta \log G_j} \times \mathbb{I} \left [ S_j \left ( \Delta G_j \right ) \in \left [ b, b+\kappa \right ) \right ]}{\sum_{\Delta G_j \in \mathcal{G}} \mathbb{I} \left [ S_j \left ( \Delta G_j \right ) \in \left [ b, b + \kappa \right ) \right ]}
	\end{equation}
	where $\kappa$ denotes the bin width and serves as a tuning parameter. This nonparametric estimator computes local averages of arc elasticities in bins determined by realizations of the simulated running variable.
	
\subsection{Statistical Inference}

To conduct statistical inference on each target elasticity, we follow an approach analogous to that in the previous section. Each estimator $\widehat{\text{AVE}}_{Z_\ell} \left ( b \right )$ is a function of the estimated parameter vectors $\widehat{\zeta}$ and $\widehat{\vartheta}$, and therefore inherits sampling variability from both. We account for this uncertainty by drawing from the known approximate distribution of the estimator of $\zeta$, namely $\mathcal{N} \left ( \widehat{\zeta}, \widehat{\Sigma}_{\zeta} \right )$. For each replication $m \in \left \{ 1, \dots, \overline{m} \right \}$, we sample a realization $\widehat{\zeta}^{(m)}$, re-estimate $\vartheta$ with maximum likelihood conditional on $\widehat{\zeta}^{(m)}$, and compute the corresponding set of average arc elasticities $\Big \{ \widehat{\text{AVE}}^{(m)}_{Z_\ell} \left ( b \right ) \Big \}_{b}$ using the parameter pair $\left ( \widehat{\zeta}^{(m)}, \widehat{\vartheta}^{(m)} \right )$.

This procedure captures the sampling uncertainty in $\widehat{\zeta}$ but treats $\widehat{\vartheta}$ as fixed within each draw. To propagate the uncertainty in $\widehat{\vartheta}$ given $\widehat{\zeta}^{(m)}$, we implement an additional parametric bootstrap\footnote{The arc elasticities $E_{Z_\ell}$ are not continuously differentiable in $\vartheta$ due to the threshold-based rule that determines referendum approval or rejection. This precludes the application of the delta method.}. Specifically, within each replication $m$, we draw $\overline{r}$ times from the known approximate distribution of the maximum likelihood estimator of $\vartheta$, given by $\mathcal{N} \left ( \widehat{\vartheta}^{(m)}, \widehat{\Sigma}^{(m)}_{\vartheta} \right )$. For each resulting pair $\left ( m,r \right )$, we recompute the average arc elasticities, yielding the set $\Big \{ \widehat{\text{AVE}}^{(m,r)}_{Z_\ell} \left ( b \right ) \Big \}_{b}$.

To construct confidence intervals around each counterfactual statistic, we compute the total variance of $\widehat{\text{AVE}}_{Z_\ell} \left ( b \right )$ by combining within- and between-replication components. For any bin of the approval vote margin $b$, we begin by calculating the within-replication variance associated with the $m$th draw:
\begin{equation}
	\widehat{\sigma}^{2 (m)} \left ( b \right ) \equiv \frac{1}{\overline{r}-1} \sum_{r=1}^{\overline{r}} \left ( \widehat{\text{AVE}}^{(m,r)}_{Z_\ell} \left ( b \right ) - \overline{\widehat{\text{AVE}}}^{(m)}_{Z_\ell} \left ( b \right ) \right )^2
\end{equation}
where the mean across inner replications is given by $\overline{\widehat{\text{AVE}}}^{(m)}_{Z_\ell} \left ( b \right ) \equiv \frac{1}{\overline{r}} \sum_{r=1}^{\overline{r}} \widehat{\text{AVE}}^{(m,r)}_{Z_\ell} \left ( b \right )$. We then average the resulting values across outer replications to obtain the within-replication component of the total variance:
\begin{equation}
	\overline{\widehat{\sigma}}^{2} \left ( b \right ) \equiv \frac{1}{\overline{m}} \sum_{m=1}^{\overline{m}} \widehat{\sigma}^{2 (m)} \left ( b \right )
\end{equation}

Next, we compute the between-replication variance, which captures the uncertainty due to sampling variation in the first-stage parameter vector $\widehat{\zeta}$:
\begin{equation}
	\widetilde{\widehat{\sigma}}^{2} \left ( b \right ) \equiv \frac{1}{\overline{m}-1} \sum_{m=1}^{\overline{m}} \left ( \widehat{\text{AVE}}^{(m)}_{Z_\ell} \left ( b \right ) - \overline{\widehat{\text{AVE}}}_{Z_\ell} \left ( b \right ) \right )^2
\end{equation}
with $\overline{\widehat{\text{AVE}}}_{Z_\ell} \left ( b \right ) \equiv \frac{1}{\overline{m}} \sum_{m=1}^{\overline{m}} \widehat{\text{AVE}}^{(m)}_{Z_\ell} \left ( b \right )$.

Finally, following \citet[pp.~76--77]{rubin1987}, we obtain the total variance of $\widehat{\text{AVE}}_{Z_\ell} \left ( b \right )$ as
\begin{equation}
	\widehat{\sigma}^{2} \left ( b \right ) = \overline{\widehat{\sigma}}^{2} \left ( b \right ) + \left ( 1 + \frac{1}{\overline{m}} \right ) \widetilde{\widehat{\sigma}}^{2} \left ( b \right )
\end{equation}

\section{Simulation}\label{sec_simulation}

In this section, we apply the methods developed in the preceding sections to simulated data. First, we confirm that it is possible to  infer the structural parameters of our spatial equilibrium model using regression discontinuity designs. Second, we verify that our maximum likelihood procedure recovers the parameters governing turnout in school expenditure referenda according to our selection model. Third, we leverage the model structure and parameter estimates to extrapolate the cutoff-specific capitalization of expenditure changes in housing values away from the threshold.

\subsection{Data Generating Process}\label{sec_sim_dgp}

Our data generating process closely mirrors the structure of the model. Households sort across school districts or choose to reside outside the metropolitan area. Given initial allocations of population masses $\left \{ N^k_j \right \}_{j,k}$, housing space $\left \{ H_j \right \}_{j}$, rental rates $\left \{ P_j \right \}_{j}$, school district expenditures $\left \{ G_j \right \}_{j}$, and property tax rates $\left \{ \tau_j \right \}_{j}$ that satisfy the equilibrium conditions described in Section \ref{sec_def_equilibrium}, one jurisdiction is randomly selected to hold a referendum involving a randomly drawn proposed change in school spending $\Delta G$. Voter turnout follows the process in Section \ref{sec_turnout_discrete}, determining a realization of the running variable $S$. If the proposal is approved, government spending in the selected jurisdiction increases by $\Delta G$, and the model is re-solved to reflect the new policy environment.

To approximate the size of our Wisconsin sample, we generate 2,000 referenda under each set of initial conditions. We combine these into a single dataset and use a regression discontinuity design to estimate the causal parameter $\text{WAVE} \left ( 0 \right )$ for each endogenous outcome. Crucially, each simulated observation features only one jurisdiction undergoing a policy change, which ensures that the Stable Unit Treatment Value Assumption (SUTVA) is not violated when comparing treated and control jurisdictions across observations. In addition to the regression discontinuity analysis, we implement the maximum likelihood estimator described in Section \ref{sec_mle_turnout} to estimate the parameters governing selection into voting.

We repeat this procedure 100 times. While structural parameters remain fixed across replications, we re-draw the unobserved location-specific amenity values $\left \{ \overline{A}_j \right \}_{j}$ and productivity shocks in the construction sector $\left \{ B_j \right \}_{j}$ in each iteration to generate new initial conditions.

\subsection{Parameterization}\label{sec_sim_params}

We partition the unit mass of households into $\mathcal{K} = 4$ distinct types, each comprising an equal mass $\sigma^k = 0.25$. We set the parameters that measure preference for school district spending to $\left ( \alpha^1, \alpha^2, \alpha^3, \alpha^4 \right ) = \left ( 0.55, 0.20, 0.15, 0.10 \right )$ and the parameters that measure preference for nonhousing goods to $\left ( \gamma^1, \gamma^2, \gamma^3, \gamma^4 \right ) = \left ( 0.35, 0.30, 0.25, 0.20 \right )$. We specify household income as a concave function of the preference for education, i.e., $\left ( y^1, y^2, y^3, y^4 \right ) = \left ( 0.45, 0.55, 0.55, 0.45 \right )$. In addition, for each type $k$, we compute the preference for housing floor space as $\beta^k = 1 - \alpha^k - \gamma^k$. We set $\chi = 1$, so that households value per capita rather than aggregate education spending, reflecting the assumption that education is a rival good. We also fix the scale parameter of the Type-I Extreme Value distributed idiosyncratic utility shock at $\theta^k = 1$ for all types.

We model the metropolitan area as consisting of $\mathcal{J} = 10$ school districts. In each jurisdiction, we set the housing supply parameters to $\gamma = 0$ and $\eta = 0.6$. To introduce unobserved heterogeneity, we draw amenity values and construction productivity shocks independently from normal distributions: $\overline{A}_j \sim \mathcal{N} \left ( 0,0.1 \right )$ and $B_j \sim \mathcal{N} \left ( -1.2, 0.05 \right )$.

When a jurisdiction holds a referendum, we draw the proposed change in log spending from a uniform distribution: $\Delta \log G \sim \mathcal{U} \left [ 0.095, 0.105 \right ]$\footnote{To estimate the parameters governing turnout, the proposed expenditure change is $\Delta G \sim \mathcal{U} \left [ 0.01, 0.40 \right ]$.}. To model turnout, we specify the expected log cost of participation $\log C_{ij}$ for each type $k \in \mathcal{K}$ using intercepts $\left ( \mu_0^1, \mu_0^2, \mu_0^3, \mu_0^4 \right ) = \left ( -3, -5, -7, -3 \right )$ and slopes $\left ( \mu_1^1, \mu_1^2, \mu_1^3, \mu_1^4 \right ) = \left ( -1, -1, 0, 0 \right )$. Finally, we fix the variance of the participation cost at $\sigma^k_0 = 3$ for all types.

\subsection{Structural Parameter Estimates}

We estimate the structural parameters $\left \{ \alpha^k \right \}_{k=1}^{4}$ and $\left \{ \gamma^k \right \}_{k=1}^{4}$, which govern households' preferences for public education expenditures and housing space, respectively, by leveraging the system of equations implied by households' choice probabilities in equation~\eqref{eq_discrete_households}. These parameters are only point identified up to scale, as the solutions to the linear system in equations~\eqref{eq_appx_popown} and~\eqref{eq_appx_popother} yield, for each $k$, the ratios $\alpha^k \slash \theta^k$ and $\gamma^k \slash \theta^k$. Although the scale parameters of the Gumbel distributed idiosyncratic utility shocks are not identified, we can compute the economically meaningful ratio $\alpha^k \slash \gamma^k$, which reflects the marginal willingness to pay for public education services relative to the marginal utility of income.

\begin{table}[H]
	\begin{center}
		\caption{Comparison of True and Estimated $\left \{ \alpha^k \slash \gamma^k \right \}_{k \in \mathcal{K}}$}\label{table_sim_results_households}
		\vspace{0mm}
		\begin{tabular}{>{\centering\arraybackslash}m{2.2cm}>{\centering\arraybackslash}p{1.8cm}>{\centering\arraybackslash}p{1.8cm}>{\centering\arraybackslash}p{1.8cm}>{\centering\arraybackslash}p{1.8cm}>{\centering\arraybackslash}p{1.8cm}>{\centering\arraybackslash}p{1.8cm}}
  \toprule\toprule
  \multirow{2}{*}{\begin{tabular}[c]{@{}c@{}}\vspace{-0.0em}Household\\Type ($k$)\vspace{-0.5em}\end{tabular}} & \multicolumn{2}{c}{$\alpha^k\slash\theta^k$} & \multicolumn{2}{c}{$\gamma^k\slash\theta^k$} & \multicolumn{2}{c}{$\alpha^k\slash\gamma^k$} \\
  \cmidrule(lr){2-3} \cmidrule(lr){4-5} \cmidrule(lr){6-7}
  & True & Estimate & True & Estimate & True & Estimate \\
  \midrule
  \addlinespace
  1 & 0.550 & 0.552 & 0.350 & 0.352 & 1.571 & 1.568 \\
   &  & (0.004) &  & (0.003) &  & (0.002) \\
  \addlinespace
  2 & 0.200 & 0.200 & 0.300 & 0.300 & 0.667 & 0.668 \\
   &  & (0.006) &  & (0.004) &  & (0.005) \\
  \addlinespace
  3 & 0.150 & 0.150 & 0.250 & 0.250 & 0.600 & 0.601 \\
   &  & (0.003) &  & (0.002) &  & (0.004) \\
  \addlinespace
  4 & 0.100 & 0.100 & 0.200 & 0.200 & 0.500 & 0.501 \\
   &  & (0.005) &  & (0.003) &  & (0.001) \\
  \addlinespace
  \bottomrule\bottomrule
\end{tabular}
	\end{center}
	\vspace{-1mm}
	\begin{footnotesize}
		\begin{spacing}{1}
			\noindent
			\textsc{Notes:} This table summarizes results from a Monte Carlo simulation based on the data generating process described in Section~\ref{sec_sim_dgp} and the parameter values specified in Section~\ref{sec_sim_params}. For each household type $k \in \left \{ 1, 2, 3, 4 \right \}$, the table reports the true parameter values alongside the corresponding estimates obtained using the regression discontinuity design and identification strategy outlined in Section~\ref{sec_ident_params_rdd}. Each entry in the ``Estimate'' columns reports the mean point estimate across 100 replications. The corresponding average standard error, obtained via the delta method, is reported in parentheses. 
		\end{spacing}
	\end{footnotesize}
\end{table}

\vspace{-2mm}

Table~\ref{table_sim_results_households} presents the results from the simulation exercise. For each household preference parameter, we report the true values alongside the corresponding estimates obtained by applying the identification strategy outlined above to regression discontinuity estimates from simulated datasets. Averaged over 100 replications, the method reliably recovers the parameters of interest, with average standard errors remaining modest. In addition, we estimate the location-specific amenity values $\left \{ \overline{A}_j \right \}_{j \in \mathcal{J}}$ and construction sector productivity shocks $\left \{ B_j \right \}_{j \in \mathcal{J}}$ by matching each household type's population mass $N^k_j$ to its model-implied counterpart.

We proceed in a similar fashion to estimate the elasticity of housing supply $\eta$. Specifically, we use equation \eqref{eq:totdiff_housing} and correctly recover the true value of the target parameter with negligible variation across replications.

\begin{table}[H]
	\begin{center}
		\caption{Comparison of True and Estimated Turnout Parameters}\label{table_sim_results_turnout}
		\vspace{0mm}
		\begin{tabular}{>{\centering\arraybackslash}m{2.2cm}>{\centering\arraybackslash}p{2.0cm}>{\centering\arraybackslash}p{2.0cm}>{\centering\arraybackslash}p{2.0cm}>{\centering\arraybackslash}p{2.0cm}}
  \toprule\toprule
  \multirow{2}{*}{\begin{tabular}[c]{@{}c@{}}\vspace{-0.0em}Household\\Type ($k$)\vspace{-0.5em}\end{tabular}} & \multicolumn{2}{c}{$\mu_0^k$} & \multicolumn{2}{c}{$\mu_1^k$} \\
  \cmidrule(lr){2-3}  \cmidrule(lr){4-5}
  & True & Estimate & True & Estimate \\
  \midrule
  \addlinespace
  1 & $-$3.000 & $-$2.995 & $-$1.000 & $-$1.020 \\
   &  & (0.007) &  & (0.037) \\
  \addlinespace
  2 & $-$5.000 & $-$4.966 & $-$1.000 & $-$1.076 \\
   &  & (0.015) &  & (0.054) \\
  \addlinespace
  3 & $-$7.000 & $-$7.021 & 0.000 & $-$0.000 \\
   &  & (0.011) &  & (0.074) \\
  \addlinespace
  4 & $-$3.000 & $-$3.002 & 0.000 & $-$0.002 \\
   &  & (0.018) &  & (0.048) \\
  \addlinespace
  \bottomrule\bottomrule
\end{tabular}
	\end{center}
	\vspace{-1mm}
	\begin{footnotesize}
		\begin{spacing}{1}
			\noindent
			\textsc{Notes:} This table summarizes results from a Monte Carlo simulation based on the data generating process described in Section~\ref{sec_sim_dgp} and the parameter values specified in Section~\ref{sec_sim_params}. For each household type $k \in \left \{ 1, 2, 3, 4 \right \}$, the table reports the true parameter values alongside the corresponding estimates obtained using the maximum likelihood procedure detailed in Section~\ref{sec_mle_turnout}. Each entry in the ``Estimate'' columns reports the mean point estimate across 100 replications. The corresponding average standard error, obtained via a parametric bootstrap with 100 iterations, is reported in parentheses. 
		\end{spacing}
	\end{footnotesize}
\end{table}

\vspace{-3mm}

Finally, we apply maximum likelihood to recover the intercept and slope coefficients that characterize the expected log cost of participating in a referendum. Specifically, we estimate $\left \{ \mu^k_0, \mu^k_1 \right \}_{k=1}^{4}$. Table~\ref{table_sim_results_turnout} summarizes the simulation results. As in the previous table, we report the true parameter values alongside their corresponding average estimates and standard errors across 100 replications. The results indicate that the proposed method accurately recovers the underlying parameters governing selection into political participation. This also holds for the variance intercept $\sigma_0$, whose average point estimate and standard error across replications are 3.008 and 0.012, respectively.

\subsection{Extrapolation of Average Housing Price Arc Elasticities}

Having established that the model's structural parameters can be estimated using regression discontinuity designs, we now implement the extrapolation procedure described in Section~\ref{sec_extrap_algorithm}.

We begin by considering the average treatment effect at the threshold. Under the model implied by the data generating process in Section~\ref{sec_sim_dgp} and the parameterization in Section~\ref{sec_sim_params}, referendum approval causes rental rates to increase at the cutoff. In addition, both $\E \left [ \log P_j(0) | S_j = s \right ]$ and $\E \left[ \log P_j (\Delta G_j) | S_j = s \right]$ are decreasing in $s$ on the left and right sides of the threshold, respectively.

Delving into the extrapolation algorithm, we use the samples of 2,000 ``observed'' referenda from each Monte Carlo replication that were employed to estimate cutoff-specific average effects and recover the structural parameters. For each of these referenda, we consider the spatial equilibrium prior to the vote and simulate 20 referenda featuring proposed changes in log school district expenditures in the $\left [ 0.01, 0.40 \right ]$ interval. 

\begin{figure}[t!]
	\caption{Average Extrapolated Effects of Referendum Approval}\label{fig_simul_extrap_effect}
	\vspace{0mm}
	\begin{minipage}[b]{0.495\textwidth}
		\begin{center}
			\includegraphics[width=\textwidth]{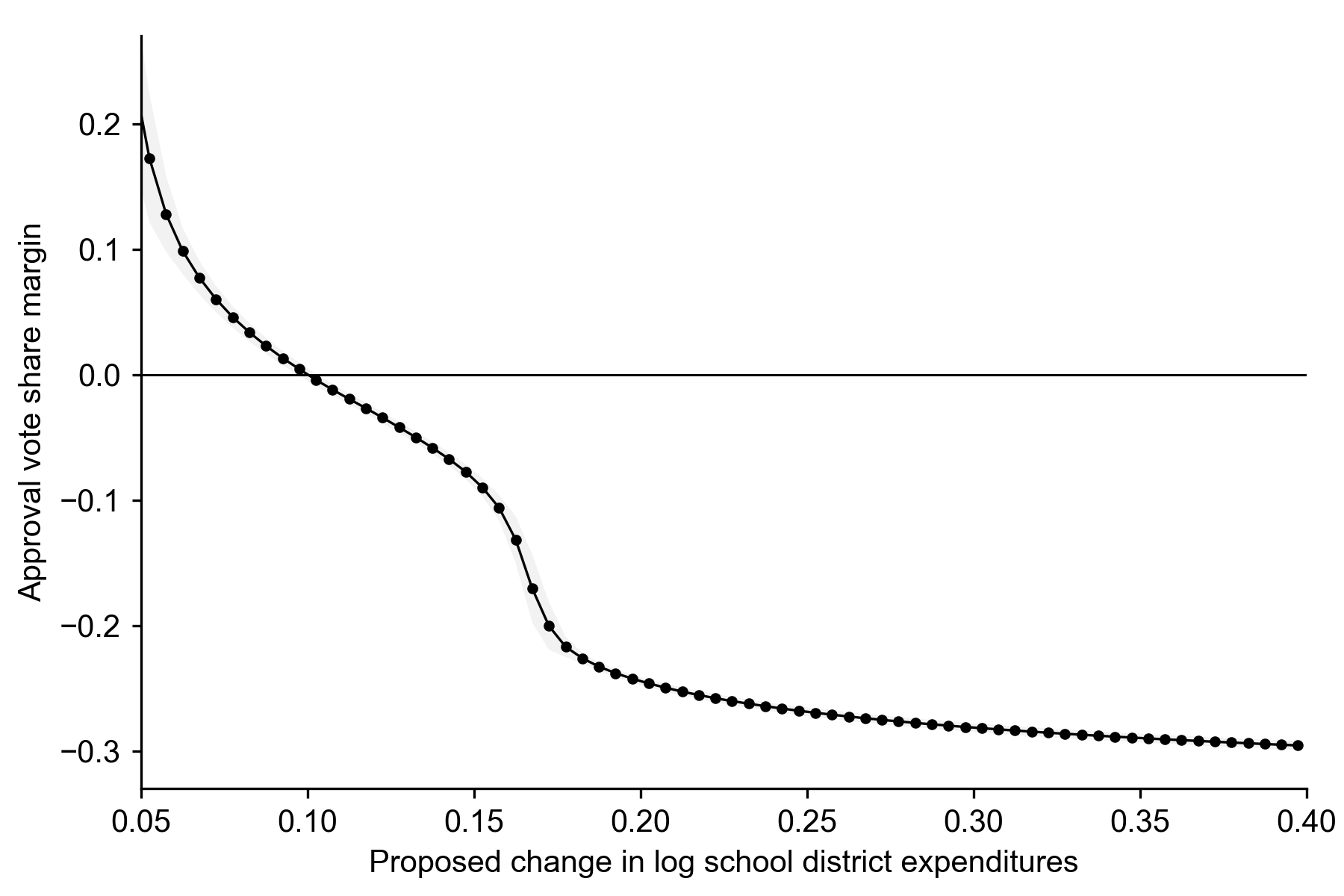}
			\subcaption{$\E \left [ S_j \left ( \Delta G_j \right ) | \Delta \log G_j \right ]$}
		\end{center}
	\end{minipage}
	\hfill
	\begin{minipage}[b]{0.495\textwidth}
		\begin{center}
			\includegraphics[width=\textwidth]{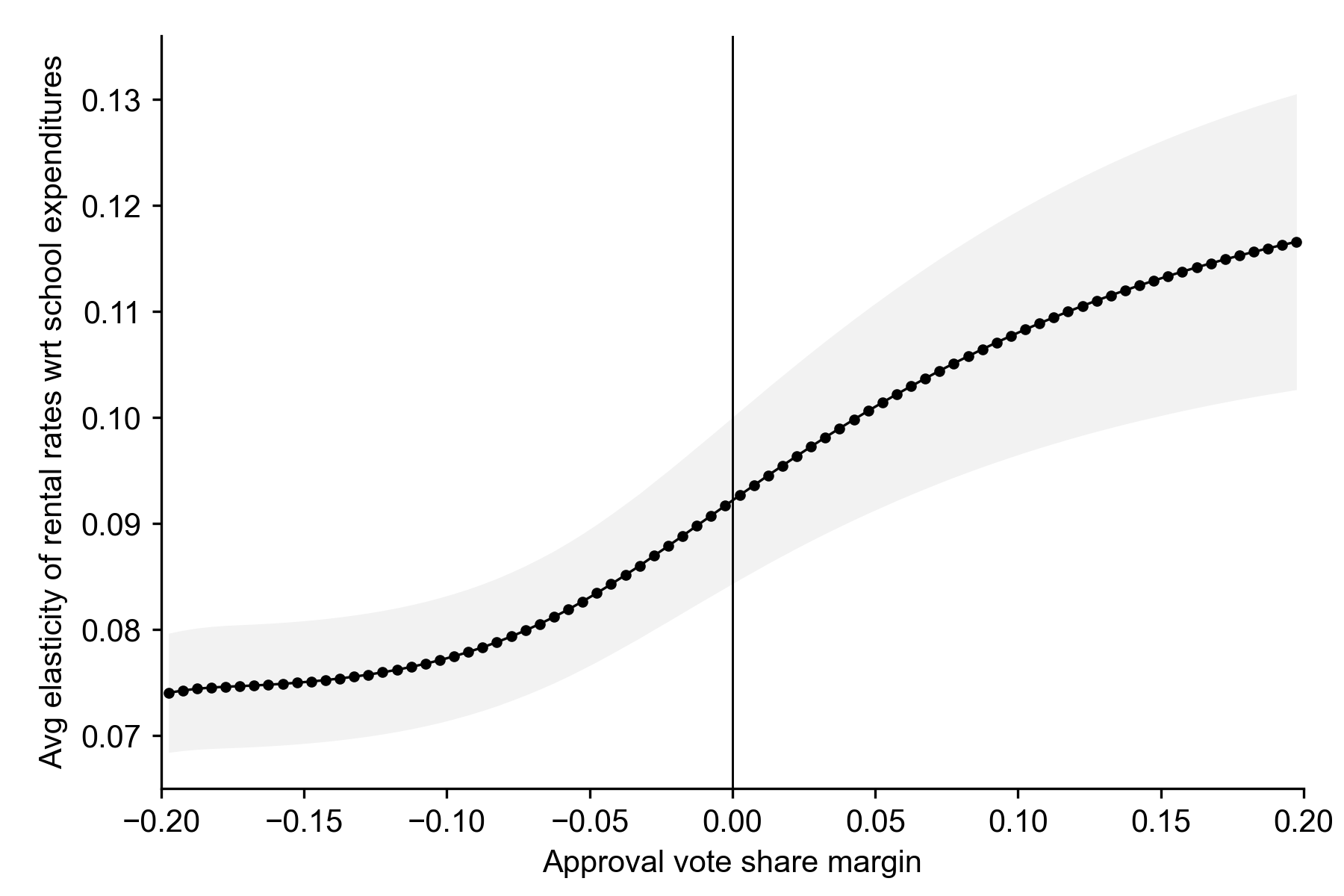}
			\subcaption{$\E \left [ \Delta \log P_j \slash \Delta \log G_j | S_j \left ( \Delta G_j \right ) \right ]$}
		\end{center}
	\end{minipage}
	\begin{minipage}[b]{0.495\textwidth}
		\begin{center}
			\includegraphics[width=\textwidth]{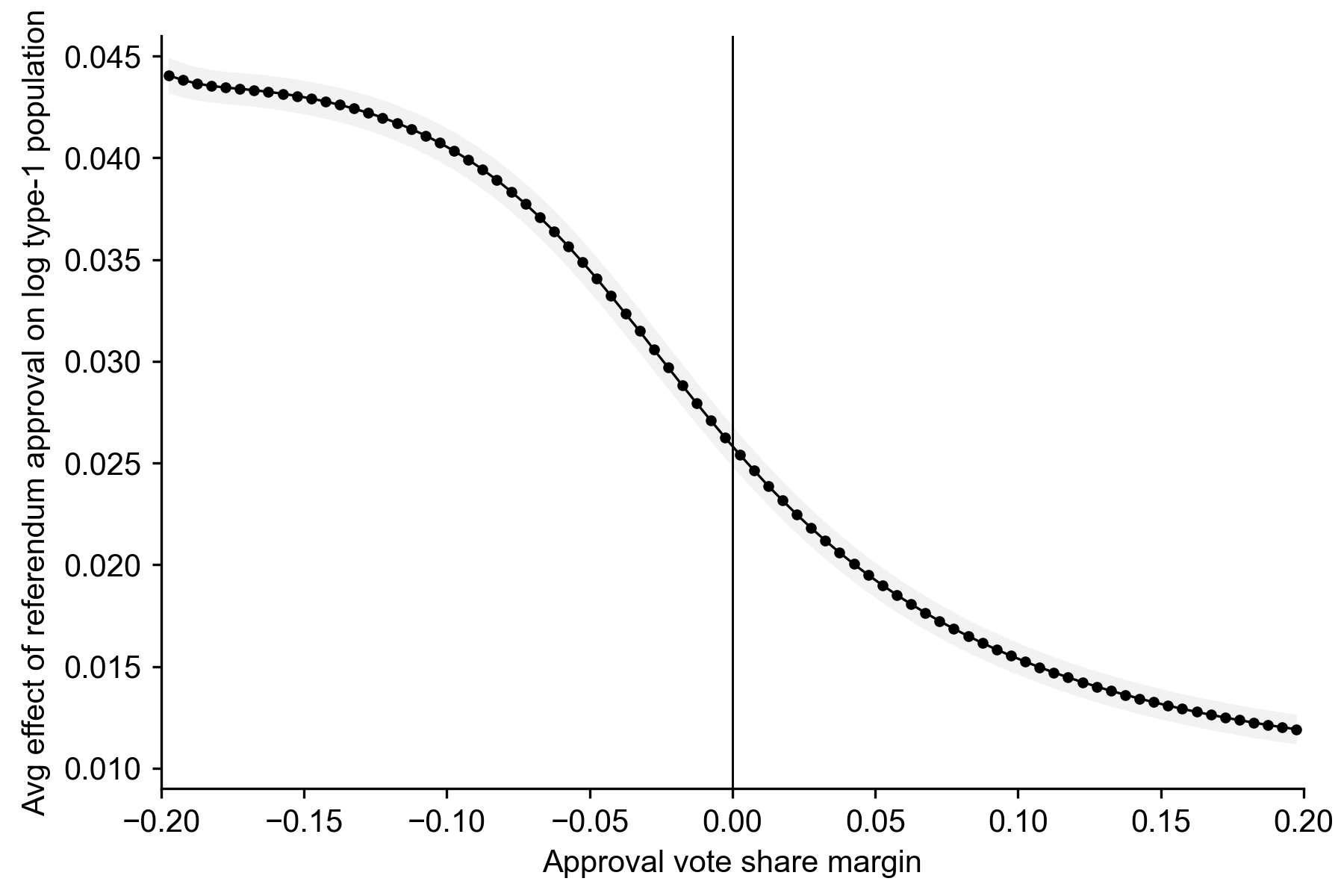}
			\subcaption{$\E \left [ \log N^1_j \left ( \Delta G_j \right ) - \log N^1_j \left ( 0 \right ) \big | S_j \left ( \Delta G_j \right ) \right ]$}
		\end{center}
	\end{minipage}
	\hfill
	\begin{minipage}[b]{0.495\textwidth}
		\begin{center}
			\includegraphics[width=\textwidth]{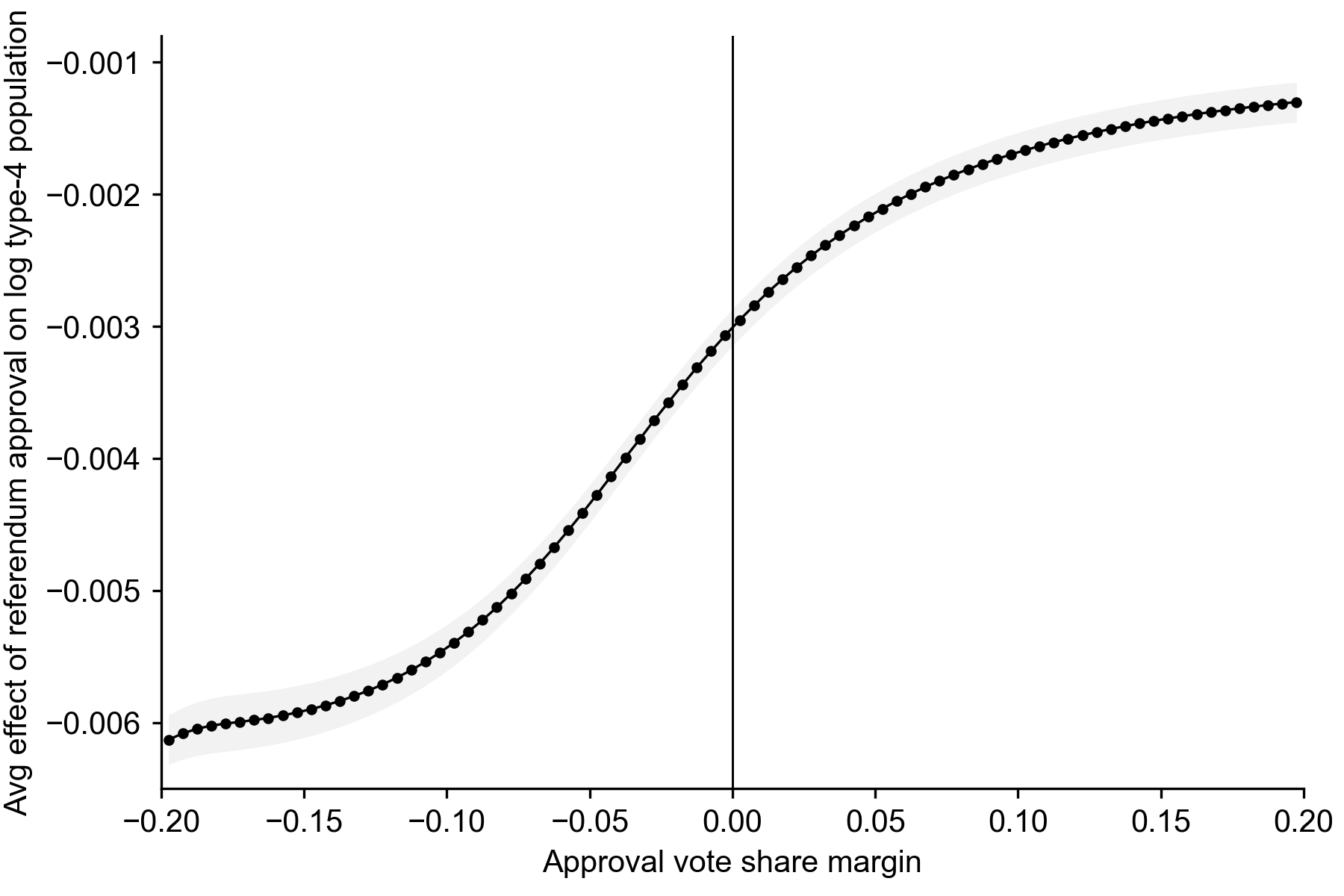}
			\subcaption{$\E \left [ \log N^4_j \left ( \Delta G_j \right ) - \log N^4_j \left ( 0 \right ) \big | S_j \left ( \Delta G_j \right ) \right ]$}
		\end{center}
	\end{minipage}
	\vspace{-2mm}
	\begin{footnotesize}
		\begin{spacing}{1}
			\noindent
			\textsc{Notes:} 
			This figure presents results from applying the extrapolation procedure described in Section~\ref{sec_extrap_algorithm} to synthetic data generated according to the process outlined in Section~\ref{sec_sim_dgp}. For each of the 2,000 referenda and each of the 100 Monte Carlo replications used to produce the estimates in Tables~\ref{table_sim_results_households} and~\ref{table_sim_results_turnout}, the extrapolation sample consists of 20 simulated referenda with proposed log expenditure changes in the $\left [ 0.01,0.40 \right ]$ interval. Panel (a) presents nonparametric estimates of the average approval vote share margin in bins of the proposed change in log school district expenditure. Panel (b) reports nonparametric estimates of the average arc elasticity of housing rents with respect to education spending in bins of the approval vote share margin. Panels (c) and (d) display analogous estimates for the average effect of referendum approval on the log mass of type $k=1$ and type $k=4$ households, respectively. In all panels, the bin width is 0.005. Bin-specific estimates are averaged across Monte Carlo replications. Shaded gray regions depict the corresponding pointwise average standard errors, computed using a two-step parametric bootstrap procedure with 100 outer replications and 20 inner replications per outer draw.
		\end{spacing}
	\end{footnotesize}
\end{figure}

Figure~\ref{fig_simul_extrap_effect} presents the results. Panel (a) displays nonparametric estimates of $\E \left [ S_j | \Delta \log G_j \right ]$, computed in bins of fixed width using a uniform kernel. The approval vote share margin appears to be, on average, a monotonically decreasing function of the proposed expenditure change. This pattern is intuitive: in a setting where the metropolitan area is in spatial equilibrium before the vote, referenda that entail smaller deviations from the status quo naturally garner broader support.

Panel (b) presents nonparametric estimates of $\E \left [ \Delta \log P_j \slash \Delta \log G_j | S_j \right ]$, also obtained using a uniform kernel and fixed-width bins. This plot represents the core result of the extrapolation exercise: the average arc elasticity of rental rates with respect to education spending is positive and increasing in the vote share margin near the cutoff. However, the slope attenuates for realizations of the running variable away from the threshold.

\par

This nonlinearity can be understood by examining Panels (c) and (d), which display analogous nonparametric estimates of $\E \left [ \log N^1_j \left( \Delta G_j \right) - \log N^1_j \left( 0 \right) | S_j \right ]$ and $\E \left [ \log N^4_j \left( \Delta G_j \right) - \log N^4_j \left( 0 \right) | S_j \right ]$, respectively. The average effect of referendum approval on the mass of $k=1$ households is positive, reflecting their high valuation of public school spending, whereas the effect on the mass of $k=4$ households is negative, consistent with their lower willingness to pay. For both groups, the magnitude of the average effect declines monotonically with the vote share margin, mirroring the pattern in Panel (a): referenda with a higher probability of passage typically involve smaller proposed expenditure changes and thus generate more modest mobility responses. As long as the net inflow of households remains positive, increased demand exerts upward pressure on rental rates, but this effect diminishes as the vote share margin moves farther from the cutoff.

\section{The Effects of School Expenditure Authorizations on Housing Prices for Nonmarginal Referenda}\label{sec_empappl}
	
In this section, we apply our method to estimate the effects of school expenditure authorization on housing prices in Wisconsin away from the referendum approval threshold.

\subsection{Spatial Partitioning and Household Heterogeneity}

Wisconsin comprises 28 Core-Based Statistical Areas (CBSAs), of which 15 are Metropolitan Statistical Areas (MSAs) and 13 are Micropolitan Statistical Areas ($\mu$SAs). We consider each of these CBSAs as a region in our spatial equilibrium model, meaning that each CBSA is partitioned into several school districts and households choose where to live within said CBSA or opt for the outside option, which we model as the combined areas of Wisconsin located outside CBSAs\footnote{According to the 2019-2023 American Community Survey, 13.4 percent of Wisconsin families live outside Core-Based Statistical Areas.}. Since CBSAs vary significantly in terms of population, we do not normalize their population to a unit mass and instead, for any household type $k$, interpret $\sigma^k$ and $N^k_j$ as population counts, rather than expected masses.

CBSAs naturally vary in their number of school districts $\mathcal{J}$\footnote{For the purpose of estimating the effect of referendum approval in jurisdictions other than the school district holding the referendum, we aggregate those school districts into an ``outer'' jurisdiction. Clearly, the definition of this outside area varies depending on the school district holding the referendum.}, while we set $\mathcal{K} = 4$ across the board. Specifically, we consider households whose income is above or below the Wisconsin median, further distinguished based on whether they have zero or a positive number of children aged less than 18 years old. Given our focus on location choice based on school district spending, we wish to differentiate families by their willingness to pay for public K-12 education services ($\alpha^k \slash \gamma^k$ in our model), and presence of children jointly with income are two likely salient factors for this parameter.
	
\subsection{Data}

In Section \ref{sec_data_avgval}, we described how we construct a panel on average housing values in Wisconsin school districts. We complement housing price data with household-type-specific population counts from the 2000 Decennial Census and the five-year American Community Surveys (ACSs) ranging from 2005-2009 to 2019-2023\footnote{Because family counts based on presence of dependent children and income are not available prior to the 2000 Decennial Census, these outcomes cannot be measured exactly five years after each referendum, as we instead can do for housing prices. We then adopt the following solution. For referenda that occurred between 1990 and 1995, population count outcomes are measured in the 2000 Decennial Census. For referenda that occurred between 1996 and 2000, population count outcomes are measured in the 2005-2009 ACS. Starting from 2001, referenda are linked to the five-year American Community Survey that begins exactly five years later. That is, we use the 2006-2010 ACS for referenda in 2001, the 2007-2011 ACS for referenda in 2002, and so on until referenda that took place in 2014, for which we use the last available, 2019-2023 ACS.}.

In addition, we collect data on school district finances provided by the Wisconsin Department of Public Instruction. Specifically, we draw each district's revenue from all sources, property tax revenue, and property tax rate. We use the revenue from all sources, including grants from the federal and state governments, to measure $G_j$ in our model, effectively imposing that jurisdictions balance their budget. In doing so, we assume that intergovernmental transfers are fixed for each jurisdiction and are not adjusted in response to changes in property tax rates\footnote{In Wisconsin, the primary source of state aid to school districts is the State Equalization Aid program. This program allocates funds through a three-tier formula under which the share of a district’s costs covered by state aid declines as the district’s property tax base per pupil increases. As a result, when a school district approves a referendum to raise expenditures, its state aid is mechanically reduced, though by less than one dollar for each additional dollar of authorized spending. In addition, this offset does not apply to referenda authorizing the issuance of general obligation bonds to finance capital expenditures, which often involve the largest proposed increases in spending. Approximately 75 percent of capital outlays are financed by school districts using local revenue (\citealt{filardo2016}), and the distribution of these expenditures varies substantially within the state (\citealt{biasi2021}; \citealt{bilaschon2024}).}. This choice does not invalidate our identification strategy because changes in school expenditure authorized by referenda are repaid entirely with revenue from property taxes. Finally, we use the property tax rate to infer the number of housing units $H_j$ by dividing each district's property tax revenue by the average property tax liability, namely the product of the average housing price $P_j$ by $\tau_j$.

\subsection{Structural Parameter Estimates}

We begin by revisiting the indirect utility function $V_{ij}$ and applying two standard normalizations. First, we divide all terms in the utility function by the strictly positive parameter $\gamma^k$. This transformation allows us to express the preference parameter $\alpha^k \slash \gamma^k$ as the marginal utility of school district expenditure in units of income rather than in utils, thereby facilitating interpretation. With a slight abuse of notation, we denote the rescaled indirect utility of household $i$ in district $j$ as
{\setlength{\abovedisplayskip}{7pt plus 2pt minus 2pt}
	\setlength{\belowdisplayskip}{7pt plus 2pt minus 2pt}
\begin{equation}
	V_{ij} = \frac{\overline{A}_j}{\gamma^k} + \frac{\alpha^k}{\gamma^k} \log G_j + \log \left [ y^k - P_j \left ( 1 + \tau_j \right ) \right ] + U_{ij}
\end{equation}}
where the idiosyncratic component $U_{ij}$ follows a Gumbel distribution with scale parameter $\theta^k \slash \gamma^k$. Second, we impose the normalization $\theta^k \slash \gamma^k = 1$, which, while affecting the scale of utility, does not alter the choice probabilities. Since our analysis does not involve computing welfare measures expressed in utils, this normalization is without loss of generality.

We are now ready to estimate the structural parameters $\left \{ \alpha^k / \gamma^k \right \}_{k=1}^{4}$ by leveraging the system of equations implied by the choice probabilities. As detailed in equations~\eqref{eq_appx_popown} and~\eqref{eq_appx_popother}, this step involves, for each household type, the estimation of 18 regression discontinuity coefficients, each of which identifies the $\text{WAVE} \left ( 0 \right )$ of a distinct outcome with respect to school expenditures, i.e., a weighted average of arc elasticities with respect to the underlying policy variable. We adopt a similar approach to estimate the elasticity of housing supply $\eta$, which, as discussed in Section~\ref{sec_ident_params_rdd}, can be recovered from just two RDD estimates.

Table~\ref{table_appl_params} reports the estimated structural parameters. Across the four household groups, the marginal willingness to pay for K–12 education expenditures---captured by $\alpha / \gamma$---is below one. Although the standard errors of pairwise differences are not small enough to support formal statistical comparisons, the point estimates display meaningful heterogeneity. In particular, $\alpha / \gamma$ is highest among households with children under the age of 18 and income above the median, and lowest among households with children and income below the median. This pattern is consistent with the findings of \cite{bilaschon2024}, which shows that the approval of school expenditure referenda affects the composition of the student body, reducing the share of Hispanic students, increasing the share of Asian students, and decreasing the proportion of pupils eligible for free or reduced-price lunch (FRPL). Taken together, this evidence suggests that household sorting across school districts is a salient margin of adjustment in response to changes in local public spending. Finally, our estimate of the elasticity of housing supply is 0.44, a value we consider plausible given our focus on relatively urbanized areas.

\begin{table}[H]
	\begin{center}
		\caption{Estimates of $\left \{ \alpha^k \slash \gamma^k \right \}_{k=1}^{4}$ and $\eta$}\label{table_appl_params}
		\vspace{0mm}
		\begin{tabular}{>{\centering\arraybackslash}p{2cm}>{\raggedright\arraybackslash}p{7.5cm}>{\centering\arraybackslash}p{2cm}}
  \toprule\toprule
  Parameter & Group & Estimate \\
  \midrule
  \addlinespace
  $\alpha^1 \slash \gamma^1$ & With Children, Below Median Income & 0.693 \\
  &  & (0.119) \\
  \addlinespace
  $\alpha^2 \slash \gamma^2$ & With Children, Above Median Income & 0.868 \\
  &  & (0.195) \\
  \addlinespace
  $\alpha^3 \slash \gamma^3$ & Without Children, Below Median Income & 0.709 \\
  &  & (0.171) \\
  \addlinespace
  $\alpha^4 \slash \gamma^4$ & Without Children, Above Median Income & 0.830 \\
  &  & (0.209) \\
  \addlinespace
  $\eta$ &  & 0.439 \\
  &  & (0.079) \\
  \bottomrule\bottomrule
\end{tabular}
	\end{center}
	\vspace{-1mm}
	\begin{footnotesize}
		\begin{spacing}{1}
			\noindent
			\textsc{Notes:} This table presents estimates of $\left \{ \alpha^k \slash \gamma^k \right \}_{k=1}^{4}$, which measure each household group's marginal willingness to pay for public education expenditure in units of income, and $\eta$, the elasticity of housing supply. Point estimates are obtained by solving the systems of equations implied by household choice probabilities \eqref{eq_discrete_households} and the housing supply equation \eqref{eq_discrete_housingsupply}, using regression discontinuity (RDD) estimates as inputs. Standard errors are computed via the delta method.
		\end{spacing}
	\end{footnotesize}
\end{table}

\vspace{-3mm}

Having completed the estimation of the parameters identified solely through RDD coefficients, we proceed to estimate the location-type-specific intercepts $\left \{ \overline{A}_j \slash \gamma^k \right \}_{j,k}$ for all school districts located within CBSAs in Wisconsin. For each CBSA, we solve the system of equations that set the model-implied conditional population masses $N^k_j \slash \sigma^k$ equal to their observed counterparts in the year prior to each referendum. We follow an analogous procedure to estimate the location-specific productivity terms $\left \{ B_j \right \}_{j}$ in the construction sector, along with the common intercept $\lambda$. To achieve point identification, we impose the normalization that the mean of $B_j$ across locations is zero.

Finally, we estimate the parameter vector $\vartheta$, which governs the unobserved log cost of participation in local referenda, using the procedure described in Section~\ref{sec_mle_turnout}. To reduce the dimensionality of the parameter space, we impose a common slope $\mu_1$ with respect to the proposed spending change $\Delta \log G_j$ and a common standard deviation $\sigma_0$. Table~\ref{table_appl_params_turnout} presents the results. As expected, the estimated value of $\mu_1$ is negative, consistent with the notion that participation costs decline in higher-stakes referenda, potentially due to lower informational or attentional barriers when the proposed policy is more salient (e.g., the construction or renovation of a school). The household-group-specific intercepts exhibit a notable pattern: participation costs are highest among households with children and income below the median, and lowest among those without children---a group that likely includes most retirees. These findings align with patterns of voter selection in U.S. local elections documented by \cite{berry2024report}.

\begin{table}[H]
	\begin{center}
		\caption{Estimates of Turnout Parameters}\label{table_appl_params_turnout}
		\vspace{0mm}
		\begin{tabular}{>{\centering\arraybackslash}p{2cm}>{\raggedright\arraybackslash}p{7.5cm}>{\centering\arraybackslash}p{2cm}}
  \toprule\toprule
  Parameter & Group & Estimate \\
  \midrule
  \addlinespace
  $\mu_0^1$ & With Children, Below Median Income & $2.49$ \\
  &  & (0.75) \\
  \addlinespace
  $\mu_0^2$ & With Children, Above Median Income & $-1.56$ \\
  &  & (0.48) \\
  \addlinespace
  $\mu_0^3$ & Without Children, Below Median Income & $-4.52$ \\
  &  & (1.33) \\
  \addlinespace
  $\mu_0^4$ & Without Children, Above Median Income & $-4.94$ \\
  &  & (1.48) \\
  \addlinespace
  $\mu_1$ &  & $-1.52$ \\
  &  & (0.69) \\
  \addlinespace
  $\sigma_0$ &  & $3.42$ \\
  &  & (0.35) \\
  \bottomrule\bottomrule
\end{tabular}
	\end{center}
	\vspace{-1mm}
	\begin{footnotesize}
		\begin{spacing}{1}
			\noindent
			\textsc{Notes:} This table reports estimates of the following parameters: $\left \{ \mu^k_0 \right \}_{k=1}^{4}$, the set of household-group-specific intercepts in the average log cost of participation in local referenda; $\mu_1$, the common slope with respect to the proposed spending change $\Delta \log G_j$; and $\sigma_0$, the common standard deviation of the unobserved log cost. Point estimates are obtained via the maximum likelihood procedure outlined in Section~\ref{sec_mle_turnout}, conditioning on the estimated parameter vector $\widehat{\zeta}$ reported in Table~\ref{table_appl_params}. Standard errors are computed using a parametric bootstrap procedure with 500 replications.
		\end{spacing}
	\end{footnotesize}
\end{table}

\vspace{-5mm}

\subsection{Extrapolation of Average Housing Price Arc Elasticities}

We now implement the extrapolation algorithm described in Section~\ref{sec_extrap_algorithm} using the full dataset of referenda held by Wisconsin school districts. For each observed referendum, we generate 75 counterfactual consultations involving proposed changes in log education expenditure drawn from a discrete set $\mathcal{G}$ comprising uniformly spaced values over the interval $\left [ 0.01, 0.16 \right ]$. For each simulated policy scenario, we compute the realization of the potential approval vote share margin $S_j \left ( \Delta G_j \right )$, the potential rental rate of housing under referendum rejection $\log P_j \left ( 0 \right )$, and the potential rental rate of housing under referendum approval $\log P_j \left ( \Delta G_j \right )$. Based on these, we calculate the arc elasticity of rental rates with respect to education spending as
\begin{equation}
	E_{P_j} \left ( \Delta G_j \right ) \equiv \frac{\log P_j \left ( \Delta G_j \right ) -  \log P_j \left ( 0 \right )}{\Delta \log G_j}
\end{equation}
To examine how these elasticities vary with the degree of local political consensus, we partition the support of the simulated approval margin $S_j \left ( \Delta G_j \right )$ into 200 bins of equal width $\kappa = 0.02$ and compute the average arc elasticity within each bin. Specifically, for each $b \in \left \{ -0.50, -0.48, \dots, 0.48 \right \}$, we estimate
\begin{equation}
	\widehat{\text{AVE}}_{P_j} \left ( b \right ) \equiv \frac{\sum_{\Delta G_j \in \mathcal{G}} E_{P_j} \left ( \Delta G_j \right ) \times \mathbb{I} \left [ S_j \left ( \Delta G_j \right ) \in \left [ b, b+\kappa \right ) \right ]}{\sum_{\Delta G_j \in \mathcal{G}} \mathbb{I} \left [ S_j \left ( \Delta G_j \right ) \in \left [ b, b+\kappa \right ) \right ]}
\end{equation}
The resulting set of estimates $\Big \{ \widehat{\text{AVE}}_{P_j} \left ( b \right ) \Big \}_{b}$ comprises average extrapolated effects of education spending on housing prices over a broad range of approval margins, extending beyond the local parameter identified at the cutoff in Section \ref{sec_identification_cutoff}.

\begin{figure}[H]
	\caption{Average Extrapolated Effects of Referendum Approval}\label{fig_appl_extrap_effect}
	\vspace{1mm}
	\begin{minipage}[b]{0.495\textwidth}
		\begin{center}
			\includegraphics[width=\textwidth]{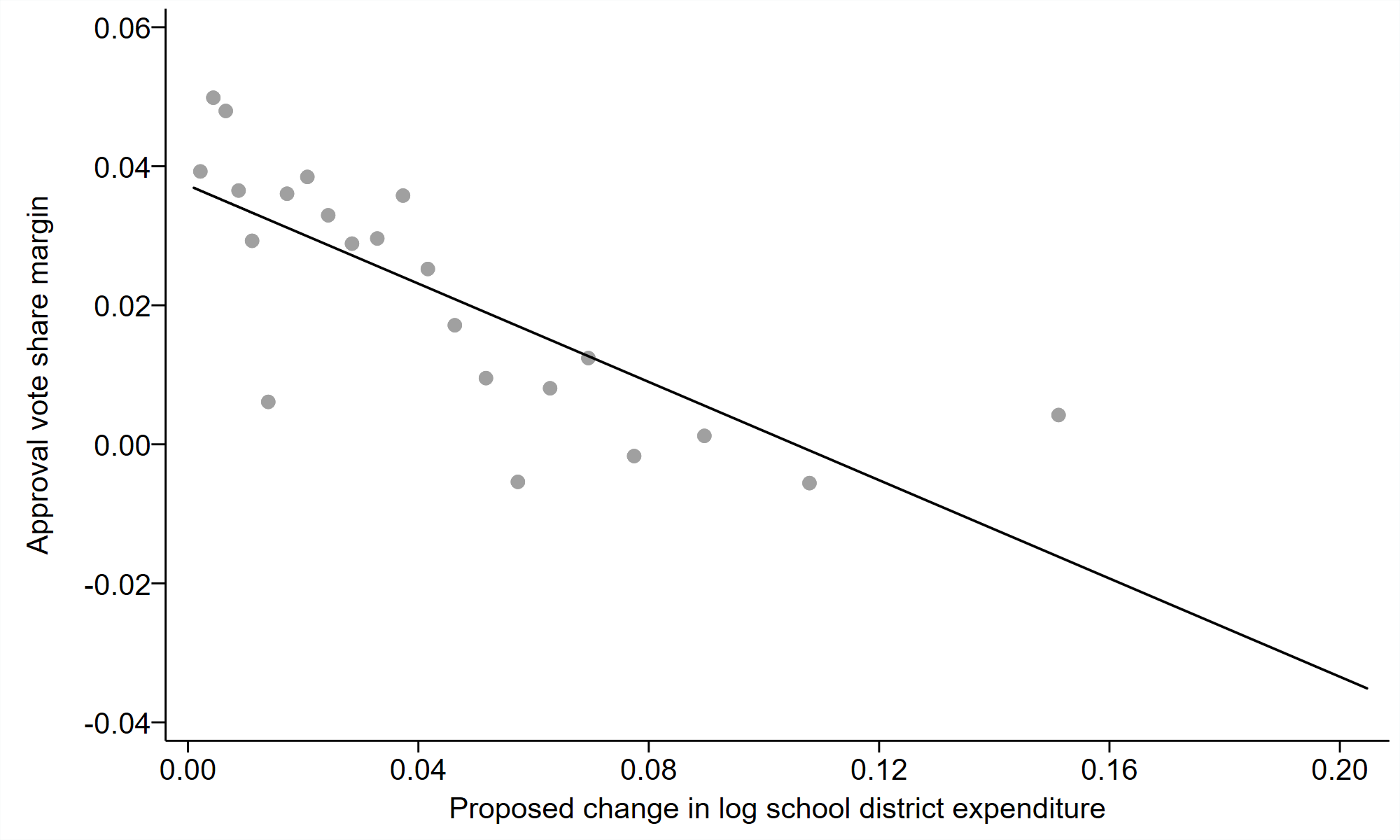}
			\subcaption{$\E \left [ S_j | \Delta \log G_j \right ]$}
		\end{center}
	\end{minipage}
	\hfill
	\begin{minipage}[b]{0.495\textwidth}
		\begin{center}
			\includegraphics[width=\textwidth]{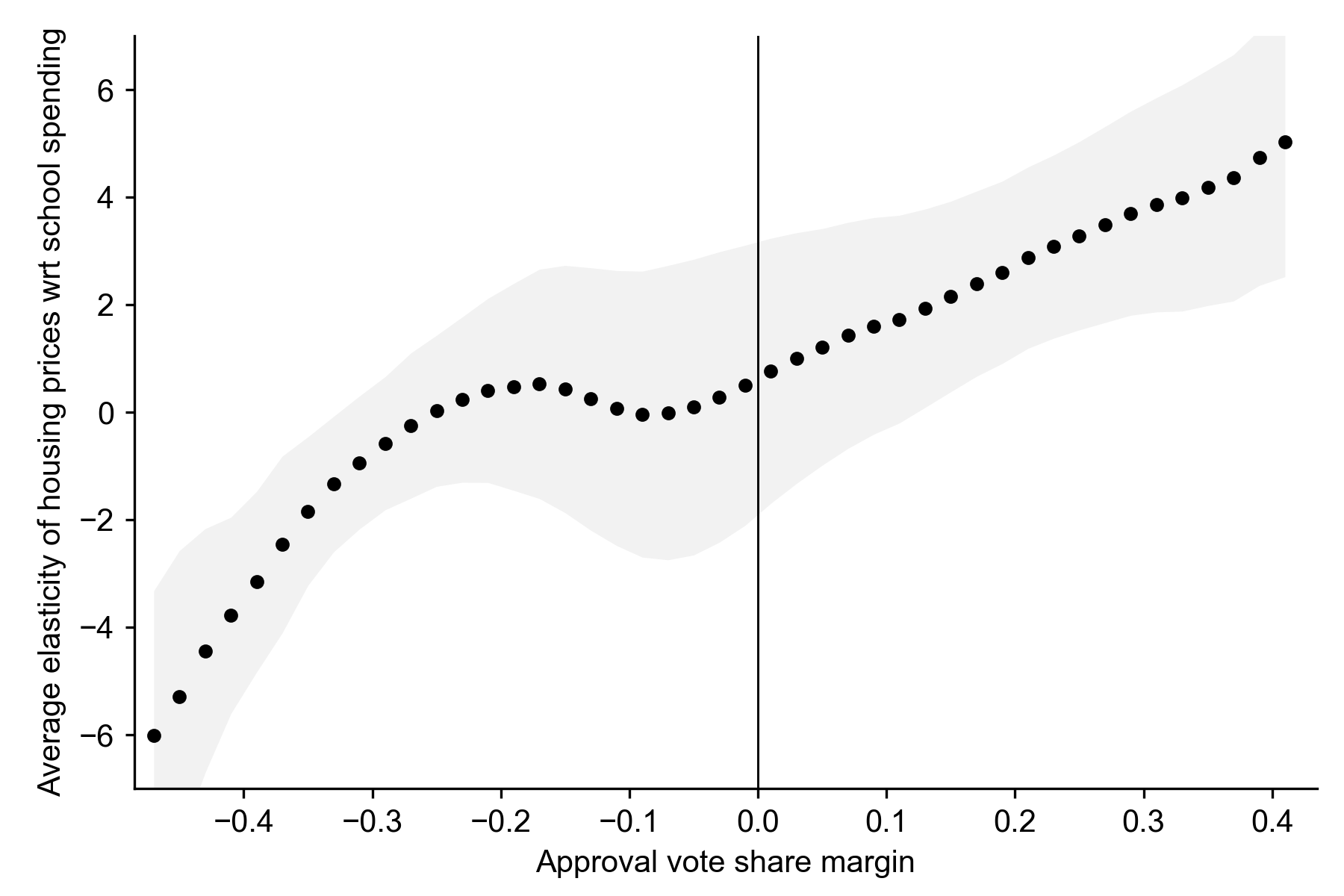}
			\subcaption{$\E \left [ \Delta \log P_j \slash \Delta \log G_j | S_j \left ( \Delta G_j \right ) \right ]$}
		\end{center}
	\end{minipage}
	\vspace{-2mm}
	\begin{footnotesize}
		\begin{spacing}{1}
			\noindent
			\textsc{Notes:} 
			This figure presents results from applying the extrapolation procedure described in Section~\ref{sec_extrap_algorithm} to a dataset of 3,528 school district referenda held in Wisconsin between 1990 and 2022. For each of these local consultations, the extrapolation sample consists of 75 simulated referenda with proposed log expenditure changes selected from a grid of values spanning the interval $\left [ 0.01, 0.16 \right ]$. Based solely on observed referenda, Panel (a) displays nonparametric estimates of the average approval vote share margin in bins of the proposed change in log school district expenditure. Estimates are adjusted for a vector of covariates that includes referendum year indicators and school district indicators. Using the extrapolation sample, Panel (b) presents nonparametric estimates of the average arc elasticity of housing prices with respect to school district spending in bins of the approval vote share margin. Shaded gray regions denote 90 percent confidence intervals. Standard errors are computed using a two-step parametric bootstrap procedure with 500 outer replications and 50 inner replications per outer draw. In Panel (a), the number of bins is selected to minimize the integrated mean squared error (IMSE) of the semi-linear covariate-adjusted estimator of the conditional outcome mean (\citealt{cattaneobinscatter2024}). In Panel (b), bins have a fixed width of 0.02.
		\end{spacing}
	\end{footnotesize}
\end{figure}

Figure~\ref{fig_appl_extrap_effect} presents the results of the extrapolation procedure. Panel (a), based exclusively on observed referenda, corroborates the theoretical prediction that proposals involving larger increases in education expenditure tend to be approved by narrower margins. Panel (b) displays the principal empirical finding of our analysis: the average arc elasticity of housing prices with respect to education spending varies substantially across the approval vote margin. To the right of the cutoff, the average elasticity increases steadily, implying that the positive capitalization effect estimated at the threshold, and documented in prior work (\citealt{cfr2010}, \citealt{bilaschon2024}), extends to referenda supported by a larger share of voters. To the left of the cutoff, the average elasticity initially remains close to zero, but then declines and turns negative for ballot measures that garnered limited approval. This pattern indicates that some rejected proposals would have reduced local housing demand through negative net sorting responses. That is, the aggregate willingness to pay for education services would have fallen sufficiently to depress equilibrium prices, primarily due to net outmigration of household groups with comparatively low valuation of public education relative to income. Taken together, these findings underscore that estimates at the approval threshold are informative but may not capture housing market responses across a broader range of vote shares.

\vspace{-6mm}

\section{Conclusion}\label{sec_concl}

Regression discontinuity designs (RDDs) are widely used in program evaluation due to their high internal validity, which stems from the relatively weak continuity assumptions required to identify average effects for units at the margin between treated and control arms. However, the extent to which these threshold-specific estimates generalize to nonmarginal units remains uncertain, and is often critical for informing policy. Most existing extrapolation methods rely on statistical assumptions about the joint distribution of outcomes, the running variable, treatment assignment, and covariates.

In this paper, we adopt a different approach by embedding the RDD in a model of economic behavior grounded in first principles. This framework provides a microfoundation for the running variable and its joint determination with economically related outcomes. By formally linking reduced-form RDD estimates to structural parameters, we recover primitives and use them to analyze counterfactual scenarios in which the running variable takes values away from the cutoff. Our approach allows us to infer the levels of relevant outcomes under both treatment states at nonmarginal realizations of the running variable—thereby identifying average partial effects beyond the threshold.

Using data from Wisconsin, we estimate that the average arc elasticity of housing prices with respect to education expenditures is approximately equal to one at the referendum approval threshold. However, our extrapolation method reveals substantial heterogeneity away from the cutoff. To the right of the threshold, the average elasticity increases steadily, implying that the positive capitalization effect estimated at the margin extends to referenda supported by a larger share of voters. To the left, the average elasticity declines and becomes negative for ballot measures that garnered limited approval. This pattern indicates that some rejected proposals would have reduced local housing demand, primarily due to a net outflow of households with a relatively low willingness to pay for enhanced education services. Taken together, these findings suggest that housing market responses to locally determined changes in government spending may vary systematically with the degree of voter support, underscoring the importance of accounting for nonmarginal variation when evaluating the incidence and efficiency of local fiscal policies.


\newpage

\setstretch{1.0}
\renewcommand\refname{References}
\bibliography{../../references/references}


\setstretch{1.5}

\clearpage

\appendix

\renewcommand{\thesubsection}{\Alph{section}.\arabic{subsection}}
\renewcommand{\thesubsubsection}{\Alph{section}.\arabic{subsection}.\arabic{subsubsection}}

\setcounter{figure}{0} \renewcommand{\thefigure}{\Alph{section}\arabic{figure}}
\setcounter{table}{0} \renewcommand{\thetable}{\Alph{section}\arabic{table}}
\setcounter{equation}{0} \renewcommand{\theequation}{\Alph{section}.\arabic{equation}}
\setcounter{theorem}{0} \renewcommand{\thetheorem}{\Alph{section}.\arabic{theorem}}

\addcontentsline{toc}{section}{Online Appendix}

\listofappendixcontents

\appsection{Causal Interpretation of the Target Estimand}\label{appx_proof_estimand}

Consider the reduced-form regression discontinuity estimand that uses $\log P_j$ as the outcome:
\begin{align}
	& \lim_{s \downarrow 0} \E \left [ \log P_j | S_j = s \right ] - \lim_{s \uparrow 0} \E \left [ \log P_j| S_j = s \right ] \notag \\
	& = \lim_{s \downarrow 0} \E \left [ \log P_j | S_j = s, D_j = 1 \right ] - \lim_{s \uparrow 0} \E \left [ \log P_j| S_j = s, D_j = 0 \right ] \\
	& = \lim_{s \downarrow 0} \E \left [ \log P_j \left ( \Delta G_j \right ) | S_j = s, D_j = 1 \right ] - \lim_{s \uparrow 0} \E \left [ \log P_j \left ( 0 \right ) | S_j = s, D_j = 0 \right ] \\
	& = \lim_{s \downarrow 0} \E \left [ \log P_j \left ( \Delta G_j \right ) | S_j = s \right ] - \lim_{s \uparrow 0} \E \left [ \log P_j \left ( 0 \right ) | S_j = s \right ] \\
	& = \E \left [ \log P_j \left ( \Delta G_j \right ) | S_j = 0 \right ] - \E \left [ \log P_j \left ( 0 \right ) | S_j = 0 \right ] \\
	& = \E \left [ \log P_j \left ( \Delta G_j \right ) - \log P_j \left ( 0 \right ) | S_j = 0 \right ] \\
	& = \E \left [ \frac{\log P_j \left ( \Delta G_j \right ) - \log P_j \left ( 0 \right )}{\Delta \log G_j} \times \Delta \log G_j \bigg | S_j = 0 \right ]
\end{align}
The first and third equalities rely on the definition of the referendum approval indicator, which allows the conditioning set to be expanded to include the event $D_j = 1$ or $D_j = 0$, depending on whether $S_j$ lies above or below the threshold, respectively. The second equality follows from the fact that, conditional on $D_j = 1$, the potential outcome $P_j \left( \Delta G_j \right)$ is observed, and similarly, $P_j (0)$ is observed when $D_j = 0$. The fourth equality uses the standard assumption that, for $d \in \left \{ 0,1 \right \}$, $\E \left [ P_j \left ( d \times \Delta G_j \right ) | S_j = s \right ]$ is a continuous function of $s$ at $s = 0$. The fifth equality exploits the linearity of the expectation operator. The sixth equality follows from multiplying and dividing by $\Delta \log G_j$. Analogously, consider the first-stage regression discontinuity estimand that uses $D_j \times \Delta \log G_j$ as the outcome:
\begin{align}
	& \lim_{s \downarrow 0} \E \left [ D_j \times \Delta \log G_j | S_j = s \right ] - \lim_{s \uparrow 0} \E \left [ D_j \times \Delta \log G_j | S_j = s \right ] \notag \\
	& = \lim_{s \downarrow 0} \E \left [ D_j \times \Delta \log G_j | S_j = s, D_j = 1 \right ] - \lim_{s \uparrow 0} \E \left [ D_j \times \Delta \log G_j | S_j = s, D_j = 0 \right ] \\
	& = \lim_{s \downarrow 0} \E \left [ \Delta \log G_j | S_j = s, D_j = 1 \right ] \\
	& = \lim_{s \downarrow 0} \E \left [ \Delta \log G_j | S_j = s \right ] \\ 
	& = \E \left [ \Delta \log G_j | S_j = 0 \right ]
\end{align}
The first and third equalities rely on the definition of the referendum approval indicator, which allows the conditioning set to be expanded to include the event $D_j = 1$ or $D_j = 0$, depending on whether $S_j$ lies above or below the threshold, respectively. The second equality exploits the fact that $D_j = 1$ and $D_j = 0$ are inside both the integrand and the conditioning set. The fourth equality uses the assumption that $\E \left [ \Delta \log G_j | S_j = s \right ]$ is a continuous function of $s$ at $s = 0$. Combining these two identification results yields the fuzzy regression discontinuity estimand
\begin{align}
	& \frac{\lim_{s \downarrow 0} \E \left [ \log P_j | S_j = s \right ] - \lim_{s \uparrow 0} \E \left [ \log P_j| S_j = s \right ]}{\lim_{s \downarrow 0} \E \left [ D_j \times \Delta \log G_j | S_j = s \right ] - \lim_{s \uparrow 0} \E \left [ D_j \times \Delta \log G_j | S_j = s \right ]} \notag \\
	& = \E \left [ \frac{\log P_j \left ( \Delta G_j \right ) - \log P_j \left ( 0 \right )}{\Delta \log G_j} \times \Delta \log G_j \bigg | S_j = 0 \right ] \frac{1}{\E \left [ \Delta \log G_j | S_j = 0 \right ]} \\
	& = \E \left [ \frac{\log P_j \left ( \Delta G_j \right ) - \log P_j \left ( 0 \right )}{\Delta \log G_j} \frac{\Delta \log G_j}{\E \left [ \Delta \log G_j | S_j = 0 \right ]} \bigg | S_j = 0 \right ] \\
	& = \E \left [ \omega_j \times \frac{\log P_j \left ( \Delta G_j \right ) - \log P_j \left ( 0 \right )}{\Delta \log G_j} \bigg | S_j = 0 \right ]
\end{align}
with the weight defined as
\begin{equation}
	\omega_j \equiv \frac{\Delta \log G_j}{\E \left [ \Delta \log G_j | S_j = 0 \right ]}
\end{equation}
Clearly, $\E \left [ \omega_j | S_j = 0 \right ] = 1$.

\appsection{Model Derivations}

\appsubsection{Household Utility Maximization}

Household $i$ faces the following utility maximization problem in location $j$:
\begin{align}
	& \max_{H,X} \left \{ A_{ij} + \alpha_i \log \frac{G_j}{N_j^\chi} + \beta_i \log H + \gamma_i \log X \right \} \notag \\
	& \ \text{s.t.} \quad X + P_j H \left ( 1+\tau_j \right ) \leq Y_i \ \text{ and } \ H = 1
\end{align}
The Lagrangian associated with this maximization problem is
\begin{align}
	\mathcal{L} \left ( H, X; \lambda \right ) & = A_{ij} + \alpha_i \log \frac{G_j}{N_j^\chi} + \beta_i \log H + \gamma_i \log X \notag \\
	& - \lambda \left ( X + P_j H \left ( 1+\tau_j \right ) - Y_i \right ) - \mu \left ( H - 1 \right )
\end{align}
The first-order necessary conditions are
\begin{align}
	\frac{\partial \mathcal{L} \left ( H, X; \lambda \right )}{\partial H_{ij}} & = \frac{\beta_i}{H_{ij}} - \lambda P_j \left ( 1 + \tau_j \right ) - \mu = 0 \\
	\frac{\partial \mathcal{L} \left ( H, X; \lambda \right )}{\partial X_{ij}} & =
	\frac{\gamma_i}{X_{ij}} - \lambda = 0 \\
	\frac{\partial \mathcal{L} \left ( H, X; \lambda \right )}{\partial \lambda} & = - X_{ij} - P_j H_{ij} \left ( 1+\tau_j \right ) + Y_i = 0 \\
	\frac{\partial \mathcal{L} \left ( H, X; \lambda \right )}{\partial \mu} & = - H_{ij} + 1 = 0
\end{align}
The fourth first-order condition ensures that $H_{ij} = 1$. Then the budget constraint is
\begin{equation}
	X_{ij} + P_{j} \left ( 1 + \tau_j \right ) = Y_i \iff X_{ij} = Y_i - P_{j} \left ( 1 + \tau_j \right )
\end{equation}
The second first-order condition implies that the first Lagrange multiplier is
\begin{equation}
	\lambda = \frac{\gamma_i}{Y_i - P_{j} \left ( 1 + \tau_j \right )}
\end{equation}
which is positive since $\gamma_i > 0$ and households retain positive disposable income. Finally, the first first-order condition entails that
\begin{equation}
	\mu = \beta_i - \frac{\gamma_i P_{j} \left ( 1 + \tau_j \right )}{Y_i - P_{j} \left ( 1 + \tau_j \right )}
\end{equation}
The second Lagrange multiplier is positive provided that
\begin{equation}
	\frac{\beta_i}{\gamma_i} > \frac{P_{j} \left ( 1 + \tau_j \right )}{Y_i - P_{j} \left ( 1 + \tau_j \right )}
\end{equation}
Plugging the Marshallian demands back into the utility function yields household $i$'s indirect utility function:
\begin{align}
	V_{ij} & = A_{ij} + \alpha_i \log \frac{G_j}{N_j^\chi} + \beta_i \log 1 + \gamma_i \log \left [ Y_i - P_{j} \left ( 1 + \tau_j \right ) \right ] \notag \\
	& = A_{ij} + \alpha_i \log \frac{G_j}{N_j^\chi} + \gamma_i \log \left [ Y_i - P_{j} \left ( 1 + \tau_j \right ) \right ]
\end{align}
Furthermore, household $i$'s valuation of exogenous amenities is $A_{ij} \equiv \overline{A}_{j} + U_{ij}$, with $U_{ij} \sim \text{Gumbel} \left ( 0, \theta \right )$. The indirect utility function can thus be re-expressed as follows:
\begin{equation}
	V_{ij} = \underbrace{\overline{A}_j + \alpha_i \log \frac{G_j}{N_j^\chi} + \gamma_i \log \left [ Y_i - P_{j} \left ( 1 + \tau_j \right ) \right ]}_{\equiv v_{ij}} + \, U_{ij}
\end{equation}
where $v_{ij}$ indicates the non-idiosyncratic component of utility. Each household chooses the location that maximizes their indirect utility. Given the parametric assumption on the random component of amenity shocks, the probability that household $i$ chooses location $j$ is
\begin{equation}
	N_{ij} = \frac{\exp \left ( v_{ij} \slash \theta \right )}{1 + \sum_{\ell \in \mathcal{J}} \exp \left ( v_{i \ell} \slash \theta \right )}
\end{equation}
Let $\delta_i \equiv \left [ \alpha_i, \gamma_i, Y_i \right ]'$ be a random vector whose joint probability distribution and support are denoted with $F$ and $\mathcal{D}$, respectively. Integrating choice probabilities over $F$ yields the expected mass of households who choose location $j$:
\begin{equation}
	N_{j} = \int_{\mathcal{D}} N_{ij} \left ( \delta_i \right ) d F \left ( \delta_i \right )
\end{equation}

\appsubsection{Equilibrium in the Housing Market}

The housing supply equation is
\begin{equation}
	\log H^{\text{S}}_j = \lambda + \eta \log P_j + B_j
\end{equation}
Since each household consumes exactly one unit of housing, the aggregate demand for housing in location $j$ is
\begin{align}
	H^{\text{D}}_j & = \int_{\mathcal{D}} N_{ij} \left ( \delta_i \right ) H_{ij} \left ( \delta_i \right ) d F \left ( \delta_i \right ) \\
	& = \int_{\mathcal{D}} N_{ij} \left ( \delta_i \right ) d F \left ( \delta_i \right ) \\
	& = N_j
\end{align}
Taking logarithms yields
\begin{equation}
	\log H^{\text{D}}_{j} = \log N_j
\end{equation}
The equilibrium rental rate of housing equates log-demand and log-supply of housing:
\begin{align}
	\log H^{\text{S}}_{j} = \log H^{\text{D}}_{j} \iff & \lambda + \eta \log P_j + B_j = \log N_j \\
	\iff & \log P_j = \frac{1}{\eta} \log N_j - \widetilde{\lambda} - \widetilde{B_j}
\end{align}
where $\widetilde{\lambda} \equiv \frac{\lambda}{\eta}$ and $\widetilde{B}_j \equiv \frac{B_j}{\eta}$. Plugging the equilibrium rental rate of housing into the equation for the log-supply of housing yields the equilibrium level of housing space:
\begin{equation}
	\log H_j = \lambda + \eta \log P_j + B_j = \lambda + \log N_j - \lambda - B_j + B_j = \log N_j
\end{equation}
Finally, the equilibrium level of housing expenditure in location $j$ is
\begin{align}
	\log P_j + \log H_j & = \frac{1}{\eta} \log N_j - \widetilde{\lambda} - \widetilde{B_j} + \log N_j \\
	& = \frac{1+\eta}{\eta} \log N_j - \widetilde{\lambda} - \widetilde{B_j}
\end{align}

\appsubsection{The Government Possibility Frontier}

Consider a voter who resides in district $j$ and chooses their preferred level of government spending $G_j$. The system of equations implied by the housing market clearing and government balanced budget conditions is
\begin{align}
	J_j \left ( G_j, P_j, 1+\tau_j \right ) & = \log H^{\text{S}}_j - \log H^{\text{D}}_{j} = 0 \label{appx_totdiff_housing}\\
	K_j \left ( G_j, P_j, 1+\tau_j \right ) & = \log \tau_j + \log P_j + \log H^{S}_j - \log G_j = 0 \label{appx_totdiff_budget}
\end{align}
The goal of this section is to compute the partial derivatives required to solve this system in its general form. Recall that
\begin{align}
	& J_{j} \equiv \lambda + \eta \log P_j + B_j - \log N_j \\
	& K_{j} \equiv \log \tau_{j} + \lambda + \left ( 1 + \eta \right ) \log P_j + B_j - \log G_{j}
\end{align}

\appsubsubsection{Sum of Exponentials}

Recall that the non-idiosyncratic component of utility is
\begin{equation}
	v_{ij} \equiv \overline{A}_j + \alpha_i \log G_j - \alpha_i \chi \log N_j + \gamma_i \log \left [ Y_i - P_{j} \left ( 1 + \tau_j \right ) \right ]
\end{equation}
The probability of household $i$ choosing location $j$ is
\begin{equation}
	N_{ij} = \frac{\exp \left ( v_{ij} \slash \theta \right )}{1 + \sum_{\ell \in \mathcal{J}} \exp \left ( v_{i \ell} \slash \theta \right )}
\end{equation}
and the expected mass of households choosing location $j$ is
\begin{equation}
	N_{j} = \int_{\mathcal{D}} N_{ij} \left ( \delta_i \right ) d F \left ( \delta_i \right )
\end{equation}
For convenience, define
\begin{equation}
	\phi_i \equiv \frac{1}{1 + \sum_{\ell \in \mathcal{J}} \exp \left ( v_{i \ell} \slash \theta \right )}
\end{equation}
As a preliminary step, we compute the partial derivatives of $\phi_i$ keeping $N_j$ constant:
\begin{align}
	\frac{\partial \phi_i}{\partial \log G_j} \bigg |_{N_j} & = - \left ( 1 + \sum_{\ell \in \mathcal{J}} \exp \left ( v_{i\ell} \slash \theta \right ) \right )^{-2} \left ( \frac{\alpha_i}{\theta} \exp \left ( v_{ij} \slash \theta \right ) \right ) \\
	& = - \frac{\alpha_i}{\theta} \phi_i \frac{\exp \left ( v_{ij} \slash \theta \right )}{1 + \sum_{\ell} \exp \left ( v_{i\ell} \slash \theta \right )} \\
	& = - \frac{\alpha_i}{\theta} \phi_i N_{ij} \\
	\frac{\partial \phi_i}{\partial \log P_j} \bigg |_{N_j} & = - \left ( 1 + \sum_{\ell \in \mathcal{J}} \exp \left ( v_{i\ell} \slash \theta \right ) \right )^{-2} \left ( - \frac{\gamma_i}{\theta} \frac{P_{j} \left ( 1 + \tau_j \right )}{Y_i - P_{j} \left ( 1 + \tau_j \right )} \exp \left ( v_{ij} \slash \theta \right ) \right ) \\
	& = \frac{\gamma_i}{\theta} \frac{P_{j} \left ( 1 + \tau_j \right )}{Y_i - P_{j} \left ( 1 + \tau_j \right )} \phi_i \frac{\exp \left ( v_{ij} \slash \theta \right )}{1 + \sum_{\ell \in \mathcal{J}} \exp \left ( v_{i\ell} \slash \theta \right )} \\
	& = \frac{\gamma_i}{\theta} \frac{P_{j} \left ( 1 + \tau_j \right )}{Y_i - P_{j} \left ( 1 + \tau_j \right )} \phi_i N_{ij} \\
	\frac{\partial \phi_i}{\partial \log \left ( 1 + \tau_j \right )} \bigg |_{N_j} & = - \left ( 1 + \sum_{\ell \in \mathcal{J}} \exp \left ( v_{i\ell} \slash \theta \right ) \right )^{-2} \left ( - \frac{\gamma_i}{\theta} \frac{P_{j} \left ( 1 + \tau_j \right )}{Y_i - P_{j} \left ( 1 + \tau_j \right )} \exp \left ( v_{ij} \slash \theta \right ) \right ) \\
	& = \frac{\gamma_i}{\theta} \frac{P_{j} \left ( 1 + \tau_j \right )}{Y_i - P_{j} \left ( 1 + \tau_j \right )} \phi_i \frac{\exp \left ( v_{ij} \slash \theta \right )}{1 + \sum_{\ell \in \mathcal{J}} \exp \left ( v_{i\ell} \slash \theta \right )} \\
	& = \frac{\gamma_i}{\theta} \frac{P_{j} \left ( 1 + \tau_j \right )}{Y_i - P_{j} \left ( 1 + \tau_j \right )} \phi_i N_{ij}
\end{align}

\appsubsubsection{Location Choice Probability}

We now compute the partial derivatives of $N_{ij}$ keeping $N_j$ constant:
\begin{align}
	& \frac{\partial N_{ij}}{\partial \log G_j} \bigg |_{N_j} \notag \\
	& = \frac{\partial N_{ij} \slash \partial G_j}{\partial \log G_j \slash \partial G_j} \bigg |_{N_j} \\
	& = G_j \frac{\partial N_{ij}}{\partial G_j} \bigg |_{N_j} \\
	& = G_j \left ( \frac{\partial \phi_i}{\partial G_j} \bigg |_{N_j} \exp \left ( v_{ij} \slash \theta \right ) + \phi_i \exp \left ( v_{ij} \slash \theta \right ) \frac{\alpha_i}{\theta} \frac{1}{G_j} \right ) \\
	& = G_j \left ( \frac{\partial \phi_i}{\partial \log G_j} \bigg |_{N_j} \frac{\partial \log G_j}{\partial G_j} \exp \left ( v_{ij} \slash \theta \right ) + N_{ij} \frac{\alpha_i}{\theta} \frac{1}{G_j} \right ) \\
	& = G_j \left ( - \frac{\alpha_i}{\theta} \phi_i N_{ij} \frac{1}{G_j} \exp \left ( v_{ij} \slash \theta \right ) + N_{ij} \frac{\alpha_i}{\theta} \frac{1}{G_j} \right ) \\
	& = G_j \left ( - \frac{\alpha_i}{\theta} N_{ij} \frac{1}{G_j} N_{ij} + N_{ij} \frac{\alpha_i}{\theta} \frac{1}{G_j} \right ) \\
	& = \left ( - \frac{\alpha_i}{\theta} N_{ij} N_{ij} + N_{ij} \frac{\alpha_i}{\theta} \right )\\
	& = \frac{\alpha_i}{\theta} N_{ij} \left ( 1 - N_{ij} \right )
\end{align}
Similarly,
\begin{align}
	& \frac{\partial N_{ij}}{\partial \log P_j} \bigg |_{N_j} \notag \\
	& = \frac{\partial N_{ij} \slash \partial P_j}{\partial \log P_j \slash \partial P_j} \bigg |_{N_j} \\
	& = P_j \frac{\partial N_{ij}}{\partial P_j} \bigg |_{N_j} \\
	& = P_j \left ( \frac{\partial \phi_i}{\partial P_j} \bigg |_{N_j} \exp \left ( v_{ij} \slash \theta \right ) - \phi_i \exp \left ( v_{ij} \slash \theta \right ) \frac{\gamma_i}{\theta} \frac{P_{j} \left ( 1 + \tau_j \right )}{Y_i - P_{j} \left ( 1 + \tau_j \right )} \frac{1}{P_j} \right ) \\
	& = P_j \left ( \frac{\partial \phi_i}{\partial \log P_j} \bigg |_{N_j} \frac{\partial \log P_j}{\partial P_j} \exp \left ( v_{ij} \slash \theta \right ) - N_{ij} \frac{\gamma_i}{\theta} \frac{P_{j} \left ( 1 + \tau_j \right )}{Y_i - P_{j} \left ( 1 + \tau_j \right )} \frac{1}{P_j} \right ) \\
	& = P_j \left ( \frac{\gamma_i}{\theta} \frac{P_{j} \left ( 1 + \tau_j \right )}{Y_i - P_{j} \left ( 1 + \tau_j \right )} \phi_i N_{ij} \frac{1}{P_j} \exp \left ( v_{ij} \slash \theta \right ) - N_{ij} \frac{\gamma_i}{\theta} \frac{P_{j} \left ( 1 + \tau_j \right )}{Y_i - P_{j} \left ( 1 + \tau_j \right )} \frac{1}{P_j} \right ) \\
	& = P_j \left ( \frac{\gamma_i}{\theta} \frac{P_{j} \left ( 1 + \tau_j \right )}{Y_i - P_{j} \left ( 1 + \tau_j \right )} N_{ij} \frac{1}{P_j} N_{ij} - N_{ij} \frac{\gamma_i}{\theta} \frac{P_{j} \left ( 1 + \tau_j \right )}{Y_i - P_{j} \left ( 1 + \tau_j \right )} \frac{1}{P_j} \right ) \\
	& = \left ( \frac{\gamma_i}{\theta} \frac{P_{j} \left ( 1 + \tau_j \right )}{Y_i - P_{j} \left ( 1 + \tau_j \right )} N_{ij} N_{ij} - N_{ij} \frac{\gamma_i}{\theta} \frac{P_{j} \left ( 1 + \tau_j \right )}{Y_i - P_{j} \left ( 1 + \tau_j \right )} \right )\\
	& = - \frac{\gamma_i}{\theta} \frac{P_{j} \left ( 1 + \tau_j \right )}{Y_i - P_{j} \left ( 1 + \tau_j \right )} N_{ij} \left ( 1 - N_{ij} \right )
\end{align}
And
\begin{equation}
	\frac{\partial N_{ij}}{\partial \log \left ( 1 + \tau_j \right )} \bigg |_{N_j} = - \frac{\gamma_i}{\theta} \frac{P_{j} \left ( 1 + \tau_j \right )}{Y_i - P_{j} \left ( 1 + \tau_j \right )} N_{ij} \left ( 1 - N_{ij} \right )
\end{equation}
Analogously, the partial derivative of $N_{ij}$ with respect to $\log N_j$ is
\begin{equation}
	\frac{\partial N_{ij}}{\partial \log N_j} = - \frac{\alpha_i \chi}{\theta} N_{ij} \left ( 1 - N_{ij} \right )
\end{equation}
Then the partial derivatives of $N_{ij}$ are
\begin{align}
	\frac{\partial N_{ij}}{\partial \log G_j} & = \frac{\partial N_{ij}}{\partial \log G_j} \bigg |_{N_j} + \frac{\partial N_{ij}}{\partial \log N_j} \frac{\partial \log N_{j}}{\partial \log G_j} \\
	& = \frac{\partial N_{ij}}{\partial \log G_j} \bigg |_{N_j} + \frac{\partial N_{ij}}{\partial \log N_j} \frac{\partial \log N_{j}}{\partial N_j} \frac{\partial N_{j}}{\partial \log G_j} \\
	& = \frac{\alpha_i}{\theta} N_{ij} \left ( 1 - N_{ij} \right ) - \frac{\alpha_i \chi}{\theta} N_{ij} \left ( 1 - N_{ij} \right ) \frac{1}{N_j} \frac{\partial N_{j}}{\partial \log G_j} \\ 
	& = \frac{1}{\theta} N_{ij} \left ( 1 - N_{ij} \right ) \left ( \alpha_i - \alpha_i \chi \frac{1}{N_j} \frac{\partial N_{j}}{\partial \log G_j} \right ) \\
	\frac{\partial N_{ij}}{\partial \log P_j} & = \frac{1}{\theta} N_{ij} \left ( 1 - N_{ij} \right ) \left ( - \frac{\gamma_i}{\theta} \frac{P_{j} \left ( 1 + \tau_j \right )}{Y_i - P_{j} \left ( 1 + \tau_j \right )} - \alpha_i \chi \frac{1}{N_j} \frac{\partial N_{j}}{\partial \log P_j} \right ) \\
	\frac{\partial N_{ij}}{\partial \log \left ( 1 + \tau_j \right )} & = \frac{1}{\theta} N_{ij} \left ( 1 - N_{ij} \right ) \left ( - \frac{\gamma_i}{\theta} \frac{P_{j} \left ( 1 + \tau_j \right )}{Y_i - P_{j} \left ( 1 + \tau_j \right )} - \alpha_i \chi \frac{1}{N_j} \frac{\partial N_{j}}{\partial \log \left ( 1 + \tau_j \right )} \right )
\end{align}

\appsubsubsection{Expected Mass of Households in a Location}

Recall that $N_j = \int_{\mathcal{D}} N_{ij} d F$. Thus,
\begin{align}
	\frac{\partial N_j}{\partial \log G_j} & = \int_{\mathcal{D}} \frac{\partial N_{ij}}{\partial \log G_j} d F \\
	\frac{\partial N_j}{\partial \log P_j} & = \int_{\mathcal{D}} \frac{\partial N_{ij}}{\partial \log P_j} d F \\
	\frac{\partial N_j}{\partial \log \left ( 1 + \tau_j \right )} & = \int_{\mathcal{D}} \frac{\partial N_{ij}}{\partial \log \left ( 1 + \tau_j \right )} d F
\end{align}
Replacing the expressions for the partial derivatives of $N_{ij}$ yields
\begin{align}
	\frac{\partial N_j}{\partial \log G_j} & = \int_{\mathcal{D}} \frac{1}{\theta} N_{ij} \left ( 1 - N_{ij} \right ) \left ( \alpha_i - \frac{\alpha_i \chi}{N_j} \frac{\partial N_{j}}{\partial \log G_j} \right ) d F \\
	\frac{\partial N_j}{\partial \log P_j} & = \int_{\mathcal{D}} \frac{1}{\theta} N_{ij} \left ( 1 - N_{ij} \right ) \left ( - \frac{\gamma_i}{\theta} \frac{P_{j} \left ( 1 + \tau_j \right )}{Y_i - P_{j} \left ( 1 + \tau_j \right )} - \frac{\alpha_i \chi}{N_j} \frac{\partial N_{j}}{\partial \log P_j} \right ) d F \\
	\frac{\partial N_j}{\partial \log \left ( 1 + \tau_j \right )} & = \int_{\mathcal{D}} \frac{1}{\theta} N_{ij} \left ( 1 - N_{ij} \right ) \left ( - \frac{\gamma_i}{\theta} \frac{P_{j} \left ( 1 + \tau_j \right )}{Y_i - P_{j} \left ( 1 + \tau_j \right )} - \frac{\alpha_i \chi}{N_j} \frac{\partial N_{j}}{\partial \log \left ( 1 + \tau_j \right )} \right ) d F
\end{align}
Rearranging terms,
\begin{align}
	& \frac{\partial N_j}{\partial \log G_j} + \chi \int_{\mathcal{D}} \frac{\alpha_i}{\theta} N_{ij} \left ( 1 - N_{ij} \right )  \frac{1}{N_j} \frac{\partial N_{j}}{\partial \log G_j} d F = \int_{\mathcal{D}} \frac{\alpha_i}{\theta} N_{ij} \left ( 1 - N_{ij} \right ) d F \\
	& \iff \frac{\partial N_j}{\partial \log G_j} \left ( 1 + \frac{\chi}{N_j} \int_{\mathcal{D}} \frac{\alpha_i}{\theta} N_{ij} \left ( 1 - N_{ij} \right ) d F \right ) = \int_{\mathcal{D}} \frac{\alpha_i}{\theta} N_{ij} \left ( 1 - N_{ij} \right ) d F \\
	& \iff \frac{\partial N_j}{\partial \log G_j} = \frac{\int_{\mathcal{D}} \frac{\alpha_i}{\theta} N_{ij} \left ( 1 - N_{ij} \right ) d F}{1 + \frac{\chi}{N_j} \int_{\mathcal{D}} \frac{\alpha_i}{\theta} N_{ij} \left ( 1 - N_{ij} \right ) d F}
\end{align}
Similarly,
\begin{align}
	& \frac{\partial N_j}{\partial \log P_j} + \chi \int_{\mathcal{D}} \frac{\alpha_i}{\theta} N_{ij} \left ( 1 - N_{ij} \right )  \frac{1}{N_j} \frac{\partial N_{j}}{\partial \log P_j} d F \\
	& = - \int_{\mathcal{D}} \frac{\gamma_i}{\theta} \frac{P_{j} \left ( 1 + \tau_j \right )}{Y_i - P_{j} \left ( 1 + \tau_j \right )} N_{ij} \left ( 1 - N_{ij} \right ) d F \\
	& \iff \frac{\partial N_j}{\partial \log P_j} \left ( 1 + \frac{\chi}{N_j} \int_{\mathcal{D}} \frac{\alpha_i}{\theta} N_{ij} \left ( 1 - N_{ij} \right ) d F \right ) \\
	& = - \int_{\mathcal{D}} \frac{\gamma_i}{\theta} \frac{P_{j} \left ( 1 + \tau_j \right )}{Y_i - P_{j} \left ( 1 + \tau_j \right )} N_{ij} \left ( 1 - N_{ij} \right ) d F \\
	& \iff \frac{\partial N_j}{\partial \log P_j} = - \frac{\int_{\mathcal{D}} \frac{\gamma_i}{\theta} \frac{P_{j} \left ( 1 + \tau_j \right )}{Y_i - P_{j} \left ( 1 + \tau_j \right )} N_{ij} \left ( 1 - N_{ij} \right ) d F}{1 + \frac{\chi}{N_j} \int_{\mathcal{D}} \frac{\alpha_i}{\theta} N_{ij} \left ( 1 - N_{ij} \right ) d F}
\end{align}
Finally,
\begin{align}
	& \frac{\partial N_j}{\partial \log \left ( 1 + \tau_j \right )} + \chi \int_{\mathcal{D}} \frac{\alpha_i}{\theta} N_{ij} \left ( 1 - N_{ij} \right )  \frac{1}{N_j} \frac{\partial N_{j}}{\partial \log \left ( 1 + \tau_j \right )} d F \\
	& = - \int_{\mathcal{D}} \frac{\gamma_i}{\theta} \frac{P_{j} \left ( 1 + \tau_j \right )}{Y_i - P_{j} \left ( 1 + \tau_j \right )} N_{ij} \left ( 1 - N_{ij} \right ) d F \\
	& \iff \frac{\partial N_j}{\partial \log \left ( 1 + \tau_j \right )} \left ( 1 + \frac{\chi}{N_j} \int_{\mathcal{D}} \frac{\alpha_i}{\theta} N_{ij} \left ( 1 - N_{ij} \right ) d F \right ) \\
	& = - \int_{\mathcal{D}} \frac{\gamma_i}{\theta} \frac{P_{j} \left ( 1 + \tau_j \right )}{Y_i - P_{j} \left ( 1 + \tau_j \right )} N_{ij} \left ( 1 - N_{ij} \right ) d F \\
	& \iff \frac{\partial N_j}{\partial \log \left ( 1 + \tau_j \right )} = - \frac{\int_{\mathcal{D}} \frac{\gamma_i}{\theta} \frac{P_{j} \left ( 1 + \tau_j \right )}{Y_i - P_{j} \left ( 1 + \tau_j \right )} N_{ij} \left ( 1 - N_{ij} \right ) d F}{1 + \frac{\chi}{N_j} \int_{\mathcal{D}} \frac{\alpha_i}{\theta} N_{ij} \left ( 1 - N_{ij} \right ) d F}
\end{align}
To summarize,
\begin{align}
	\frac{\partial N_j}{\partial \log G_j} & = \frac{\int_{\mathcal{D}} \frac{\alpha_i}{\theta} N_{ij} \left ( 1 - N_{ij} \right ) d F}{1 + \frac{\chi}{N_j} \int_{\mathcal{D}} \frac{\alpha_i}{\theta} N_{ij} \left ( 1 - N_{ij} \right ) d F} \\
	\frac{\partial N_j}{\partial \log P_j} & = - \frac{\int_{\mathcal{D}} \frac{\gamma_i}{\theta} \frac{P_{j} \left ( 1 + \tau_j \right )}{Y_i - P_{j} \left ( 1 + \tau_j \right )} N_{ij} \left ( 1 - N_{ij} \right ) d F}{1 + \frac{\chi}{N_j} \int_{\mathcal{D}} \frac{\alpha_i}{\theta} N_{ij} \left ( 1 - N_{ij} \right ) d F} \\
	\frac{\partial N_j}{\partial \log \left ( 1 + \tau_j \right )} & = - \frac{\int_{\mathcal{D}} \frac{\gamma_i}{\theta} \frac{P_{j} \left ( 1 + \tau_j \right )}{Y_i - P_{j} \left ( 1 + \tau_j \right )} N_{ij} \left ( 1 - N_{ij} \right ) d F}{1 + \frac{\chi}{N_j} \int_{\mathcal{D}} \frac{\alpha_i}{\theta} N_{ij} \left ( 1 - N_{ij} \right ) d F}
\end{align}
For compactness, define $\rho_{ij} \equiv \frac{P_{j} \left ( 1 + \tau_j \right )}{Y_i - P_{j} \left ( 1 + \tau_j \right )}$. Then the partial derivatives of interest can be expressed as
\begin{align}
	& \frac{\partial N_{ij}}{\partial \log G_j} \notag \\
	& = \frac{1}{\theta} N_{ij} \left ( 1 - N_{ij} \right ) \left ( \alpha_i - \alpha_i \frac{\frac{\chi}{N_j} \int_{\mathcal{D}} \frac{\alpha_i}{\theta} N_{ij} \left ( 1 - N_{ij} \right ) d F}{1 + \frac{\chi}{N_j} \int_{\mathcal{D}} \frac{\alpha_i}{\theta} N_{ij} \left ( 1 - N_{ij} \right ) d F} \right ) \\
	& = \frac{1}{\theta} N_{ij} \left ( 1 - N_{ij} \right ) \left ( \frac{\alpha_i}{1 + \frac{\chi}{N_j} \int_{\mathcal{D}} \frac{\alpha_i}{\theta} N_{ij} \left ( 1 - N_{ij} \right ) d F} \right ) \\
	& = \frac{\frac{\alpha_i}{\theta} N_{ij} \left ( 1 - N_{ij} \right )}{1 + \frac{\chi}{N_j} \int_{\mathcal{D}} \frac{\alpha_i}{\theta} N_{ij} \left ( 1 - N_{ij} \right ) d F} \\
	& \frac{\partial N_{ij}}{\partial \log P_j} \notag \\
	& = \frac{1}{\theta} N_{ij} \left ( 1 - N_{ij} \right ) \left ( - \gamma_i \rho_{ij} + \alpha_i \frac{\frac{\chi}{N_j} \int_{\mathcal{D}} \frac{\gamma_i}{\theta} \rho_{ij} N_{ij} \left ( 1 - N_{ij} \right ) d F}{1 + \frac{\chi}{N_j} \int_{\mathcal{D}} \frac{\alpha_i}{\theta} N_{ij} \left ( 1 - N_{ij} \right ) d F} \right ) \\
	& = \frac{1}{\theta} N_{ij} \left ( 1 - N_{ij} \right ) \left ( \frac{-\gamma_i \rho_{ij} + \left ( \alpha_i - \gamma_i \rho_{ij} \right ) \frac{\chi}{N_j} \int_{\mathcal{D}} \frac{\gamma_i \rho_{ij}}{\theta} N_{ij} \left ( 1 - N_{ij} \right ) d F}{1 + \frac{\chi}{N_j} \int_{\mathcal{D}} \frac{\alpha_i}{\theta} N_{ij} \left ( 1 - N_{ij} \right ) d F} \right ) \\
	& \frac{\partial N_{ij}}{\partial \log \left ( 1 + \tau_j \right )} \notag \\
	& = \frac{1}{\theta} N_{ij} \left ( 1 - N_{ij} \right ) \left ( -\gamma_i \rho_{ij} + \alpha_i \frac{\frac{\chi}{N_j} \int_{\mathcal{D}} \frac{\gamma_i \rho_{ij}}{\theta} N_{ij} \left ( 1 - N_{ij} \right ) d F}{1 + \frac{\chi}{N_j} \int_{\mathcal{D}} \frac{\alpha_i}{\theta} N_{ij} \left ( 1 - N_{ij} \right ) d F} \right ) \\
	& = \frac{1}{\theta} N_{ij} \left ( 1 - N_{ij} \right ) \left ( \frac{-\gamma_i \rho_{ij} + \left ( \alpha_i - \gamma_i \rho_{ij} \right ) \frac{\chi}{N_j} \int_{\mathcal{D}} \frac{\gamma_i \rho_{ij}}{\theta} N_{ij} \left ( 1 - N_{ij} \right ) d F}{1 + \frac{\chi}{N_j} \int_{\mathcal{D}} \frac{\alpha_i}{\theta} N_{ij} \left ( 1 - N_{ij} \right ) d F} \right )
\end{align}

\appsubsubsection{System of Equations for the Government Possibility Frontier}

As a consequence, the partial derivatives associated with the original system of equations can be rewritten as follows:
\begin{align}
	\frac{\partial J_j}{\partial \log G_j} & = - \int_{\mathcal{D}} \frac{\frac{\alpha_i}{\theta} N_{ij} \left ( 1 - N_{ij} \right )}{1 + \frac{\chi}{N_j} \int_{\mathcal{D}} \frac{\alpha_i}{\theta} N_{ij} \left ( 1 - N_{ij} \right ) d F} d F \label{eq:j_first}\\
	\frac{\partial J_j}{\partial \log P_j} & = \eta - \int_{\mathcal{D}} \frac{1}{\theta} N_{ij} \left ( 1 - N_{ij} \right ) \notag \\
	& \left ( \frac{-\gamma_i \rho_{ij} + \left ( \alpha_i - \gamma_i \rho_{ij} \right ) \frac{\chi}{N_j} \int_{\mathcal{D}} \frac{\gamma_i \rho_{ij}}{\theta} N_{ij} \left ( 1 - N_{ij} \right ) d F}{1 + \frac{\chi}{N_j} \int_{\mathcal{D}} \frac{\alpha_i}{\theta} N_{ij} \left ( 1 - N_{ij} \right ) d F} \right ) d F \\
	\frac{\partial J_j}{\partial \log \left ( 1 + \tau_j \right )} & = - \int_{\mathcal{D}} \frac{1}{\theta} N_{ij} \left ( 1 - N_{ij} \right ) \notag \\
	& \left ( \frac{-\gamma_i \rho_{ij} + \left ( \alpha_i - \gamma_i \rho_{ij} \right ) \frac{\chi}{N_j} \int_{\mathcal{D}} \frac{\gamma_i \rho_{ij}}{\theta} N_{ij} \left ( 1 - N_{ij} \right ) d F}{1 + \frac{\chi}{N_j} \int_{\mathcal{D}} \frac{\alpha_i}{\theta} N_{ij} \left ( 1 - N_{ij} \right ) d F} \right ) d F
\end{align}
In addition,
\begin{align}
	\frac{\partial K_j}{\partial \log G_j} & = -1 \\
	\frac{\partial K_j}{\partial \log P_j} & = 1 + \eta\\
	\frac{\partial K_j}{\partial \log \left ( 1 + \tau_j \right )} & = \frac{1+\tau_j}{\tau_j} \label{eq:k_last}
\end{align}
For compactness, define the following terms:
\begin{align}
	\widehat{\alpha}_j & \equiv \int_{\mathcal{D}} \frac{\alpha_i}{\theta} N_{ij} \left ( 1 - N_{ij} \right ) d F \\
	\widehat{\gamma}_j & \equiv \int_{\mathcal{D}} \frac{\gamma_i \rho_{ij}}{\theta} N_{ij} \left ( 1 - N_{ij} \right ) d F
\end{align}
The partial derivatives in \eqref{eq:j_first}-\eqref{eq:k_last} can be rewritten as follows:
\begin{align}
	\frac{\partial J_j}{\partial \log G_j} & = - \frac{\widehat{\alpha}_j}{1 + \frac{\chi}{N_j} \widehat{\alpha}_j} \\
	\frac{\partial J_j}{\partial \log P_j} & = \eta + \frac{\widehat{\gamma}_j - \frac{\chi}{N_j} \widehat{\gamma}_j \left ( \widehat{\alpha}_j - \widehat{\gamma}_j \right )}{1 + \frac{\chi}{N_j} \widehat{\alpha}_j} \\
	\frac{\partial J_j}{\partial \log \left ( 1 + \tau_j \right )} & = \frac{\widehat{\gamma}_j - \frac{\chi}{N_j} \widehat{\gamma}_j \left ( \widehat{\alpha}_j - \widehat{\gamma}_j \right )}{1 + \frac{\chi}{N_j} \widehat{\alpha}_j} \\
	\frac{\partial K_j}{\partial \log G_j} & = -1 \\
	\frac{\partial K_j}{\partial \log P_j} & = 1 + \eta \\
	\frac{\partial K_j}{\partial \log \left ( 1 + \tau_j \right )} & = \frac{1+\tau_j}{\tau_j}
\end{align}

\appsubsubsection{Partial Derivatives with Myopic Voting}

The assumption of myopic voting entails that voters perceive jurisdiction boundaries as fixed and do not account for the mobility implications of a change in local expenditures and taxes. As a consequence, all of the terms involving a partial derivative of $N_{j}$ are set to zero. The resulting partial derivatives from the previous section change as follows. For any district $j$,
	\begin{align}
	\frac{\partial J_j}{\partial \log G_j} & = 0 \\
	\frac{\partial J_j}{\partial \log P_j} & = \eta\\
	\frac{\partial J_j}{\partial \log \left ( 1 + \tau_j \right )} & = 0
\end{align}
In addition,
\begin{align}
	\frac{\partial K_j}{\partial \log G_j} & = -1 \\
	\frac{\partial K_j}{\partial \log P_j} & = 1 + \eta\\
	\frac{\partial K_j}{\partial \log \left ( 1 + \tau_j \right )} & = \frac{1+ \tau_j}{\tau_j}
\end{align}

\appsubsubsection{The Slope of the Government Possibility Frontier}

Totally differentiating both equations with respect to their three common arguments yields
\begin{align}
	\frac{\partial J_j}{\partial \log G_j} d \log G_j + \frac{\partial J_j}{\partial \log P_j} d \log P_j + \frac{\partial J_j}{\partial \log \left ( 1 + \tau_j \right )} d \log \left ( 1 + \tau_j \right ) = 0 \\
	\frac{\partial K_j}{\partial \log G_j} d \log G_j + \frac{\partial K_j}{\partial \log P_j} d \log P_j + \frac{\partial K_j}{\partial \log \left ( 1 + \tau_j \right )} d \log \left ( 1 + \tau_j \right ) = 0
\end{align}
For compactness, introduce the following notation:
\begin{align*}
	J_g \equiv \frac{\partial J_j}{\partial \log G_j} \qquad J_p \equiv \frac{\partial J_j}{\partial \log P_j} \qquad J_\tau \equiv \frac{\partial J_j}{\partial \log \left ( 1 + \tau_j \right )} \\
	K_g \equiv \frac{\partial K_j}{\partial \log G_j} \qquad K_p \equiv \frac{\partial K_j}{\partial \log P_j} \qquad K_\tau \equiv \frac{\partial K_j}{\partial \log \left ( 1 + \tau_j \right )} \\
	dg \equiv d \log G_j \qquad dp \equiv d \log P_j \qquad d\tau \equiv d \log \left ( 1 + \tau_j \right )
\end{align*}
The system of equations can thus be rewritten as
\begin{align}
	& J_g dg + J_p dp + J_\tau d\tau = 0 \label{eq_totdiff_J} \\
	& K_g dg + K_p dp + K_\tau d\tau = 0 \label{eq_totdiff_K}
\end{align}
First, we compute the total derivative of the rental rate of housing with respect to government spending. To begin with, we divide both sides of \eqref{eq_totdiff_K} by $dg$ and rearrange terms:
\begin{equation}
	K_g + K_p \frac{dp}{dg} + K_\tau \frac{d\tau}{dg} = 0 \iff \frac{d\tau}{dg} = \frac{-K_g -K_p \frac{dp}{dg}}{K_\tau} \label{eq_totdiff_K_solved}
\end{equation}
Then we divide \eqref{eq_totdiff_J} by $dg$ and plug \eqref{eq_totdiff_K_solved} into it:
\begin{align}
	J_g + J_p \frac{dp}{dg} + J_\tau \frac{d\tau}{dg} = 0 & \iff J_g + J_p \frac{dp}{dg} + J_\tau \frac{-K_g -K_p \frac{dp}{dg}}{K_\tau} = 0 \\
	& \iff  J_g + J_p \frac{dp}{dg} - \frac{J_\tau K_g}{K_\tau} - \frac{J_\tau K_p}{K_\tau} \frac{dp}{dg} = 0 \\
	& \iff \frac{dp}{dg} \left ( J_p - \frac{J_\tau K_p}{K_\tau} \right ) = -J_g + \frac{J_\tau K_g}{K_\tau} \\
	& \iff \frac{dp}{dg} = - \frac{J_g - \frac{J_\tau K_g}{K_\tau}}{J_p - \frac{J_\tau K_p}{K_\tau}} \\
	& \iff \frac{dp}{dg} = - \frac{J_g K_\tau - J_\tau K_g}{J_p K_\tau - J_\tau K_p}
\end{align}
We follow similar steps to compute the total derivative of the property tax rate with respect to government spending. As above, we divide both sides of \eqref{eq_totdiff_K} by $dg$ and rearrange terms to isolate $\frac{dp}{dg}$:
\begin{equation}
	K_g + K_p \frac{dp}{dg} + K_\tau \frac{d\tau}{dg} = 0 \iff \frac{dp}{dg} = \frac{-K_g -K_\tau \frac{d\tau}{dg}}{K_p}\label{eq_totdiff_K_solved2}
\end{equation}
Then we divide \eqref{eq_totdiff_J} by $dg$ and plug \eqref{eq_totdiff_K_solved2} into it:
\begin{align}
	J_g + J_p \frac{dp}{dg} + J_\tau \frac{d\tau}{dg} = 0 & \iff J_g + J_p \frac{-K_g -K_\tau \frac{d\tau}{dg}}{K_p} + J_\tau \frac{d\tau}{dg} = 0 \\
	& \iff  J_g - \frac{J_p K_g}{K_p} - \frac{J_p K_\tau}{K_p} \frac{d\tau}{dg} + J_\tau \frac{d\tau}{dg} = 0 \\
	& \iff \frac{d\tau}{dg} \left ( J_\tau - \frac{J_p K_\tau}{K_p} \right ) = -J_g + \frac{J_p K_g}{K_p} \\
	& \iff \frac{d\tau}{dg} = - \frac{J_g - \frac{J_p K_g}{K_p}}{J_\tau - \frac{J_p K_\tau}{K_p}} \\
	& \iff \frac{d\tau}{dg} = - \frac{J_g K_p - J_p K_g}{J_\tau K_p - J_p K_\tau}
\end{align}

\appsubsection{Preferred Property Tax Rates}

The goal of this section is to derive the property tax rate preferred by any household type $k$ residing in any area $a$ for any jurisdiction $j$.

\appsubsubsection{First-Order Conditions}

Consider a household in district $j$ choosing their preferred level of government spending. Recall that household $i$'s indirect utility stemming from choosing location $j$ is
\begin{equation}
	V_{ij} = \overline{A}_{j} + \alpha_i \log G_j - \alpha_i \chi \log N_j + \gamma_i \log \left [ Y_i - P_j \left ( 1 + \tau_j \right ) \right ] + U_{ij}
\end{equation}
The derivative of this indirect utility function with respect to government spending is
\begin{equation}
	\frac{d V_{ij}}{d \log G_j} = \alpha_i - \alpha_i \chi \frac{d \log N_j}{d \log G_j} - \gamma_i \rho_{ij} \frac{d \log P_j}{d \log G_j} - \gamma_i \rho_{ij} \frac{d \log \left ( 1 + \tau_j \right )}{d \log G_j} \label{firstdev}
\end{equation}
Denoting the amount of government spending preferred by household $i$ as $G_{ij}$, the first-order condition associated with the implied maximization problem is
\begin{equation}
	\frac{d \log P_j}{d \log G_j} \Bigg |_{G_j = G_{ij}} + \frac{d \log \left ( 1 + \tau_j \right )}{d \log G_j} \Bigg |_{G_j = G_{ij}} = \frac{\alpha_i}{\gamma_i \rho_{ij}} \left ( 1 - \chi \frac{d \log N_j}{d \log G_j} \Bigg |_{G_j = G_{ij}} \right ) \label{eq_foc}
\end{equation}

\appsubsubsection{Preferred Property Tax Rates with Myopic Voting}

Under the assumption that voters are myopic, the total derivative of the rental rate of housing with respect to government spending becomes
\begin{equation}
	\frac{dp}{dg} = - \frac{J_g K_\tau - J_\tau K_g}{J_p K_\tau - J_\tau K_p} = 0
\end{equation}
Similarly, the total derivative of the tax rate with respect to government spending is
\begin{equation}
	\frac{d \tau}{dg} = - \frac{J_g K_p - J_p K_g}{J_\tau K_p - J_p K_\tau} = \frac{\tau_j}{1 + \tau_j}
\end{equation}
We can finally compute the slope of the Government Possibility Frontier:
\begin{equation}
	\frac{d \log P_j}{d \log G_j} + \frac{d \log \left ( 1 + \tau_j \right )}{d \log G_j} = 0 + \frac{\tau_j}{1 + \tau_j} = \frac{\tau_j}{1 + \tau_j}
\end{equation}
Since $\frac{d \log N_j}{d \log G_j} = 0$ by assumption, household $i$'s preferred property tax rate $\tau_{ij}$ solves the first-order condition \eqref{eq_foc}:
\begin{equation}
	\frac{\tau_{ij}}{1 + \tau_{ij}} = \frac{\alpha_i}{\gamma_i \rho_{ij}} \iff \tau_{ij} = \max \left \{ \frac{\alpha_i}{\gamma_i \rho_{ij} - \alpha_i}, 0 \right \}
\end{equation}

\appsubsubsection{Second-Order Conditions}\label{appx_soc}

The goal of this section is to determine whether $\tau_{ij}$ is indeed a maximizer of $V_{ij}$. When voters are assumed to be myopic, the first derivative in \eqref{firstdev} is
\begin{equation}
	\frac{d V_{ij}}{d \log G_j} = \alpha_i - \gamma_i \rho_{ij} \frac{\tau_j}{1 + \tau_j}
\end{equation}
where $\tau_j$ is implicitly a function of $\log G_j$. The second derivative of the objective function then is
\begin{equation}
	\frac{d^2 V_{ij}}{d \log G_j^2} = - \gamma_i \rho_{ij} \frac{d \tau_j}{d \log G_j} \frac{1}{1+\tau_j^2}
\end{equation}
By an application of the chain rule,
\begin{equation}
	\frac{d \log \left ( 1 + \tau_j \right )}{d \log G_j} = \frac{d \log \left ( 1 + \tau_j \right )}{d \tau_j} \frac{d \tau_j}{d \log G_j} = \frac{1}{1+\tau_j} \frac{d \tau_j}{d \log G_j}
\end{equation}
Rearranging terms,
\begin{equation}
	\frac{d \tau_j}{d \log G_j} = \left ( 1 + \tau_j \right ) \frac{d \log \left ( 1 + \tau_j \right )}{d \log G_j} = \left ( 1+\tau_j \right ) \frac{\tau_j}{1 + \tau_j} = \tau_j
\end{equation}
Combining previous derivations, the second derivative of the indirect utility with respect to (log) government spending is
\begin{equation}
	\frac{d^2 V_{ij}}{d \log G_j^2} = - \gamma_i \rho_{ij} \frac{\tau_j}{1+\tau_j^2}
\end{equation}
which is strictly negative because $\gamma_i$, $\rho_{ij}$, and $\tau_j$ are strictly positive, implying that the indirect utility $V_{ij}$ is a strictly concave function of $\log G_j$. Thus, $\tau_{ij}$ attains the unique global maximum of $V_{ij}$ provided that it is an interior solution.

\appsection{Identification of Model Parameters}\label{appx_identification}

This section outlines how we identify the structural parameters of our spatial equilibrium model using regression discontinuity designs.

\appsubsection{Outcome Elasticities with respect to Expenditure Changes}

First, we compute the elasticity of any equilibrium variable at location $\ell \in \mathcal{J}$ with respect to school district $j$'s expenditure change $G_j$. Unlike derivations pertaining to the Government Possibility Frontier, we consider the response of all equilibrium variables to a discrete change in government spending.

\appsubsubsection{Household Supply}

The expected mass of households who choose location $j$ is
\begin{equation}
	N^k_j = \sigma^k \frac{\exp \left ( v^{k}_{j} \slash \theta^k \right )}{1 + \sum_{\ell \in \mathcal{J}} \exp \left ( v^{k}_{\ell} \slash \theta^k \right )}
\end{equation}
where $\sigma^k$ denotes the mass of type-$k$ households in the economy and
\begin{equation}
	v^k_\ell \equiv \overline{A}_\ell + \alpha^k \log G_{\ell} - \chi \alpha^k \log N_{\ell} + \gamma^k \log \left [ y^k - P_\ell \left ( 1 + \tau_{\ell} \right ) \right ]
\end{equation}
We wish to derive an expression for the difference between logged population mass with and without referendum approval:
\begin{equation}
	\Delta \log N^k_j \equiv \log N^k_j \left ( \Delta G_j \right ) - \log N^k_j \left ( 0 \right ) 
\end{equation}
where $\Delta G_j$ is the proposed expenditure hike on which residents vote. To keep notation compact, we express potential outcomes as functions of a binary treatment state indicating referendum approval, so that $\Delta \log N^k_j \equiv \log N^k_j \left ( 1 \right ) - \log N^k_j \left ( 0 \right )$. Then
\begin{align}
	\Delta \log N^k_j & = \log \sigma^k + \frac{v^{k}_{j} \left ( 1 \right )}{\theta^k} - \log \left ( 1 + \sum_{\ell \in \mathcal{J}} \exp \left ( \frac{v^{k}_{\ell} \left ( 1 \right )}{\theta^k} \right ) \right ) \notag \\
	& - \log \sigma^k - \frac{v^{k}_{j} \left ( 0 \right )}{\theta^k} + \log \left ( 1 + \sum_{\ell \in \mathcal{J}} \exp \left ( \frac{v^{k}_{\ell} \left ( 0 \right )}{\theta^k} \right ) \right ) \\
	& = \frac{\Delta v^k_j}{\theta^k} \notag \\
	& - \left ( \log \left ( 1 + \sum_{\ell \in \mathcal{J}} \exp \left ( \frac{v^{k}_{\ell} \left ( 1 \right )}{\theta^k} \right ) \right ) - \log \left ( 1 + \sum_{\ell \in \mathcal{J}} \exp \left ( \frac{v^{k}_{\ell} \left ( 0 \right )}{\theta^k} \right ) \right ) \right ) \\
	& = \frac{\Delta v^k_j}{\theta^k} - \left ( \log Z^k \left ( 1 \right ) - \log Z^k \left ( 0 \right ) \right )
\end{align}
First, 
\begin{equation}
	\frac{\Delta v^k_j}{\theta^k} = \frac{\alpha^k}{\theta^k} \Delta \log G_j - \frac{\chi \alpha^k}{\theta^k} \Delta \log N_j - \frac{\gamma^k \rho^k_j}{\theta^k} \Delta \log P_j - \frac{\gamma^k \rho^k_j}{\theta^k} \Delta \log \left ( 1 + \tau_j \right )
\end{equation}
where $\rho^k_j \equiv \frac{P_j \left ( 1 + \tau_j \right )}{y^k - P_j \left ( 1 + \tau_j \right )}$. Second, for any $t \in \left [ 0,1 \right ]$,
\begin{equation}
	v^k_{\ell,t} =  v^k_{\ell} \left ( 0 \right ) + t \left [ v^k_{\ell} \left ( 1 \right ) - v^k_{\ell} \left ( 0 \right ) \right ]
\end{equation}
and define
\begin{equation}
	Z^k_{t} \equiv 1 + \sum_{\ell \in \mathcal{J}} \exp \left ( \frac{v^{k}_{\ell,t}}{\theta^k} \right )
\end{equation}
Clearly, $Z^k_{t} = Z^k \left ( 0 \right )$ if $t=0$ and $Z^k_{t} = Z^k \left ( 1 \right )$ if $t=1$. Then
\begin{align}
	& \log Z^k \left ( 1 \right ) - \log Z^k \left ( 0 \right ) \notag \\
	& = \int_{0}^{1} \frac{d}{dt} \log Z^k_t dt \\
	& = \int_{0}^{1} \frac{1}{Z^k_t} \left ( 1 + \sum_{\ell \in \mathcal{J}} \exp \left ( \frac{v^{k}_{\ell,t}}{\theta^k} \right ) \right ) \frac{v^k_{\ell} \left ( 1 \right ) - v^k_{\ell} \left ( 0 \right )}{\theta^k} dt \\
	& = \int_{0}^{1} \sum_{\ell} \frac{N^k_{\ell,t}}{\sigma^k} \frac{v^k_{\ell} \left ( 1 \right ) - v^k_{\ell} \left ( 0 \right )}{\theta^k} dt \\
	& = \sum_{\ell} \frac{v^k_{\ell} \left ( 1 \right ) - v^k_{\ell} \left ( 0 \right )}{\theta^k} \int_{0}^{1} \frac{N^k_{\ell,t}}{\sigma^k}  dt \\
	& = \sum_{\ell} \frac{\Delta v^k_{\ell}}{\theta^k} \int_{0}^{1} \frac{N^k_{\ell,t}}{\sigma^k}  dt
\end{align}
The first equality exploits the Fundamental Theorem of Calculus. The second equality follows from an application of the chain rule. The third equality defines $N^k_{\ell,t} \equiv \sigma^k \frac{\exp \left ( \frac{v^{k}_{\ell,t}}{\theta^k} \right )}{1 + \sum_{m \in \mathcal{J}} \exp \left ( \frac{v^{k}_{m,t}}{\theta^k} \right )}$. In addition, we define the mean-value population mass in location $\ell$ as
\begin{equation}
	\overline{N}^{k}_{\ell} \equiv \int_{0}^{1} N^k_{\ell,t} dt
\end{equation}
To summarize,
\begin{equation}
	\log Z^k \left ( 1 \right ) - \log Z^k \left ( 0 \right ) = \sum_{\ell} \frac{\overline{N}^{k}_{\ell}}{\sigma^k} \frac{\Delta v^k_{\ell}}{\theta^k} 
\end{equation}
Because $N^k_{\ell,t}$ is continuous on $\left [ 0,1 \right ]$, the mean-value theorem for integrals states that there exists a point $t^*_\ell \in \left ( 0,1 \right )$ such that $N^k_{\ell,t} = N^k_{\ell,t^*_\ell}$. A solution that is both second-order-accurate and pragmatic is the mid-point value:
\begin{equation}
	\overline{N}^k_{\ell} \approx \frac{N^k_{\ell} \left ( 0 \right ) + N^k_{\ell} \left ( 1 \right )}{2} \equiv \widetilde{N}^k_{\ell}
\end{equation}
Combining previous derivations,
\begin{align}
	& \log Z^k \left ( 1 \right ) - \log Z^k \left ( 0 \right ) \notag \\
	& \approx \sum_{\ell} \frac{N^k_{\ell} \left ( 0 \right ) + N^k_{\ell} \left ( 1 \right )}{2 \sigma^k} \frac{\Delta v^k_{\ell}}{\theta^k} \\
	& = \sum_{\ell} \frac{\widetilde{N}^k_{\ell}}{\sigma^k} \left ( \frac{\alpha^k}{\theta^k} \Delta \log G_\ell - \frac{\chi \alpha^k}{\theta^k} \Delta \log N_\ell - \frac{\gamma^k \rho^k_\ell}{\theta^k} \Delta \log P_\ell - \frac{\gamma^k \rho^k_\ell}{\theta^k} \Delta \log \left ( 1 + \tau_\ell \right ) \right )
\end{align}
Finally, the difference between log household supply in the two treatment states is
\begin{align}
	& \Delta \log N^k_j \notag \\
	& \approx \frac{\alpha^k}{\theta^k} \Delta \log G_j - \frac{\chi \alpha^k}{\theta^k} \Delta \log N_j - \frac{\gamma^k \rho^k_j}{\theta^k} \Delta \log P_j - \frac{\gamma^k \rho^k_j}{\theta^k} \Delta \log \left ( 1 + \tau_j \right ) \notag \\
	& - \sum_{\ell} \frac{\widetilde{N}^k_{\ell}}{\sigma^k}  \left ( \frac{\alpha^k}{\theta^k} \Delta \log G_\ell - \frac{\chi \alpha^k}{\theta^k} \Delta \log N_\ell - \frac{\gamma^k \rho^k_\ell}{\theta^k} \Delta \log P_\ell - \frac{\gamma^k \rho^k_\ell}{\theta^k} \Delta \log \left ( 1 + \tau_\ell \right ) \right )
\end{align}
Finally, we divide both sides by the proposed change in log school district spending:
\begin{align}
	& \frac{\Delta \log N^k_j}{\Delta \log G_j} \notag \\
	& \approx \frac{\alpha^k}{\theta^k} - \frac{\chi \alpha^k}{\theta^k} \frac{\Delta \log N_j}{\Delta \log G_j} - \frac{\gamma^k \rho^k_j}{\theta^k} \frac{\Delta \log P_j}{\Delta \log G_j} - \frac{\gamma^k \rho^k_j}{\theta^k} \frac{\Delta \log \left ( 1 + \tau_j \right )}{\Delta \log G_j} \notag \\
	& - \sum_{\ell} \frac{\widetilde{N}^k_{\ell}}{\sigma^k}  \left ( \frac{\alpha^k}{\theta^k} \frac{\Delta \log G_\ell}{\Delta \log G_j} - \frac{\chi \alpha^k}{\theta^k} \frac{\Delta \log N_\ell}{\Delta \log G_j} - \frac{\gamma^k \rho^k_\ell}{\theta^k} \frac{\Delta \log P_\ell}{\Delta \log G_j} - \frac{\gamma^k \rho^k_\ell}{\theta^k} \frac{\Delta \log \left ( 1 + \tau_\ell \right )}{\Delta \log G_j} \right ) \\
	& = \left ( 1 - \frac{\widetilde{N}^k_{j}}{\sigma^k} \right ) \left ( \frac{\alpha^k}{\theta^k} - \frac{\chi \alpha^k}{\theta^k} \frac{\Delta \log N_j}{\Delta \log G_j} - \frac{\gamma^k \rho^k_j}{\theta^k} \frac{\Delta \log P_j}{\Delta \log G_j} - \frac{\gamma^k \rho^k_j}{\theta^k} \frac{\Delta \log \left ( 1 + \tau_j \right )}{\Delta \log G_j} \right ) \notag \\
	& - \sum_{\ell \neq j} \frac{\widetilde{N}^k_{\ell}}{\sigma^k}  \left ( \frac{\alpha^k}{\theta^k} \frac{\Delta \log G_\ell}{\Delta \log G_j} - \frac{\chi \alpha^k}{\theta^k} \frac{\Delta \log N_\ell}{\Delta \log G_j} - \frac{\gamma^k \rho^k_\ell}{\theta^k} \frac{\Delta \log P_\ell}{\Delta \log G_j} - \frac{\gamma^k \rho^k_\ell}{\theta^k} \frac{\Delta \log \left ( 1 + \tau_\ell \right )}{\Delta \log G_j} \right )
\end{align}
For any location $j' \neq j$, analogous derivations yield
\begin{align}
	& \frac{\Delta \log N^k_{j'}}{\Delta \log G_j} \notag \\
	& \approx \left ( 1 - \frac{\widetilde{N}^k_{j'}}{\sigma^k} \right ) \left ( \frac{\alpha^k}{\theta^k} \frac{\Delta \log G_{j'}}{\Delta \log G_j} - \frac{\chi \alpha^k}{\theta^k} \frac{\Delta \log N_{j'}}{\Delta \log G_j} - \frac{\gamma^k \rho^k_{j'}}{\theta^k} \frac{\Delta \log P_{j'}}{\Delta \log G_j} - \frac{\gamma^k \rho^k_{j'}}{\theta^k} \frac{\Delta \log \left ( 1 + \tau_{j'} \right )}{\Delta \log G_j} \right ) \notag \\
	& - \sum_{\ell \neq j'} \frac{\widetilde{N}^k_{\ell}}{\sigma^k}  \left ( \frac{\alpha^k}{\theta^k} \frac{\Delta \log G_\ell}{\Delta \log G_j} - \frac{\chi \alpha^k}{\theta^k} \frac{\Delta \log N_\ell}{\Delta \log G_j} - \frac{\gamma^k \rho^k_\ell}{\theta^k} \frac{\Delta \log P_\ell}{\Delta \log G_j} - \frac{\gamma^k \rho^k_\ell}{\theta^k} \frac{\Delta \log \left ( 1 + \tau_\ell \right )}{\Delta \log G_j} \right ) \label{eq_logpopmass_own}
\end{align}

\appsubsubsection{Rental Rate of Housing}

In any location $\ell$, the equilibrium rental rate of housing is
\begin{equation}
	\log P_\ell = \frac{1}{\eta} \log \sum_{k} N^{k}_\ell - \frac{\lambda}{\eta} - \frac{B_\ell}{\eta}
\end{equation}
We wish to compute $\Delta \log P_\ell \equiv \log P_\ell \left ( 1 \right ) - \log P_\ell \left ( 0 \right )$. To begin with,
\begin{equation}
	\Delta \log P_\ell = \frac{1}{\eta} \left ( \log \sum_{k} N^{k}_\ell \left ( 1 \right ) - \log \sum_{k} N^{k}_\ell \left ( 0 \right ) \right )
\end{equation}
Now define
\begin{equation}
	M_\ell \left ( 0 \right ) \equiv \sum_{k} N^{k}_\ell \left ( 0 \right ) \qquad M_\ell \left ( 1 \right ) \equiv \sum_{k} N^{k}_\ell \left ( 1 \right )
\end{equation}
For any $t \in \left [ 0,1 \right ]$,
\begin{equation}
	N^k_{\ell,t} =  N^k_{\ell} \left ( 0 \right ) + t \left [ N^k_{\ell} \left ( 1 \right ) - N^k_{\ell} \left ( 0 \right ) \right ]
\end{equation}
and define
\begin{equation}
	M_{\ell,t} \equiv \sum_{k} N^{k}_{\ell,t}
\end{equation}
Clearly, $M_{\ell,t} = M_\ell \left ( 0 \right )$ if $t=0$ and $M_{\ell,t} = M_\ell \left ( 1 \right )$ if $t=1$. Then
\begin{align}
	& \log M_\ell \left ( 1 \right ) - \log M_\ell \left ( 0 \right ) \notag \\
	& = \int_{0}^{1} \frac{d \log M_{\ell,t}}{dt} dt \\
	& = \int_{0}^{1} \frac{1}{M_{\ell,t}} \sum_{k} \left [ N^{k}_{\ell} \left ( 1 \right ) - N^{k}_{\ell} \left ( 0 \right ) \right ] dt \\
	& = \int_{0}^{1} \sum_{k} \frac{N^{k}_{\ell,t}}{M_{\ell,t}} \frac{ N^{k}_{\ell} \left ( 1 \right ) - N^{k}_{\ell} \left ( 0 \right )}{N^{k}_{\ell,t}} dt \\
	& = \int_{0}^{1} \sum_{k} \frac{N^{k}_{\ell,t}}{M_{\ell,t}} \frac{d \log N^{k}_{\ell,t}}{dt} dt \\
	& = \sum_{k} \int_{0}^{1} \frac{N^{k}_{\ell,t}}{M_{\ell,t}} \frac{d \log N^{k}_{\ell,t}}{dt} dt \\
	& = \sum_{k} \Delta \log N^{k}_{\ell} \int_{0}^{1} \frac{N^{k}_{\ell,t}}{M_{\ell,t}} \frac{d \log N^{k}_{\ell,t}}{dt} \frac{1}{\Delta \log N^{k}_{\ell}} dt
\end{align}
The first equality uses the Fundamental Theorem of Calculus. The second and fourth equalities apply the chain rule. The third equality multiplies and divides by $N^{k}_{\ell,t}$. Now define the mean-value weight as
\begin{align}
	\overline{L}^{k}_{\ell} & \equiv \int_{0}^{1} \frac{N^{k}_{\ell,t}}{M_{\ell,t}} \frac{d \log N^{k}_{\ell,t}}{dt} \frac{1}{\Delta \log N^{k}_{\ell}} dt \\
	& = \int_{0}^{1} \frac{N^{k}_{\ell,t}}{M_{\ell,t}} \frac{ \Delta N^{k}_{\ell}}{N^{k}_{\ell,t}} \frac{1}{\Delta \log N^{k}_{\ell}} dt \\
	& = \int_{0}^{1} \frac{1}{M_{\ell,t}} \frac{\Delta N^{k}_{\ell}}{\Delta \log N^{k}_{\ell}} dt \\
	& = \frac{\Delta N^{k}_{\ell}}{\Delta \log N^{k}_{\ell}} \int_{0}^{1} \frac{1}{M_{\ell,t}} dt \\
	& = \frac{\Delta N^{k}_{\ell}}{\Delta \log N^{k}_{\ell}} \frac{\Delta \log M_{\ell}}{\Delta M_{\ell}}
\end{align}
Thus,
\begin{equation}
	\overline{L}^{k}_{\ell} = \frac{ \Delta N^{k}_{\ell}}{\Delta M_{\ell}} \frac{\Delta \log M_{\ell}}{ \Delta \log N^{k}_{\ell}}
\end{equation}
To summarize,
\begin{equation}
	\log M_\ell \left ( 1 \right ) - \log M_\ell \left ( 0 \right ) = \sum_{k} \overline{L}^{k}_{\ell} \Delta \log N^{k}_{\ell}
\end{equation}
Because $N^k_{\ell,t}$ is continuous on $\left [ 0,1 \right ]$, the mean-value theorem for integrals states that there exists a point $t^*_\ell \in \left ( 0,1 \right )$ such that $\frac{N^{k}_{\ell,t}}{M_{\ell,t}} = \frac{ N^{k}_{\ell,t^*_\ell}}{M_{\ell,t^*_\ell}}$. A solution that is both second-order-accurate and pragmatic is the mid-point value:
\begin{equation}
	\overline{L}^k_{\ell} \approx \frac{N^{k}_{\ell} \left ( 0 \right ) + N^{k}_{\ell} \left ( 1 \right )}{\sum_{m} \left [ N^{m}_{\ell} \left ( 0 \right ) + N^{m}_{\ell} \left ( 1 \right ) \right ]} \equiv \widetilde{L}^k_{\ell}
\end{equation}
Combining previous derivations,
\begin{equation}
	\log M_\ell \left ( 1 \right ) - \log M_\ell \left ( 0 \right ) \approx \sum_{k} \widetilde{L}^{k}_{\ell} \Delta \log N^{k}_{\ell}
\end{equation}
Finally, the difference between log inverse housing demand in the two treatment states is
\begin{equation}
	\Delta \log P_\ell \approx \frac{1}{\eta} \sum_{k} \widetilde{L}^{k}_{\ell} \Delta \log N^{k}_{\ell}
\end{equation}
Finally, we divide both sides by the proposed change in log school district spending:
\begin{equation}
	\frac{\Delta \log P_\ell}{\Delta \log G_j} \approx \frac{1}{\eta} \sum_{k} \widetilde{L}^{k}_{\ell} \frac{\Delta \log N^{k}_{\ell}}{\Delta \log G_j} \label{eq_housingdemand}
\end{equation}

\appsubsubsection{Housing Units}

In any location $\ell$, the equilibrium number of housing units is
\begin{equation}
	\log H_\ell = \lambda + \eta \log P_\ell + B_\ell
\end{equation}
We wish to compute $\Delta \log H_\ell \equiv \log H_\ell \left ( 1 \right ) - \log H_\ell \left ( 0 \right )$. Trivially,
\begin{equation}
	\Delta \log H_\ell = \eta \Delta \log P_\ell
\end{equation}
Finally, we divide both sides by the proposed change in log school district spending:
\begin{equation}
	\frac{\Delta \log H_\ell}{\Delta \log G_j} = \eta \frac{\Delta \log P_\ell}{\Delta \log G_j} \label{eq_housingsupply}
\end{equation}

\appsubsubsection{Balanced Budget}

In any location $\ell$, the balanced budget condition is
\begin{equation}
	\log G_\ell = \log \tau_\ell + \log P_\ell + \log H_\ell
\end{equation}
We wish to compute $\Delta \log G_\ell \equiv \log G_\ell \left ( 1 \right ) - \log G_\ell \left ( 0 \right )$. Trivially,
\begin{equation}
	\Delta \log G_\ell = \Delta \log \tau_\ell + \Delta \log P_\ell + \Delta \log H_\ell
\end{equation}
Finally, we divide both sides by the proposed change in log school district spending:
\begin{equation}
	\frac{\Delta \log G_\ell}{\Delta \log G_j} = \frac{\Delta \log \tau_\ell}{\Delta \log G_j} + \frac{\Delta \log P_\ell}{\Delta \log G_j} + \frac{\Delta \log H_\ell}{\Delta \log G_j} \label{eq_balancedbudget}
\end{equation}

\appsubsection{Identification with Regression Discontinuity Estimands}

We now translate the elasticities obtained above into a system of linear equations, where the unknowns are structural parameters and the known terms correspond to regression discontinuity estimands. This mapping is obtained by taking expectations with respect to the joint distribution of the model's unobservables and conditioning on $S_j = 0.5$, under which regression discontinuity estimands identify weighted averages of elasticities.

\appsubsubsection{Household Supply}

The elasticity of household supply in location $j$ with respect to a change in school district expenditures in location $j$ (equation \ref{eq_logpopmass_own}) is
\begin{align}
	& \frac{\Delta \log N^k_j}{\Delta \log G_j} \notag \\
	& \approx \left ( 1 - \frac{\widetilde{N}^k_{j}}{\sigma^k} \right ) \left ( \frac{\alpha^k}{\theta^k} - \frac{\chi \alpha^k}{\theta^k} \frac{\Delta \log N_j}{\Delta \log G_j} - \frac{\gamma^k \rho^k_j}{\theta^k} \frac{\Delta \log P_j}{\Delta \log G_j} - \frac{\gamma^k \rho^k_j}{\theta^k} \frac{\Delta \log \left ( 1 + \tau_j \right )}{\Delta \log G_j} \right ) \notag \\
	& - \sum_{\ell \neq j} \frac{\widetilde{N}^k_{\ell}}{\sigma^k}  \left ( \frac{\alpha^k}{\theta^k} \frac{\Delta \log G_\ell}{\Delta \log G_j} - \frac{\chi \alpha^k}{\theta^k} \frac{\Delta \log N_\ell}{\Delta \log G_j} - \frac{\gamma^k \rho^k_\ell}{\theta^k} \frac{\Delta \log P_\ell}{\Delta \log G_j} - \frac{\gamma^k \rho^k_\ell}{\theta^k} \frac{\Delta \log \left ( 1 + \tau_\ell \right )}{\Delta \log G_j} \right )
\end{align}
Taking expectations of both sides with respect to the joint probability distribution of the unobservables and conditioning on the running variable being equal to the cutoff yields the following equation:
\begin{align}
		\E \left [ \frac{\Delta \log N^k_j}{\Delta \log G_j} \Bigg | S_j = 0.5 \right ] & = \frac{\alpha^k}{\theta^k} \times \E \left [ \left ( 1 - \frac{\widetilde{N}^k_{j}}{\sigma^k} \right ) \Bigg | S_j = 0.5 \right ] \notag \\
		& - \frac{\chi \alpha^k}{\theta^k} \times \E \left [ \left ( 1 - \frac{\widetilde{N}^k_{j}}{\sigma^k} \right ) \frac{\Delta \log N_j}{\Delta \log G_j} \Bigg | S_j = 0.5 \right ] \notag \\
		& - \frac{\gamma^k}{\theta^k} \times \E \left [ \rho^k_j \left ( 1 - \frac{\widetilde{N}^k_{j}}{\sigma^k} \right ) \frac{\Delta \log P_j}{\Delta \log G_j} \Bigg | S_j = 0.5 \right ] \notag \\
		& - \frac{\gamma^k}{\theta^k} \times \E \left [ \rho^k_j \left ( 1 - \frac{\widetilde{N}^k_{j}}{\sigma^k} \right ) \frac{\Delta \log \left ( 1 + \tau_j \right )}{\Delta \log G_j} \Bigg | S_j = 0.5 \right ] \notag \\
		& - \frac{\alpha^k}{\theta^k} \times \sum_{\ell \neq j} \E \left [ \frac{\widetilde{N}^k_{\ell}}{\sigma^k} \frac{\Delta \log G_\ell}{\Delta \log G_j} \Bigg | S_j = 0.5 \right ] \notag \\
		& + \frac{\chi \alpha^k}{\theta^k} \times \sum_{\ell \neq j} \E \left [ \frac{\widetilde{N}^k_{\ell}}{\sigma^k} \frac{\Delta \log N_\ell}{\Delta \log G_j} \Bigg | S_j = 0.5 \right ] \notag \\
		& + \frac{\gamma^k}{\theta^k} \times \sum_{\ell \neq j} \E \left [ \rho^k_\ell \frac{\widetilde{N}^k_{\ell}}{\sigma^k} \frac{\Delta \log P_\ell}{\Delta \log G_j} \Bigg | S_j = 0.5 \right ] \notag \\
		& + \frac{\gamma^k}{\theta^k} \times \sum_{\ell \neq j} \E \left [ \rho^k_\ell \frac{\widetilde{N}^k_{\ell}}{\sigma^k} \frac{\Delta \log \left ( 1 + \tau_\ell \right )}{\Delta \log G_j} \Bigg | S_j = 0.5 \right ] \label{eq_appx_popown}
\end{align}
For any location $j' \neq j$, analogous derivations yield the following equation:
\begin{align}
		\E \left [ \frac{\Delta \log N^k_{j'}}{\Delta \log G_j} \Bigg | S_j = 0.5 \right ] & = \frac{\alpha^k}{\theta^k} \times \E \left [ \left ( 1 - \frac{\widetilde{N}^k_{j'}}{\sigma^k} \right ) \frac{\Delta \log G_{j'}}{\Delta \log G_j} \Bigg | S_j = 0.5 \right ] \notag \\
		& - \frac{\chi \alpha^k}{\theta^k} \times \E \left [ \left ( 1 - \frac{\widetilde{N}^k_{j'}}{\sigma^k} \right ) \frac{\Delta \log N_{j'}}{\Delta \log G_j} \Bigg | S_j = 0.5 \right ] \notag \\
		& - \frac{\gamma^k}{\theta^k} \times \E \left [ \rho^k_{j'} \left ( 1 - \frac{\widetilde{N}^k_{j'}}{\sigma^k} \right ) \frac{\Delta \log P_{j'}}{\Delta \log G_j} \Bigg | S_j = 0.5 \right ] \notag \\
		& - \frac{\gamma^k}{\theta^k} \times \E \left [ \rho^k_{j'} \left ( 1 - \frac{\widetilde{N}^k_{j'}}{\sigma^k} \right ) \frac{\Delta \log \left ( 1 + \tau_{j'} \right )}{\Delta \log G_j} \Bigg | S_j = 0.5 \right ] \notag \\
		& - \frac{\alpha^k}{\theta^k} \times \E \left [ \frac{\widetilde{N}^k_{j}}{\sigma^k} \Bigg | S_j = 0.5 \right ] \notag \\
		& - \frac{\alpha^k}{\theta^k} \times \sum_{\ell \neq j,j'} \E \left [ \frac{\widetilde{N}^k_{\ell}}{\sigma^k} \frac{\Delta \log G_\ell}{\Delta \log G_j} \Bigg | S_j = 0.5 \right ] \notag \\
		& + \frac{\chi \alpha^k}{\theta^k} \times \sum_{\ell \neq j'} \E \left [ \frac{\widetilde{N}^k_{\ell}}{\sigma^k} \frac{\Delta \log N_\ell}{\Delta \log G_j} \Bigg | S_j = 0.5 \right ] \notag \\
		& + \frac{\gamma^k}{\theta^k} \times \sum_{\ell \neq j'} \E \left [ \rho^k_{\ell} \frac{\widetilde{N}^k_{\ell}}{\sigma^k} \frac{\Delta \log P_\ell}{\Delta \log G_j} \Bigg | S_j = 0.5 \right ] \notag \\
		& + \frac{\gamma^k}{\theta^k} \times \sum_{\ell \neq j'} \E \left [ \rho^k_{\ell} \frac{\widetilde{N}^k_{\ell}}{\sigma^k} \frac{\Delta \log \left ( 1 + \tau_\ell \right )}{\Delta \log G_j} \Bigg | S_j = 0.5 \right ] \label{eq_appx_popother}
\end{align}

\appsubsubsection{Rental Rate of Housing}

The elasticity of housing demand in location $\ell \in \mathcal{J}$ with respect to a change in school district expenditures in location $j$ (equation \ref{eq_housingdemand}) is
\begin{equation}
	\frac{\Delta \log P_\ell}{\Delta \log G_j} \approx \frac{1}{\eta} \sum_{k} \widetilde{L}^{k}_{\ell} \frac{\Delta \log N^{k}_{\ell}}{\Delta \log G_j}
\end{equation}
Taking expectations of both sides with respect to the joint probability distribution of the unobservables and conditioning on the running variable being equal to the cutoff yields the following equation:
\begin{equation}
		\E \left [ \frac{\Delta \log P_\ell}{\Delta \log G_j} \Bigg | S_j = 0.5 \right ] = \frac{1}{\eta} \times \sum_{k} \E \left [ \widetilde{L}^{k}_{\ell} \frac{\Delta \log N^{k}_{\ell}}{\Delta \log G_j} \Bigg | S_j = 0.5 \right ] 
\end{equation}

\appsubsubsection{Housing Units}

The elasticity of housing supply in location $\ell \in \mathcal{J}$ with respect to a change in school district expenditures in location $j$ (equation \ref{eq_housingsupply}) is
\begin{equation}
	\frac{\Delta \log H_\ell}{\Delta \log G_j} = \eta \frac{\Delta \log P_\ell}{\Delta \log G_j}
\end{equation}
Taking expectations of both sides with respect to the joint probability distribution of the unobservables and conditioning on the running variable being equal to the cutoff yields the following equation:
\begin{equation}
		\E \left [ \frac{\Delta \log H_\ell}{\Delta \log G_j} \Bigg | S_j = 0.5 \right ] = \eta \times \E \left [ \frac{\Delta \log P_\ell}{\Delta \log G_j} \Bigg | S_j = 0.5 \right ] \label{eq_appx_housingsupply}
\end{equation}

\appsubsubsection{Balanced Budget}

The elasticity of school district expenditures in location $\ell \in \mathcal{J}$ with respect to a change in school district expenditures in location $j$ (equation \ref{eq_balancedbudget}) is
\begin{equation}
	\frac{\Delta \log G_\ell}{\Delta \log G_j} = \frac{\Delta \log \tau_\ell}{\Delta \log G_j} + \frac{\Delta \log P_\ell}{\Delta \log G_j} + \frac{\Delta \log H_\ell}{\Delta \log G_j}
\end{equation}
Taking expectations of both sides with respect to the joint probability distribution of the unobservables and conditioning on the running variable being equal to the cutoff yields the following equation:
\begin{align}
		\E \left [ \frac{\Delta \log G_\ell}{\Delta \log G_j} \Bigg | S_j = 0.5 \right ] & = \E \left [ \frac{\Delta \log \tau_\ell}{\Delta \log G_j} \Bigg | S_j = 0.5 \right ] \notag \\
		& + \E \left [ \frac{\Delta \log P_\ell}{\Delta \log G_j} \Bigg | S_j = 0.5 \right ] + \E \left [ \frac{\Delta \log H_\ell}{\Delta \log G_j} \Bigg | S_j = 0.5 \right ]
\end{align}

\appsection{Statistical Inference on Structural Parameters}\label{appx_sec_inference}

In this section, we provide details on statistical inference for the structural parameters that govern household preferences and the elasticity of housing supply.

\appsubsection{Household Preferences}

Assume there are two jurisdictions ($| \mathcal{J} | = 2$) and set $\chi = 1$. For compactness, let $\theta_n$ denote the $n$th regression discontinuity estimand, with the numbering following the order of appearance in equations \eqref{eq_appx_popown}-\eqref{eq_appx_popother}. The system of equations can then be written as
\begin{align}
\begin{cases}
	\theta_{1}  & = \alpha \left ( \theta_{2}-\theta_{3}-\theta_{6}+\theta_{7} \right ) + \gamma \left ( -\theta_{4}-\theta_{5}+\theta_{8}+\theta_{9} \right )\\
	\theta_{10} & = \alpha \left ( \theta_{11}-\theta_{12}-\theta_{15}+\theta_{16} \right ) + \gamma \left ( -\theta_{13}-\theta_{14}+\theta_{17}+\theta_{18} \right )
\end{cases}
\end{align}
Define the intermediate sums
\begin{align}
	\psi_{1} & \equiv \theta_{2}-\theta_{3}-\theta_{6}+\theta_{7} & \xi_{1} & \equiv -\theta_{4}-\theta_{5}+\theta_{8}+\theta_{9},\\
	\psi_{2} & \equiv \theta_{11}-\theta_{12}-\theta_{15}+\theta_{16} & \xi_{2} & \equiv -\theta_{13}-\theta_{14}+\theta_{17}+\theta_{18}
\end{align}
With these definitions, the system can be expressed in matrix form as
\begin{align}
	\begin{bmatrix}
		\psi_{1} & \xi_{1}\\
		\psi_{2} & \xi_{2}
	\end{bmatrix}
	\begin{bmatrix}
		\alpha\\
		\gamma
	\end{bmatrix}
	=
	\begin{bmatrix}
		\theta_{1}\\ \theta_{10}
	\end{bmatrix}
\end{align}
where the determinant of the coefficient matrix is given by $\Delta \equiv \psi_{1} \xi_{2} - \psi_{2} \xi_{1} \neq 0$. The solution to the system is then
\begin{equation}
	\alpha = \frac{\theta_{1} \xi_{2} - \theta_{10} \xi_{1}}{\Delta} \qquad \gamma  = \frac{\psi_{1} \theta_{10} - \psi_{2} \theta_{1}}{\Delta}
\end{equation}
Let $\widehat{\theta} = \left [ \widehat{\theta}_{1},\dots, \widehat{\theta}_{18} \right ]'$ denote the vector of estimated regression discontinuity parameters, with associated variance-covariance matrix $\widehat{\Sigma}_{\theta}$. We compute the Jacobian matrix of $\left [ \alpha, \gamma \right ]'$ with respect to the vector of underlying estimands, yielding
\begin{align}
\textsc{J} = \frac{1}{\Delta}
\begin{bmatrix}
	\xi_{2}                           & -\psi_{2} \\[2pt]
	-\alpha \xi_{2}                    &  \theta_{10}-\gamma \xi_{2} \\[2pt]
	\alpha \xi_{2}                    & -\theta_{10}+\gamma \xi_{2} \\[2pt]
	\theta_{10}-\alpha \psi_{2}             & -\gamma \psi_{2} \\[2pt]
	\theta_{10}-\alpha \psi_{2}             & -\gamma \psi_{2} \\[2pt]
	\alpha \xi_{2}                    & -\theta_{10}+\gamma \xi_{2} \\[2pt]
	-\alpha \xi_{2}                    &  \theta_{10}-\gamma \xi_{2} \\[2pt]
	-\theta_{10}+\alpha \psi_{2}             &  \gamma \psi_{2} \\[2pt]
	-\theta_{10}+\alpha \psi_{2}             &  \gamma \psi_{2} \\[2pt]
	-\xi_{1}                           &  \psi_{1} \\[2pt]
	\alpha \xi_{1}                    & -\theta_{1}+\gamma \xi_{1} \\[2pt]
	-\alpha \xi_{1}                    &  \theta_{1}-\gamma \xi_{1} \\[2pt]
	-\theta_{1}+\alpha \psi_{1}              &  \gamma \psi_{1} \\[2pt]
	-\theta_{1}+\alpha \psi_{1}              &  \gamma \psi_{1} \\[2pt]
	-\alpha \xi_{1}                    &  \theta_{1}-\gamma \xi_{1} \\[2pt]
	\alpha \xi_{1}                    & -\theta_{1}+\gamma \xi_{1} \\[2pt]
	\theta_{1}-\alpha \psi_{1}              & -\gamma \psi_{1} \\[2pt]
	\theta_{1}-\alpha \psi_{1}              & -\gamma \psi_{1}
\end{bmatrix}'
\end{align}
where each row corresponds to a partial derivative with respect to $\theta_n$ for $n = 1, \dots, 18$. Let $\left [ \widehat{\alpha}, \widehat{\gamma} \right ]'$ denote the estimate of $\left [ \alpha, \gamma \right ]'$. Substituting estimated parameters into the Jacobian yields the matrix $\widehat{\textsc{J}}$. By an application of the Delta method, the estimated variance-covariance matrix of $\left [ \widehat{\alpha}, \widehat{\gamma} \right ]'$ is
\begin{align}
	\begin{bmatrix}
		\Var \left [ \widehat{\alpha} \right ] & \Cov \left [ \widehat{\alpha}, \widehat{\gamma} \right ]\\
		\Cov \left [ \widehat{\alpha}, \widehat{\gamma} \right ] & \Var \left [ \widehat{\gamma} \right ]
	\end{bmatrix}
	\approx
	\widehat{\textsc{J}} \widehat{\Sigma}_{\theta} \widehat{\textsc{J}}'
\end{align}
Finally, applying the Delta method to the ratio $\alpha \slash \gamma$, the variance of the estimator $\widehat{\alpha} / \widehat{\gamma}$ is approximated by
\begin{align}
	\Var \left [ \widehat{\alpha} \slash \widehat{\gamma} \right ] \approx
	\begin{bmatrix}
		1 \slash \widehat{\gamma} & - \widehat{\alpha} \slash \widehat{\gamma}^2
	\end{bmatrix}
	\widehat{\textsc{J}} \widehat{\Sigma}_{\theta} \widehat{\textsc{J}}'
	\begin{bmatrix}
		1 \slash \widehat{\gamma} \\
		- \widehat{\alpha} \slash \widehat{\gamma}^2
	\end{bmatrix}
\end{align}

\appsubsection{Elasticity of Housing Supply}

As described in equation \eqref{eq_appx_housingsupply}, the housing supply elasticity $\eta$ is point identified as the ratio of two regression discontinuity estimands, whose outcomes are housing quantity $H$ and housing price $P$, respectively. Let $\theta_H$ and $\theta_P$ denote these estimands, with corresponding estimators $\widehat{\theta}_H$ and $\widehat{\theta}_P$. By the Delta method, the estimated variance of $\widehat{\eta}$ is
\begin{align}
	\Var \left [ \widehat{\eta} \right ] \approx
	\begin{bmatrix}
		1 \slash \widehat{\theta}_P & - \widehat{\theta}_H \slash \widehat{\theta}_P^2
	\end{bmatrix}
	\begin{bmatrix}
		\Var \left [ \widehat{\theta}_H \right ] & \Cov \left [ \widehat{\theta}_H, \widehat{\theta}_L \right ] \\
		\Cov \left [ \widehat{\theta}_H, \widehat{\theta}_L \right ] & \Var \left [ \widehat{\theta}_P \right ] 
	\end{bmatrix}
	\begin{bmatrix}
		1 \slash \widehat{\theta}_P \\
		- \widehat{\theta}_H \slash \widehat{\theta}_P^2
	\end{bmatrix}
\end{align}

\end{document}